\documentclass[a4paper,12pt]{thesis}
\usepackage[dutch,british]{babel}
\usepackage{amsmath,amsfonts,amsthm,amssymb,color,graphicx,epstopdf,tikz,standalone,cite}

\newcommand{\diff}[2]{\frac{\text{d} #1}{\text{d} #2}}
\newcommand{\ddiff}[2]{\frac{\text{d}^2 #1}{\text{d} {#2}^2}}

\newcommand{\pdiff}[2]{\frac{\partial #1}{\partial #2}}

\newcommand{\pddiff}[3]{\frac{\partial^2 #1}{\partial #2 \partial #3}}
\newcommand{\dee}{\text{d}}
\newcommand{\intdee}{\text{ d}}
\newcommand{\here}{\boldsymbol{p}}
\newcommand{\there}{\boldsymbol{q}}
\newcommand{\tg}[2]{\text{T}^{\phantom{*}}_{#1}\mathbb{#2}}
\newcommand{\ctg}[2]{\text{T}^*_{#1}\mathbb{#2}}
\newcommand{\tgb}[1]{\text{T}\mathbb{#1}}
\newcommand{\ctgb}[1]{\text{T}^*\mathbb{#1}}
\newcommand{\define}{\overset{\text{def}.}{=}}
\newcommand{\notation}{\overset{\text{not}.}{=}}

\newtheorem{Invariance}{Theorem}
\newtheorem{Hessian}[Invariance]{Theorem}
\newtheorem{Identify}[Invariance]{Theorem}

\begin{document}
\pagenumbering{gobble}
\selectlanguage{dutch}
\begin{titlepage}
\begin{center}
\vspace{0.8cm}
\includegraphics[scale=0.5]{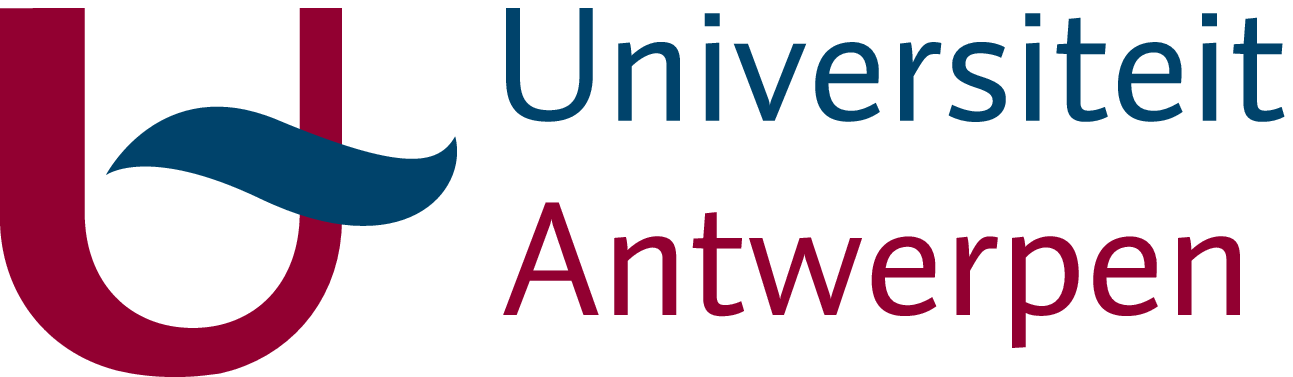}\\[0.1cm]
Faculteit Wetenschappen\\
Departement Fysica\\[2.5cm]
\Large
\textbf{Extension of information geometry for modelling non-statistical systems}\\[1cm]
\textbf{Uitbreiding van de informatiemeetkunde voor het modelleren van niet-statistische systemen}\\[2.5cm]
\normalsize
Proefschrift voorgelegd tot het behalen van de graad van\\[0.3cm]
\textbf{doctor in de wetenschappen}\\[0.3cm]
aan de Universiteit Antwerpen, te verdedigen door\\[0.3cm]
\textbf{Ben Anthonis}
\vfill
\end{center}
\textbf{Promotoren}\\
Prof.\ dr.\ Jan Naudts\\
Prof.\ dr.\ Jacques Tempere\hfill Antwerpen, 2014
\end{titlepage}
\newpage
%\null\thispagestyle{empty}
\begin{center}{\color{white}Page intentionally left blank}\end{center}
\newpage

\pagenumbering{gobble}
\section*{Dankwoord}
Hoewel het mijn naam is die in grote letters op de kaft van dit proefschrift pronkt, is ook dit doctoraat een onderneming geweest waar heel wat mensen rechtstreeks of onrechtstreeks mee te maken hadden. Daarom wil ik hen hier bedanken.\\
De eerste in deze lange lijst is mijn promotor, prof.\ Jan Naudts. Deze verhandeling is reeds de derde die ik schrijf onder zijn begeleiding. Het is dan ook van hem dat ik het meest geleerd heb over de wiskundige natuurkunde. Hoewel onze meningen over hoe een bepaald vraagstuk moest worden aangepakt meer dan eens volledig verschilden, heeft hij mij toch het vertrouwen en de vrijheid geschonken om mijn eigen ding te doen. Ook stond hij me telkens toe om tijd te investeren in het onderwijs, of het nu als onderwijs\-assistent was of als ombudspersoon. Een dergelijke promotor is niet alle doctorandi gegeven.\\
Mijn co-promotor, prof.\ Jacques Tempere, was niet even nauw betrokken bij dit onderzoek als mijn promotor, maar toch was ook zijn steun van onmiskenbaar belang. Zo was hij het die me de kans gaf om aan een doctoraat te beginnen wanneer de groep Wiskundige Natuurkunde hier de fondsen niet voor had. De aanstekelijkheid van zijn enthousiasme bleek niet beperkt te zijn tot de fysica. Hij vormt dan ook het levende bewijs dat de universiteit geen ivoren toren is en nog steeds in staat is om haar studenten meer te leren dan alleen maar vakspecifieke kennis.\\
For other interesting discussions I would like to thank prof.\ Hiroki Suyari, prof.\ Atsumi Ohara, prof.\ Hiroshi Matsuzoe and prof.\ Tatsuaki Wada, who have honoured prof.\ Naudts and me with work visits. Also the members of my jury, prof.\ Giovanni Pistone, prof.\ Christian Maes, prof.\ Sandra Van Aert and prof.\ Etienne Goovaerts, whose remarks were very useful in preparing the final version of this dissertation.\\
De collega-studenten en -doctoraatsstudenten die ik de afgelopen jaren heb gehad vormen intussen een redelijk grote groep. Toch zijn velen onder hen vrienden geworden. In de hoop niemand te vergeten zijn dit: Amy, Anne, Bert, Bert, Bob, Damiaan, Dries, Enya, Erik, Giovanni, Jeroen, Katleen, Katrijn, Maarten, Mathias, Nick, Nick, Selma, Stijn, Sven, Thomas, Tobias, Tych\'e, Wim, Winny en Wouter. Geert en Inge, wiens masterthesis ik heb mogen begeleiden, horen mee in deze categorie thuis. Ook kunnen prof.\ Michiel Wouters, prof.\ Dirk Callebaut en Onur Umucal\i lar hier niet ongenoemd gelaten worden. Drie vermeldingen zijn weggelegd voor vrienden die als ere-collega's zouden kunnen gelden: Asbj\o rn, Brent en Yannick.\\
Vanzelfsprekend zijn er nog andere mensen met wie ik heb samengewerkt de afgelopen jaren en daarom verdienen om in dit dankwoord voor te komen. Dit zijn: prof.\ Dirk Lamoen en prof.\ Jan Sijbers, voor wie ik oefeningenlessen heb mogen verzorgen; Hilde en Gwendolyn, voor het samenwerken daarbij; prof.\ Paul Scheunders, prof.\ Joke Hadermann en prof.\ Sabine Van Doorslaer van wiens onder\-wijs-, examen- en zelfevaluatie-commissies ik deel mocht uitmaken; Werner Peeters, voor de interessante discussies over wiskunde, onderwijs en de combinatie van beiden; Brigitte Brusselmans en Pieter Caris, voor al de keren dat ik op hen kon rekenen als ombudspersoon; Annick Van Son, voor de college- en examenroosters die ze telkens voor de fysicastudenten heeft gemaakt; Lieff Serrien en Hilde Evans om mij de weg te wijzen in het doolhof van de universitaire administratie.\\
Dan zijn er ook nog collega's met wie ik niet heb samengewerkt in de strikte betekenis van het woord maar die toch niet onbelangrijk zijn: (opnieuw) Lieff en alle andere mensen die hebben geholpen aan de organisatie van de jaarlijkse departementsreceptie; Quinten, die de verantwoordelijkheid daarvoor heeft overgenomen; (opnieuw) Nick Verhelst, die zijn tweede plaatsje hier kan opeisen door het verderzetten van mijn taak als ombudspersoon; Eveline Jacques, Els Grieten, Dirk Valkenborg en prof.\ Bart Goethals, die de anders zo saaie trein- en busritten telkens een pak plezieriger maakten; Sonja en Virginie, voor de goedemorgen waarmee ik de afgelopen vier jaar mijn dag mocht beginnen en voor de belangrijke maar soms ondankbare taak van het proper houden van onze kantoren.\\
De allergrootste groep van mensen die ik hier aan bod wil laten komen is die van de honderden studenten die ``mijn'' oefeningenlessen hebben bevolkt. Ik heb me geweldig geamuseerd met het lesgeven aan jullie. Een speciale vermelding behoort hier toe aan Karen, die ik sinds de afgelopen herfst tot mijn beste vrienden mag rekenen en wie me telkens een \emph{pep talk} gaf wanneer de laatste loodjes begonnen te wegen.\\
Wie zeker niet vergeten mag worden in dit dankwoord is mijn familie. In de eerste plaats houdt dit mijn ouders in: Willy en Gerda---moeke en voke voor mij---voor alle steun en aanmoediging, voor de interesse in mijn onderzoek, voor het geduld dat zij met mij hadden toen ik nog maar eens overliep van enthousiasme en natuurlijk ook voor alle dingen die ouders doen voor hun kinderen en waarvoor die laatsten hen nooit genoeg kunnen bedanken. Ook zonder Va Stanne en mijn nonkel (die ik niet zo noem) Dominiek, zou dit dankwoord niet volledig kunnen zijn.\\
Ten slotte is er nog \'e\'en iemand die hier niet vermeld is. Iemand die misschien nog harder naar het eindresultaat zou hebben uitgekeken dan ikzelf en die dolgraag aanwezig zou geweest zijn op de verdediging van dit werk. Het heeft echter niet mogen zijn. Daarom, Moe Simone, draag ik deze thesis op aan jou.

%\newpage
%\null\thispagestyle{empty}
%\begin{center}{\color{white}Page intentionally left blank}\end{center}
\newpage

\pagenumbering{gobble}
\section*{Nederlandstalige samenvatting}
In deze doctoraatsverhandeling wordt een abstract meetkundig formalisme ontwikkeld voor het modelleren van data die zeer algemeen mogen zijn. Dit wiskundig kader werd het ``data set model formalisme'' genoemd en is ge\"inspireerd op de informatie\-meet\-kunde. Het modelleren---of fitten---gebeurt met behulp van een divergentiefunctie: een veralgemeende afstandsmaat waarvan de relatieve entropie waarschijnlijk het bekendste voorbeeld is. Door te eisen dat deze modellen een differentiaalmeetkundige vari\"eteit vormen kunnen zij worden uitgerust met een meetkundige structuur dewelke volgt uit de divergentiefunctie. Het belang van deze structuur is dat zij toestaat de belangrijkste eigenschappen van het modelleringsproces kwantitatief te beschrijven. Centraal hierin staat de zogenaamde Hessiaanse structuur, die het mogelijk maakt de Riemanniaanse metriek van de modelvari\"eteit af te leiden uit een familie van scalaire functies. Dit vereist de keuze van een vlakke affiene connectie, waarvoor tevens een constructie uit de divergentiefunctie gegeven wordt.\\
Het data set model formalisme biedt een aantal voordelen ten opzichte van bestaande modelleringstechnieken. Het belangrijkste daarvan is de grote wis\-kundige flexibiliteit, die te danken is aan het gebruik van differentiaal\-meet\-kunde. Vanuit theoretisch oogpunt is vooral de mogelijkheid om een breed scala aan verschillende modellerings\-problemen in eenzelfde kader onder te brengen interessant. Zo omvat het formalisme de informatie\-meet\-kunde maar eveneens regressiemodellen en elementen van de kwantumstatistiek. Ook voor praktische toepassingen is dit werk interessant. De reden hiervoor is dat de ontwikkelde technieken toestaan gegeven data via meetkundige me\-tho\-des te modelleren, zelfs wanneer de modellen kwalitatief verschillen van de data. Dit opent perspectieven voor toepassingen binnen het vakgebied van machinaal leren (\emph{machine learning}) en voor het ontwikkelen van een uitbreiding van de informatietheorie voor kwantum\-systemen.\\
Daar het formalisme een rechtstreekse veralgemening van de informatie\-meet\-kunde inhoudt, kan het gebruik er van ook in die discipline van nut zijn. De belangrijkste innovatie is hierbij dat het nieuwe formalisme toestaat de volledige ruimte van kansverdelingen over een verzameling als model te nemen. Dit kan in de informatiemeetkunde niet op een zinvolle manier gebeuren aangezien de kansmaten daar noodzakelijkerwijs de rol van data vervullen. Deze verzameling tevens als model gebruiken zou het modelleringsproces bijgevolg redundant maken. Bijkomend wordt in dit proefschrift een eenvoudige techniek afgeleid voor het beantwoorden van de vraag of een statistisch model tot de exponenti\"ele familie behoort. De eenvoud van deze werkwijze doet vermoeden dat zij reeds eerder door andere onderzoekers ontdekt is. Desalniettemin ontbreken alle verwijzingen hiernaar in de informatie\-meet\-kundige referentiewerken. Dit contrasteert met praktische nut dat zij biedt, zodat een uiteenzetting in dit werk gerechtvaardigd lijkt.\\
De verhandeling vangt aan met een uiteenzetting van de basidee\"en van de differentiaal\-meet\-kunde. De daar besproken begrippen vormen het wiskundig kader waaraan de rest van deze tekst wordt opgehangen. Eveneens werd een kort historisch overzicht van de belangrijkste ontwikkelingen in de informatie\-meet\-kunde opgenomen. Deze twee delen vormen samen de inleiding (Hoofdstukken 2 en 3). De stand van het eigenlijke onderzoek wordt beschreven in Hoofdstuk 4. Het data set model formalisme wordt daarin in detail behandeld. Ook de meetkundige structuur wordt gedefinieerd en de studie van haar verband met het modelleringsproces vindt daar plaats. Het vijfde en laatste eigenlijke hoofdstuk richt zich tot een aantal toepassingen, dewelke vooral illustratief van aard zijn.

\newpage
\selectlanguage{british}
\tableofcontents
\newpage\pagenumbering{arabic}

\chapter{Opening Chapter}
\section{Situating the dissertation}
An often recurring class of problems in science takes the form of describing data by a simplified, parametrised model. Possibly the best-known approach in physics to this kind of modelling is through the method of maximum entropy \cite{Jaynes,Jaynes2,BarndorfNielsen,Naudts}. Its most common use is probably in the context of the thermodynamic canonical ensemble. To apply this method one defines a suitable entropy function and one or more measurable quantities (often called Hamiltonians or extensive variables) on the space of all probability distributions over the possible states of the system. %Well-known examples of Hamiltonians include the total energy and the magnetisation.
Those distributions which predict the same values for each of the Hamiltonians are grouped together in subsets, each of which are represented by their element exhibiting the highest entropy---the so-called model distributions. To identify the model distribution within such a subset is an optimisation problem with constraints, and the Lagrange multipliers appearing there serve also as labels for these distributions.
%For a physical system where the Hamiltonians are the energy and the magnetisation, the parameters (also called intensive variables) are related to the temperature and the external magnetic field.
The task of modelling the experimental data is thus reduced to determining the values of the parameters consistent with the values of the extensive variables that are measured.\\
While this dissertation does not contain a criticism of the maximum entropy method, a very different and less widely known approach to modelling is studied herein. More specifically, this work is concerned with the development of a framework meant to include a broad range of modelling problems and which is based entirely on geometric foundations: the data set model formalism. The geometry is derived from general divergence functions which quantify how well data is described by a given element of the model. An example of such a divergence function, found in statistics and statistical physics, is the relative entropy or Kullback-Leibler divergence \cite{Naudts,KullbackLeibler}.\\
The motivation to develop such an abstract geometrical theory comes from a criticism made against the mainstream formulation of information theory. This discipline is important to modelling problems as it is intimately related to the question of how to make meaningful statements regarding the quality of the modelling procedure. In particular the fact that information theory is based on probability theory has received critical attention from some authors. This school of thought appears to have started in the work of Ingarden and Urbanik. They proposed an approach to information theory based on Boolean rings and explained their motivation for doing so as \cite{IngUrban}
\begin{quote}
\emph{\ldots information seems intuitively a much simpler and more elementary notion than that of probability \ldots [it] represents a more primary step of knowledge than that of cognition of probability}.
\end{quote}
Another proponent of this idea is none other than Andrej Kolmogorov, who asserted \cite{Kolmogorov}
\begin{quote}
\emph{Information theory must precede probability theory, and not be based
on it.}
\end{quote} 
It must be remarked that the paper in which this statement can be found is devoted to arguing, amongst other things, that the basis of information theory must be combinatorial in nature. While a geometric theory does not have a close relation to combinatorics, the radically different viewpoint it offers, especially when sufficiently abstract, could present new and useful insights into the essence of information even if Kolmogorov should be definitively vindicated.\\
Such insights of information without probability might also prove to be valuable in physics. An outspoken advocate of the importance of information theory there was John Archibald Wheeler. He stated that \cite{Wheeler}
\begin{quote}
\emph{All things physical are information theoretic in origin \ldots and information gives rise to physics}.
\end{quote}
One author to have made an attempt---be it a controversial one---to express Wheeler's idea is Roy Frieden, who tries to base all of physics on an information theoretical foundation \cite{Frieden}. However, it is especially in quantum theory that Wheeler's viewpoint has found many adherents. This is testified by the numerous investigations into the question of whether or not it is possible to frame quantum theory in purely information-theoretic terms, see for example \cite{Fuchs,Zeilinger,Clifton,Chiribella,Masanes}. The characterisation of information used by these researchers needs to take into account the laws governing microscopic physics, including the laws of probability. Due to the presence of incompatible observables however, quantum theory requires the notion of conditional probability to be discarded. This implies that quantum information theory must be appreciably different from it classical counterpart, despite being drafted in similar mathematical terms \cite{NielsenChang,Petz}. One potential solution to this problem could lie precisely in the development of a sufficiently abstract formalism, such as the one introduced here, simultaneously generalising both theories of information.\\
Another argument from quantum physics is related to recent experimental advances in that field, such as those for which Serge Haroche and David Weinland were awarded the 2012 Nobel Prize in Physics. The possibility to perform so-called weak measurements \cite{Aharanov1,Haroche} has received much attention, not only in the experimental physics community but also from researchers interested in the foundations of quantum theory \cite{Aharanov2,Aharanov3} and in quantum information theory \cite{Tsang}. In such a measurement, information about the state of the system can be obtained without collapsing the wave function. Recent work even claims successful direct observations of the wave function itself through a combination of weak measurements and ordinary projective measurements \cite{LundeenNature,LundeenPRL}---a feat considered impossible by the Copenhagen interpretation of quantum mechanics \cite{Hall}. An adaptation of quantum information theory in order to accommodate these findings may therefore be required. The availability of an abstract and thereby flexible framework of information would likely be regarded as a boon for those researchers working out the details of such a transition.\\
Even when no changes to quantum information theory are shown to be needed, the contents of this dissertation may still prove to be useful to that field of study. Newly obtained results---at the time of writing still unpublished---making use of the data set model formalism indicate it may be possible to simultaneously simplify and generalise existing results such as those of Petz on positive-operator valued measures \cite{Petz}.\\
The data set model formalism is for a large part a generalisation of the results of information geometry. This discipline---which is to be introduced more elaborately in a later chapter---is a differential geometric framework for probability theory and statistical models. The generalisation performed in this research consists mainly in removing the limitation to these topics, leading to a very general picture of the modelling process. While such an approach could have the disadvantage of leaving an impression of abstraction and technicality on the reader, it is also believed to be advantageous in the long term. Indeed, it is precisely by exercising this increased level of abstraction that it is hoped that the geometric essence of information will be laid bare without referring to any context-specific properties.\\
Some additionally obtained results may be of use to researchers interested in statistics. In particular, the construction of the data set model geometry offers a convenient method to establish whether or not a parametrised family of probability distributions belongs to the exponential family. In this case, the geometric structure also facilitates the search for the canonical parameters of the family. While the simplicity of this method suggests that it is not an original finding, reference thereto seems to be missing from the information geometry literature.\\%Unlike previous theoretical work in information geometry, this dissertation is thus not a translation of ideas from statistics into the language of differential geometry in order to advance the former field through application of results and techniques of the latter. Rather, this work is a modest attempt to capture the geometric spirit of information theory and to identify some of its more fundamental properties also in a more general setting. If successful, this endeavour may come to offer a greater flexibility to scientists and other researchers in the choice of their theoretical frameworks.\\
Furthermore, a varied range of potential applications is expected in the longer term. This supposition is based on a number of advantageous innovations. Perhaps the most prominent of these is that the parametrised family of models is no longer required to be a subset of the data. It is even allowed for models to be mathematical objects qualitatively different from those data. This is to be contrasted with both the maximum entropy method and with information geometry. For this reason, a possibly less obvious but still rather promising field for applications is that of machine learning. This is a very broad area of ongoing research related to artificial intelligence, data mining and other methods for information processing which have dramatically acquired importance over the last few years. A list of example problems includes curve fitting or the estimation of probability density but also applications less---or less obviously---related to parameter estimation such as fingerprint recognition, Google's page rank algorithm, automatised translation and many more. (See for example \cite{Alpaydin,Smola} for an introduction to this very rich discipline.)

%{\color{blue} SOMETHING ON QIT/QET???}
%This is challenging in an attempt to generalise the existing theory of information geometry as many important definitions and theorems rely precisely upon the notion of conditional probability. For example, the geometrical quantities encountered in information geometry have definitions which are required to be unchanged when the underlying measurable space is mapped to another such space through a sufficient statistic. When conditional probabilities are removed from the theory, as must happen in a quantum setting, it is thus not immediately obvious how to define the necessary quantities. The original goal of this research was thus to construct quantum analogues of certain classical information theoretical notions using a geometrical framework.

\section{Structure of the dissertation}
This opening chapter has summarised the results of the research upon which this dissertation reports and it has sketched the broader context in which these are to be seen. For clarity, a brief explanation of the conventions and notations used in the rest of the text follows shortly hereafter.\\
The following two chapters form the introduction. Chapter 2 gives a short overview of the basic concepts of differential geometry which are part of this dissertation's lexicon: manifolds, vectors, differential forms, the Riemannian metric, as well as affine connections and their curvature. These notions will provide the mathematical language in which the rest of the dissertation is framed. Chapter 3 introduces information geometry along the lines of its historical development. This chapter also contains a brief introduction to divergence functions as they appear in literature and a short overview of applications of differential geometry in thermodynamics. The main chapter of this dissertation is found fourth and therein the theoretical aspects of the data set model formalism are elucidated. The basic elements and assumptions are explained, a geometrical structure for the models is erected and the essential properties of this structure are studied. Chapters 2--4 share a similar internal structure as to make the analogies between them more clear. Selected examples and applications of the formalism are found in the fifth chapter for the purpose of illustration. The end of this dissertation is formed by the conclusion and an outlook of possible future research in this context, as well as a curriculum vitae of the author and the bibliography.
\section{Conventions and notations}
There are some conventions and notations which are used throughout this dissertation. A fairly large part of these are expected to be already known or intuitively clear to the reader. Nevertheless, a short summary of these is presented here.\\
As this dissertation makes elaborate use of differential geometry, one of the most useful conventions is that of Einstein summation. This convention states that when an index appears twice within the same term of an equation, this index is implied to be summed over all of its values. In such pairs, the index will always appear once as an upper index and once as a lower index. Unpaired indices must necessarily appear in every term of an equation. It is conventional for this to mean that the equation holds for all possible values of the unpaired index. For example, it is possible to write down the affine transformation of the vector $\vec{v}$ into $\vec{w}$ by the matrix $A$ and the vector $\vec{b}$ in terms of components as
\begin{align*}
w^i = {a^i}_j v^j + b^i
\end{align*}
rather than the equivalent traditional, but much more cumbersome, notation
\begin{align*}
w^i = \sum_{j=1}^n {a^i}_j v^j + b^i \quad \forall i \in \{1\ldots n\}.
\end{align*}
The convention for unpaired indices---the expression in which they appear holds for all values the index may take---is extended to all quantities and the symbols representing them. Whenever an expression makes use of a symbol which is not specified uniquely, the expression is assumed to hold for every possible concrete object the symbol could represent given any restrictions that may apply. This may sound like a complicated and technical way of introducing quantities. Should the reader be left with this feeling, he or she is reminded that this same convention has already been invoked above when introducing the matrix $A$ and the vectors $\vec{v}$ and $\vec{b}$. A similar convention is well-known from expressions for functions but there it is traditionally obscured by speaking about ``variables''. However, the same result is obtained by agreeing that---for example---the expression
\begin{align*}
f(x) = x\ln(x)
\end{align*}
holds for all possible real numbers $x$ in the domain of the function $f$. This convention is useful to avoid what can sometimes become long enumerations of definitions mostly obvious from context, as these have a habit of making mathematical texts more tedious and less pleasant to read than really necessary.\\
Another situation where a more concise notation is used than the one the reader may be familiar with is partial derivatives with respect to parameters or coordinates. From Chapter 4 onwards, very general divergence functions, denoted by the letter $D$, will be used. These are functions taking a pair $(x,m_\theta)$ of arguments and mapping the pair to a real number $D(x||m_\theta)$. Since the second argument will be an element of a manifold and can be endowed with coordinates, it is possible to differentiate these functions with respect to the coordinates. The traditional notation for such a derivative, evaluated in the point with coordinates $\theta$, would look like
\begin{align*}
\pdiff{}{\xi^k} D(x||m_\xi) \bigg|_{\xi=\theta}.
\end{align*}
Since such expressions will appear frequently, it is preferable to use a more concise notation. Therefore, the quantity above will also be represented by the expression resembling the one for functions of the parameters only,
\begin{align*}
\partial_k D(x||m_\theta).
\end{align*}
For readers who find this notation confusing, it may be useful to keep in mind the analogy with the commonly used expression $f^\prime(a)$, which denotes evaluation of the derivative of a function $f$ in the point $a$, avoiding the use of an auxiliary variable. The traditional notation will still be used where ambiguity could arise from using the shorter alternative.\\
When specific notations are introduced on the fly, this will be denoted by using the symbol $\notation$ as the equality sign---not to be confused with the inequality $\neq$. A similar symbol $\define$ is used to indicate that the equality is in fact also the definition of the quantity being introduced.\\
A final convention has to do with measures over sets. Sometimes it is necessary to sum or integrate over such a set or a subset thereof. This will appear in a number of expressions originating in statistics, for example expressions for expectation values. For this the integration sign shall be used, even when the measurable set over which the integration takes place may be discrete. Unless explicitly stated otherwise, $\dee x$ represents the measure in the integral---whereas in the literature it would be common to use a notation like $\dee \mu(x)$ instead.

\newpage

\chapter{Elementary differential geometry}
This chapter is the first of two introductory chapters. As such, it will provide a quick introduction into differential geometry. This branch of mathematics will provide the mathematical framework for this dissertation. Before commencing the introduction of this subject in earnest, it should be noted that differential geometry is an exceptionally interesting and rich field. As such no brief introduction could possibly do it justice. Therefore this discussion is limited to those concepts strictly necessary for the understanding of this dissertation. This list of topics, which will also serve as a backbone for the following two chapters, includes manifolds, coordinate functions, the metric tensor and the affine connection.\\
Readers interested in discovering more of this most elegant field are referred to introductory works such as \cite{Pressley} and \cite{Frankel}, upon which this introduction has loosely been based and which serve as the main reference material. The book by Pressley is dedicated to the ``extrinsic'' variant of differential geometry, which means it studies curves and surfaces embedded in a larger space. Frankel's book on the other hand, devotes most of its attention to the ``intrinsic'' theory, which can be formulated independently of a containing space and which is also closer to the differential geometry applied in this dissertation. Readers with an interest in information geometry, which will be discussed in the next chapter, may also find the first chapters of Amari's books \cite{Amari1985} and \cite{AmariNagaoka} interesting introductions, although they are far less broad in scope than more dedicated works.
\section{The origins of differential geometry}
Differential geometry is a field of study opened up by Carl Friedrich Gauss originally concerned with curves and surfaces embedded in Euclidean space. Its scope was greatly expanded after the contributions of the Hungarian mathematician J\'anos Bolyai and the Russian mathematician Nikolai Ivanovich Lobachevsky, who discovered the solution to a long standing problem \cite{Adler}. Over the course of two millennia, scores of mathematicians were plagued by the nature of Euclid's so-called Parallel Postulate \cite{Euclid}, which in the Playfair formulation reads\footnote{It should be noted that this formulation is only correct in the presence of the four preceding axioms of Euclidean geometry. In the current context, however, this formulation suffices and it has the advantage of simplicity.}
\begin{quote}
\emph{In a plane, given a line $A$ and a point $\here$ not on it, at most one line $B$ parallel to the given line $A$ can be drawn through $\here$}.
\end{quote}
\begin{center}
\begin{tikzpicture}
\path[draw] (5.2,1.4) node {$A$};
\path[draw] (5.2,0.25) node {$B$};
\path[draw] (0,0) -- (5,1);
\path[fill] (2,-0.75) circle (0.0625);
\path[draw] (0,-1.15) -- (5,-0.15);
\path[draw] (2.2,-0.35) node {$\here$};
\end{tikzpicture}
\end{center}
It was generally believed that this axiom is a consequence of the other axioms of Euclidean geometry, even though no one was able to demonstrate this. Bolyai and Lobachevsky independently provided the verdict by showing the existence of mathematically consistent geometries which did not satisfy Euclid's Parallel Postulate. To achieve this, they altered the axiom as to demand that more than one straight line through $\here$ can be constructed which is parallel to $A$. The resulting geometries are known as hyperbolic geometries.\\
Two decades later, the German mathematician Georg Bernhard Riemann developed what is now known as Riemannian geometry \cite{Riemann}. This is probably the best-known class of non-Euclidean geometries in the physics community as the mathematical framework of Einstein's general theory of relativity is based upon Riemann's ideas. Through the collaboration of mathematicians such as \'Elie Cartan, Henri Poincar\'e and others, the study of Riemannian geometry eventually led to the mathematical branch now known as differential geometry, a very general framework for describing the geometry of spaces.
\section{Manifolds and (co)tangent spaces}
The stages upon which differential geometry takes place, that is the spaces of which the geometry is studied, are manifolds. Manifolds will be denoted by the double letter $\mathbb{M}$. These are sets in which every element $\here$, usually called a point, has a neighbourhood $\mathcal{N}_{\here}$ which allows for the definition of coordinates for the points in that neighbourhood. A coordinate function $\varphi$ is a homeomorphism $\varphi : \mathcal{N}_{\here} \rightarrow \mathbb{R}^n$, that is it associates with every point an ordered set of $n$ real numbers.\\
\begin{center}
\begin{tikzpicture}[scale=1.25]

\path[draw] (-2.9,0.9) node {$\mathcal{N}_{\here}$};

\path[draw] (2.5,-1) -- (2.5,1);
\path[draw] (3,-1) -- (3,1);
\path[draw] (3.5,-1) -- (3.5,1);
\path[draw] (4,-1) -- (4,1);
\path[draw] (4.5,-1) -- (4.5,1);
\path[draw] (2.25,-0.75) -- (4.75,-0.75);
\path[draw] (2.25,-0.25) -- (4.75,-0.25);
\path[draw] (2.25,0.25) -- (4.75,0.25);
\path[draw] (2.25,0.75) -- (4.75,0.75);

\path[fill] (3.5,0.25) circle (0.0625);
\path[fill,white] (3.7,0.35) rectangle (4.4,-0.2);
\path[draw] (4,0.1) node {$\varphi(\here)$};

\path[draw] (5.2,0.9) node {$\mathbb{R}^n$};

\path[draw,->] (0.25,0.6) arc (120:60:1.5);
\path[draw] (1,1.1) node {$\varphi$};

\path[draw,clip] (0,0) arc (0:360:1.4 and 0.7);
%
%\path[draw] (-2.8,-1) arc (180:120: 1 and 2);
%\path[draw] (-2.2,-1) arc (180:125: 1 and 2.2);
%\path[draw] (-1.6,-1) arc (180:130: 1 and 2.4);
%\path[draw] (-1,-1) arc (180:135: 1 and 2.6);
%\path[draw] (-0.4,-1) arc (180:140: 1 and 2.8);
%
%\path[draw] (-2.8,0.7) arc (90:15: 3 and 0.8);
%\path[draw] (-2.8,0.35) arc (90:10: 2.8 and 0.7);
%\path[draw] (-2.8,0) arc (90:15: 2.6 and 0.6);
%\path[draw] (-2.8,-0.35) arc (90:15: 2.5 and 0.5);

\path[draw,gray] (-2.8,-1) arc (180:120: 1 and 2);
\path[draw,gray] (-2.2,-1) arc (180:125: 1 and 2.2);
\path[draw,gray] (-1.6,-1) arc (180:130: 1 and 2.4);
\path[draw,gray] (-1,-1) arc (180:135: 1 and 2.6);
\path[draw,gray] (-0.4,-1) arc (180:140: 1 and 2.8);

\path[draw,gray] (-2.8,0.7) arc (90:15: 3 and 0.8);
\path[draw,gray] (-2.8,0.35) arc (90:10: 2.8 and 0.7);
\path[draw,gray] (-2.8,0) arc (90:15: 2.6 and 0.6);
\path[draw,gray] (-2.8,-0.35) arc (90:15: 2.5 and 0.5);

\path[fill] (-1.45,0.255) circle (0.0625);
\path[fill,white] (-1.4,0.2) rectangle (-1.1,-0.15);
\path[draw] (-1.2,0) node {$\here$};

\end{tikzpicture}
\end{center}
These numbers $\{\varphi^i(\here)\}$ are called the coordinates of the point $\here$. A homeomorphism is a bijective map which is continuous and which has a continuous inverse \cite{Willard}. It is important for $\varphi$ to be a homeomorphism and not just any map. The invertibility means that $\varphi$ endows only a single point of its domain with a specific set of coordinates. The continuity of the coordinate map means that points close together in $\mathbb{M}$ are associated with coordinates which are also close together in $\mathbb{R}^n$. This ensures that the topological structure of $\mathbb{M}$ is faithfully reflected in the coordinates. The natural number $n$ is called the dimension of the manifold. For practical reasons the dimension is assumed to be finite, although many interesting properties and applications do exist in situations where the dimension is infinite, see for instance \cite{Pistone,PistoneSempi}.\\
Many mathematical sets encountered in physics are manifolds, even though they are not always identified as such. The space in which classical physics takes place is a manifold and the coordinate maps provide the familiar coordinates of points. The same thing is true for space-time as it appears in the theory of relativity. Most of the terminology in the theory of manifolds is actually derived from the analogy with physical space. Other examples of manifolds include, although care should be taken in the sense that not all of these can be covered by a single coordinate function, the configuration space of classical mechanics and the state space of statistical physics, as well as all finite-dimensional vector spaces and---by extension---Hilbert spaces, which are widely known as an essential element of modern quantum theory.\\
Two very important structures which can be defined on manifolds and used in describing the geometry of a space are the metric tensor and affine connections. In order to elucidate these concepts, it is necessary to first introduce a few more elementary concepts such as vectors and differential forms.\\
After the manifold, the most basic object of differential geometry is the vector. Perhaps the physically most intuitive interpretation is derived from velocity vectors. Consider an observer who measures the value of a function $f$ everywhere on her path and who also keeps time in order to tabulate the measured value as a function of time. This observer can then compute the rate of change in the value of $f$ she observed with respect to the time passed. It is possible to define the velocity vector of this observer as a functional which returns precisely this rate of change when applied to the function $f$. More rigorously, when a coordinate function $\varphi$ is fixed, then a vector $\vec{v}$ at a point $\here$ is a linear operator on real-valued functions $f: \mathbb{M}\rightarrow \mathbb{R}$ satisfying
\begin{align}
f \mapsto \vec{v}(f) = v^i \pdiff{}{\varphi^i} (f \circ \varphi^{-1}) \Big|_{\varphi(\here)} \label{eq: defvector}
\end{align}
for all functions $f$ for which these derivatives exist. The numbers $v^i$ are called the components of $\vec{v}$. In the case of velocity vectors these components equal the derivatives of the coordinate functions with respect to time.\\
Even though the definition (\ref{eq: defvector}) depends on the choice of a coordinate function $\varphi$, it can be shown that the vector itself is indeed invariant under suitable coordinate transformations. In particular, if a second set of coordinates $\{\zeta^a\}$ is employed, these must be a diffeomorphic function, that is a smooth function with a smooth inverse, of the original coordinates $\{\varphi^i\}$. In such case, the chain rule of calculus implies the derivatives relate to each other as
\begin{align}
\label{eq: trafovec}
\pdiff{}{\zeta^a} = \pdiff{\varphi^i}{\zeta^a} \pdiff{}{\varphi^i},
\end{align}
where the first factor on the right hand side represents the components of the Jacobian matrix of the transformation. If the coordinate independence of the vector $\vec{v} = v^i \partial_i = v^a \partial_a$ is to be respected, this means the components must transform according to the inverse transformation, that is
\begin{align*}
v^a = \pdiff{\zeta^a}{\varphi^i} v^i.
\end{align*}
That this is satisfied is also a consequence of the chain rule, as it is equivalent to the expression
\begin{align*}
\diff{\zeta^a}{t} = \pdiff{\zeta^a}{\varphi^i} \diff{\varphi^i}{t}.
\end{align*}
As was hinted at before, the most common example of a vector is the velocity vector, both in classical mechanics and in the theory of relativity. Acceleration is also described as a vector. Many other quantities are described as vectors in physics, such as linear and angular momentum, force, torque\ldots\ Most of these, however, including the four listed, appear more naturally in modern theories as differential forms \cite{Frankel}. Those objects and their operations can be framed as vectors but to do this requires a few pieces of extra mathematical machinery, not all of which are contained in the scope of this introduction. It is important to remark that the particular definition of vectors used here is more restricted than the algebraic definition as simply an element of a vector space. This particular kind of vectors are properly called tangent vectors. Another type of vectors in the algebraic sense to be discussed in this dissertation are the differential forms with which this section will be concluded. Other, more general, vector structures will not play a major role in this dissertation and will therefore not be given further consideration in this introduction.\\
It is possible to verify that all tangent vectors with the same point of application $\here$ form a vector space when endowed with componentwise addition and scalar multiplication. This follows from the fact that linear combinations of such vectors are still of the form (\ref{eq: defvector}). This vector space is called the tangent space at the point $\here$, or $\tg{\here}{M}$ for short. Together, all vectors over a manifold constitute the tangent bundle\footnote{Some sources such as \cite{Taubes} reserve this name for the bundle projection map.}, abbreviated $\tgb{M}$. The tangent bundle can also be seen as a manifold---the configuration space of mechanics is probably the best known tangent bundle---but it is not a vector space as addition is only defined for pairs of vectors which share a point of application. The operations obtained by applying partial derivative operators $\{\partial_i\}$ and evaluating the result at a point $\here$ form the basis vectors of the tangent space $\tg{\here}{M}$.\\
A particularly interesting type of subsets of the tangent bundle are vector fields. A vector field can be likened to a function $F: \mathbb{M} \rightarrow \tgb{M}$ but with the additional restriction that the function value of a point $\here$ must be an element of $\tg{\here}{M}$ and not of another tangent space. The traditional but more technical formulation states that a map $F:\mathbb{M}\rightarrow \tgb{M}$ is a vector field when $\pi \circ F$, where $\pi$ is the bundle projection map associating with every vector its point of application, is the identity mapping on the domain of $F$. The basis vectors employed thus far, the differential operators $\{\partial_i\}$, naturally form $n=\dim(\mathbb{M})$ linearly independent vector fields. Such a collection of vector fields which form a basis for each tangent space (in their domain) is called a frame. It is not required that a frame consists out of differential operators with respect to coordinate functions but this dissertation will exclusively employ such so-called coordinate frames as they are probably the most familiar to readers.\\
As the tangent spaces $\tg{\here}{M}$ are vector spaces, it is meaningful to speak of the dual vector space: the space of all linear functionals on $\tg{\here}{M}$ with values in the real numbers. Where tangent vectors are probably most familiar to readers as column vectors, their duals are traditionally denoted---and better known---as row vectors. The dual space of $\tg{\here}{M}$ is called the cotangent space at $\here$ and is denoted in a concise way as $\ctg{\here}{M}$. Together these cotangent spaces make up the cotangent bundle $\ctgb{M}$. Smooth fields over this bundle are called (differential) $1$-forms. Phase space as it appears in Hamiltonian mechanics is probably the most familiar example of a cotangent bundle, even though it is not usually introduced as such in physics textbooks.\\
Just as a canonical basis $\{\partial_i\}$ of partial derivative operators exists for $\tg{\here}{M}$ given a coordinate function $\varphi$, a canonical basis $\{\sigma^i\}$ exists for $\ctg{\here}{M}$ given this same coordinate function. The functionals $\sigma^j$ making up this dual basis satisfy
\begin{align}
\label{eq: defduals}
\sigma^j(\partial_i) \define \delta^j_i.
\end{align}
A most elegant way of introducing the canonical dual basis is through the exterior derivative operator, for which the symbol ``$\dee$'' is used. This operator maps its arguments, which are differentiable functions, into differential \mbox{$1$-forms}. More in particular the exterior derivative $\dee f$ of a scalar function $f$ acts on a vector $\vec{v}$ as
\begin{align*}
(\dee f)(\vec{v}) \define \vec{v}(f).
\end{align*}
By choosing $f = \varphi^i$, the $i^{\text{th}}$ component of the coordinate function, it follows that
\begin{align*}
(\dee \varphi^i)(\partial_j)\Big|_{\here} &= \partial_j ( \varphi^i \circ \varphi^{-1} ) \Big|_{\varphi(\here)}\\
&= \delta^i_j.
\end{align*}
This, combined with the linearity of these operators, implies $\sigma^i = \dee \varphi^i$ for the duals of the basis $\{\partial_j\}$. It is often useful to choose the $1$-forms $\{\dee \varphi^i\}$ as the basis for the cotangent spaces, even though it is not strictly necessary to do so. One advantage of this choice is that the exterior derivative of a function $f$ can easily be expressed as
\begin{align}
\label{eq: defdf}
\dee f = (\partial_i f) \intdee \varphi^i.
\end{align}
This expression also shows there is a connection between the exterior derivative of a function and its gradient. The numbers $\partial_i f$ are equal to the components of the gradient of $f$ in Cartesian coordinates but not in an arbitrary curvilinear coordinate system. Expression (\ref{eq: defdf}) however, holds in all coordinate systems, which offers obvious practical advantages.\\
The $1$-forms introduced allow for the construction of a very rich algebraic structure. They can be endowed with a product, denoted as $\wedge$, which is fully antisymmetric. More specifically, given two $1$-forms $\alpha$ and $\beta$, the $2$-form $\alpha \wedge \beta$ is a bilinear mapping defined by
\begin{align*}
(\alpha \wedge \beta)(\vec{v},\vec{w}) &= \alpha(\vec{v}) \beta(\vec{w}) - \alpha(\vec{w}) \beta(\vec{v}).
\end{align*}
This operation has many uses in differential geometry, including the definition of the exterior derivative of $1$-forms. Given a $1$-form $\alpha = \alpha_i \dee \varphi^i$, where $\alpha_i = \alpha(\partial_i)$, its exterior derivative equals
\begin{align*}
\dee \alpha \define (\dee \alpha_i) \wedge \dee \varphi^i = (\partial_j \alpha_i) \intdee \varphi^j \wedge \dee \varphi^i.
\end{align*}
From the skew symmetry of $\dee \varphi^i \wedge \dee \varphi^j$, it automatically follows that $\dee(\dee f) = 0$ for all real-valued functions $f$. A very useful property is the (partial) converse to this, which is called Poincar\'e's lemma. It states that, on a manifold which can be contracted into a single point by a continuous transformation, the vanishing of a form's exterior derivative, i.\ e.\ $\dee \alpha = 0$, is a sufficient condition for the existence of a function $f$ such that $\alpha = \dee f$. Though not all manifolds in this thesis are indeed contractible, it is nevertheless often possible to apply Poincar\'e's lemma locally even when it does not hold globally. Such a function $f$ is called a potential for the $1$-form $\alpha$.\\
This structure of differential forms can be extended to define higher forms and their exterior derivatives as well. This is done in a completely analogous fashion. The use of forms in this dissertation will remain fairly limited to accommodate readers more familiar with the index-heavy tensorial notation traditionally used in the physics literature. As a consequence forms or arbitrary degree will not be treated in this introduction. Nevertheless, some results involving higher forms than those of first degree will be required in some situations. One example of such a result is the analogue of the Leibniz rule for the wedge product of two $1$-forms $\alpha$ and $\beta$:
\begin{align*}
\dee (\alpha \wedge \beta) &= (\dee \alpha) \wedge \beta - \alpha \wedge (\dee \beta).
\end{align*}
Also Poincar\'e's lemma will be used for higher forms. Its generalisation is straightforward: on a contractible manifold any differential $p$-form of which the exterior derivative vanishes can be written as the exterior derivative of some differential $(p-1)$-form.
\section{The Riemannian metric}
So far this introduction to differential geometry has not yet included any mention of distance or angles, arguably both important concepts in any study of geometry. In order to introduce these quantities, one must choose an inner product for each of the tangent spaces. Such a mapping is called a metric tensor and in particular, since it is defined on the tangent spaces, a Riemannian metric. This mapping is denoted by $g$ and so it can be written that\footnote{Where context avoids confusion, the index $\here$ from this notation will not be mentioned explicitly. Alternatively, this definition can be expanded to map pairs of vector fields into real-valued functions over $\mathbb{M}$.}
\begin{align*}
g(\cdot,\cdot)\big|_{\here} : \tg{\here}{M} \times \tg{\here}{M} &\rightarrow \mathbb{R}.
\end{align*}
An inner product must be symmetric, which means that $g(\vec{v},\vec{w}) = g(\vec{w},\vec{v})$ for all vectors $\vec{v}$ and $\vec{w}$ in the same tangent space. Another requirement is the positive definiteness, which means $g(\vec{v},\vec{v}) \geqslant 0$ for all vectors $\vec{v}$ and that the value can only be equal to zero when all the vector components equal zero. Often the value of the inner product will be written in its coordinate representation $g(\vec{v},\vec{w}) = g_{ij} v^i w^j$, where the components are given by
\begin{align}
\label{eq: defcomg}
g_{ij} = g(\partial_i,\partial_j).
\end{align}
The fact that this notation is possible follows from the property of bilinearity, which means linearity holds in both arguments separately.\\ Lengths of vectors in $\tg{\here}{M}$ and angles between two vectors in this space can be computed using the metric through the respective familiar relations
\begin{align*}
v = \sqrt{g(\vec{v},\vec{v})|_{\here}} \quad \text{and}\quad \cos(\phi) = \frac{g(\vec{v},\vec{w})|_{\here}}{vw}.
\end{align*}
Readers will no doubt be familiar with the dot product on $\mathbb{R}^3$ but this does not necessarily paint a representative image of Riemannian metrics. This is because a large part part of the physics literature describes space also as $\mathbb{R}^3$, which itself can also be seen as a vector space. Elementary textbooks traditionally make use of this extra structure when introducing vector calculus. In particular, identifying all tangent spaces with each other conceals that the dot product is actually defined on each tangent space separately. The best known examples of non-trivial Riemannian metrics are probably the solutions to the field equations which lie at the heart of Einstein's general theory of relativity. The Fisher information matrix \cite{Fisher} appearing in statistics, for example in the Cram\'er-Rao inequality \cite{Rao}, is used as a metric tensor in the field of information geometry. This will be discussed in more detail in the next chapter.
\section{Connections and curvature}
A similarly important structure on a manifold is the affine connection. The connection may seem a more technical notion than the metric but it has many practical applications nonetheless. The most common way to introduce a connection on a tangent bundle is through a covariant derivative operation, denoted by $\nabla$. Although strictly speaking an abuse of notation, this symbol is often used to refer to the connection as well. Just as derivatives of functions play a role in many branches of mathematics and science, a similar notion is often very useful for vector fields. Consider the problem of defining the directional derivative of a vector field $\vec{X}$ in the direction of the vector $\vec{v} \in \tg{\here}{M}$. It is possible to construct such a derivative and the resulting vector is denoted by $\nabla_{\vec{v}} \vec{X}$. However, some difficulty is involved in working out this notion so it is instructive to first take a look at the easy aspects and leave the hard part for last.\\
A first property that the covariant derivative must satisfy is linearity in its first argument---just like the more familiar directional derivative of scalar functions. This means
\begin{align*}
\nabla_{\vec{v}} \vec{X} = v^i \nabla_{\partial_i} \vec{X} \notation v^i \nabla_i \vec{X}.
\end{align*}
Furthermore, the covariant derivative should be linear for addition in the second argument and satisfy the Leibniz rule
\begin{align}
\label{eq: CovLeibniz}
\nabla_{\vec{v}}( f \vec{X}) = \vec{v}(f) \vec{X} + f \nabla_{\vec{v}} \vec{X},
\end{align}
where $f$ is an arbitrary differentiable function. This property---combined with the expansion of the vector field $\vec{X} = X^i \partial_i $---states that
\begin{align}
\label{eq: defnabla}
\nabla_{\vec{v}} \vec{X} &= \vec{v}(X^i) \partial_i + X^i \nabla_{\vec{v}} \partial_i.
\end{align}
This shows it suffices to determine the covariant derivatives of the basis vectors---here of course seen as fields. Because of the first linearity property, it is even enough to know at every point $\here$ the coefficients of the expansion
\begin{align}
\label{eq: defomega}
\nabla_i \partial_j = {\omega^k}_{ij} \partial_k.
\end{align}
These numbers ${\omega^k}_{ij}$ are called the coefficients of the connection at $\here$. There is no obvious way to determine these numbers in all contexts, something which is also the case with the metric tensor. In fact, the whole point of Einstein's general theory of relativity is that the geometric properties of spacetime follow from physical laws rather than from mathematical principles! The connections playing an important role in information geometry are also defined from context-specific arguments. This is elucidated in the next chapter.\\
It is interesting to note that the second term in the right hand side of equation (\ref{eq: defnabla}) is essentially a linear transformation of the vector $\vec{X}$ (at $\here$). This is even more apparent when leaving the first argument of the covariant derivative unmentioned:
\begin{align*}
\nabla \vec{X} &= \partial_i \otimes [\dee X^i  + {\omega^i}_j X^j ],
\end{align*}
where the expression between square brackets is an element of $\ctgb{M}$ and the connection $1$-forms
\begin{align}
\label{eq: defw1f}
{\omega^i}_j \define {\omega^i}_{kj} \dee \varphi^k
\end{align}
were implicitly introduced. This notation of connection forms turns out to be very convenient in some formulas. Against the general trend in this dissertation of limiting the use of forms, this notation will appear in some places. For example the proof of Riemann's theorem following shortly would be much more cumbersome when making use of the more usual notations introduced in equations (\ref{eq: defnabla}) and (\ref{eq: defomega}).\\
Apart from yielding a definition for the derivative of a vector field, a connection can also be used to define whether or not two vectors at different points are parallel. In order to do this, the concept of parallel transport must first be defined. Given a curve $C : [t_i,t_f] \rightarrow \mathbb{M}$, it is said a vector field $\vec{Y}$, defined (at least) on the curve $C$ with tangent vector field $\vec{X} = \frac{\dee}{\dee t}$, is parallel along $C$ or has undergone parallel transport along $C$ if
\begin{align*}
0 &= \nabla_{\vec{X}}\vec{Y}\\
&= \partial_j \otimes X^i [\partial_i Y^j + {\omega^j}_{ik} Y^k ]\\
&= \partial_j \otimes \left[ \diff{Y^j}{t} + {\omega^j}_{ik} \diff{\varphi^i}{t} Y^k\right].
\end{align*}
This gives a way to transport vectors from one tangent space to another: solving these $n$ ordinary first order differential equations yields the components of the vector field everywhere along the curve (parametrised by $t$). In the drawing this is demonstrated for a toy connection. The tangent vector field $\vec{X}$ to the curve (in grey) is also shown.
\begin{center}
\begin{tikzpicture}[scale=1.25]

\path[draw,-latex,gray] (0.593,0.667) -- (0.959,1.008);
\path[draw,-latex,gray] (1.271,1.101) -- (1.741,1.272);
\path[draw,-latex,gray] (2.069,1.205) -- (2.568,1.178);
\path[draw,-latex,gray] (2.952,1.074) -- (3.443,0.979);
\path[draw,-latex,gray] (3.835,0.944) -- (4.334,0.918);
\path[draw,-latex,gray] (4.633,1.048) -- (5.103,1.219);
\path[draw,-latex,gray] (5.311,1.481) -- (5.677,1.823);

\path[draw,-latex] (0.593,0.667) -- (0.429,1.140);
\path[draw,-latex] (1.271,1.101) -- (1.009,1.527);
\path[draw,-latex] (2.069,1.205) -- (1.786,1.617);
\path[draw,-latex] (2.952,1.074) -- (2.696,1.504);
\path[draw,-latex] (3.835,0.944) -- (3.608,1.390);
\path[draw,-latex] (4.633,1.048) -- (4.383,1.481);
\path[draw,-latex] (5.311,1.481) -- (4.974,1.851);

%\path[draw,-latex] (0.593,0.667) -- (0.799,1.400);
%\path[draw,-latex] (1.271,1.101) -- (1.379,1.855);
%\path[draw,-latex] (2.069,1.205) -- (2.078,1.966);
%\path[draw,-latex] (2.952,1.074) -- (2.862,1.831);
%\path[draw,-latex] (3.835,0.944) -- (3.647,1.682);
%\path[draw,-latex] (4.633,1.048) -- (4.351,1.755);
%\path[draw,-latex] (5.311,1.481) -- (4.939,2.146);

\path[draw,domain=0:2*pi,smooth,variable=\t] plot( {\t*cos(20) - sin(20)*0.6*sin(\t r)},{0.6*sin(\t r)*cos(20) + \t*sin(20)} );

\end{tikzpicture}
\end{center}
In general it should be noted that transporting a vector over different curves with identical end points need not yield the same result. Equivalently, parallel transport of a vector around a closed loop will in general not make the final vector coincide with the first one. The difference between the original vector and a vector transported around a closed loop can be quantified using the curvature of the connection. If the curvature vanishes, the connection is said to be flat and this is equivalent to saying that parallel transport is independent of the chosen curve.\\
Most introductory texts introduce curvature through the commutativity of covariant derivatives, which is equivalent with the vanishing of the curvature. While this is correct, such derivation usually does not include the reasoning leading up to the most interesting result, which is the existence of covariant constant vector fields. A vector field $\vec{X}$ is said to be covariant constant if parallel transport of the value of the vector field at a point $\here$ to a point $\there$ is equal to the value of the field at $\there$. This definition requires the parallel transport to be independent of the path and so the vanishing of the connection is a necessary condition for covariant constant vector fields. The next two pages will deal with showing that this is also a sufficient condition by proving the slightly stronger result known as Riemann's theorem.\\
The best-known property of manifolds endowed with a flat connection is that they allow for global Cartesian coordinates. However, the connection may arbitrary. Rather it must be the metric connection of Levi-Civita, which has coefficients given by
\begin{align}
\label{eq: LCcon}
{\omega^k}_{ij} = {\Gamma^k}_{ij} \define \frac12 g^{ks} \left( \partial_i g_{sj} + \partial_j g_{is} - \partial_s g_{ij}\right),
\end{align}
where the numbers $g^{ij}$ are the components of the matrix inverse of the metric tensor. It will be shown that when this particular connection exhibits vanishing curvature, there exists on the manifold a local set of coordinates in which the metric tensor everywhere takes the Euclidean form
\begin{align}
\label{eq: gEuclid}
g(\vec{v},\vec{w}) = \sum_{j=1}^n v^j w^j.
\end{align}
The property that the metric tensor can be written in this form when the metric connection (\ref{eq: LCcon}) is flat, is Riemann's theorem. Note that this is not a trivial statement. The metric can take this form at any point by choosing an appropriate set of coordinates called normal coordinates. However, nothing guarantees that this form will be valid also outside this one point while still using the same set of coordinates. It is only when the curvature vanishes that coordinates exist in which this property holds everywhere.\\
First, it will be shown that for any connection the parallel displacement of vector fields is independent of the path followed if and only if its curvature vanishes. Only then will it be shown how this gives rise to the Euclidean form of the metric tensor.\\
A vector field $\vec{X}$ is said to be parallel along a curve $C$ parametrised by $t$ if its covariant derivative vanishes along that curve or, equivalently,
\begin{align}
0 %&= \nabla_{\vec{X}} \vec{Y}\nonumber\\
&= \partial_i \otimes\left[ \diff{X^i}{t} + {\omega^i}_{jk} \diff{\varphi^j}{t} X^k\right]\nonumber \\
&= \partial_i \otimes \left[ \pdiff{X^i}{\varphi^j} \diff{\varphi^j}{t} + {\omega^i}_{jk} \diff{\varphi^j}{t} X^k\right]\nonumber\\
&= \partial_i \otimes [ \dee X^i + {\omega^i}_k X^k]\left( \diff{\varphi^j}{t} \partial_j\right).\label{eq: curveind}
\end{align}
From the condition (\ref{eq: curveind}) for the vanishing of the covariant derivative it can be seen that parallel transport yields the same result independent of the curve $C$ if and only if 
\begin{align}
\label{eq: mu}
\dee X^i + {\omega^i}_k X^k = 0
\end{align}
since this left hand side is the curve-independent part of the condition. Frobenius' theorem states that a sufficient condition for the equations (\ref{eq: mu}) to have a solution is then for the $2$-forms
\begin{align*}
\dee ( \dee X^i + {\omega^i}_k X^k)
\end{align*}
to vanish \cite{Frankel}. It is possible to rewrite, using (\ref{eq: mu}) again in the second equality
\begin{align*}
\dee (\dee X^i + {\omega^i}_k X^k) &= \dee^2 X^i + (\dee {\omega^i}_k) X^k  +\dee X^k \wedge {\omega^i}_k \\
&= (\dee {\omega^i}_j) X^j + ( - {\omega^k}_j X^j)\wedge {\omega^i}_k  \\
&= ( \dee {\omega^i}_j + {\omega^i}_k \wedge {\omega^k}_j) X^j.
\end{align*}
This means parallel transport of any vector field $\vec{X}$ is path-independent when the curvature $2$-forms
\begin{align}
\label{eq: defOmega}
{\Omega^l}_k &\define \dee {\omega^l}_k + {\omega^l}_s \wedge {\omega^s}_k
\end{align}
vanish identically, which concludes the proof of the first statement made above. In a more index-heavy notation, this can be expressed as
\begin{align*}
0 &= {\Omega^l}_{kij} = \partial_i {\omega^l}_{jk} - \partial_j {\omega^l}_{ik} + {\omega^l}_{is} {\omega^s}_{jk} - {\omega^l}_{js} {\omega^s}_{ik}.
\end{align*}
When the curvature vanishes, parallel transport is path independent and so covariant constant vector fields can be constructed by parallel transporting a vector at a single point to other points of $\mathbb{M}$.\\
The proof of Riemann's theorem is continued by showing that these covariant constant vector fields are partial derivatives with respect to some coordinate functions $\theta^i$. This is useful since in these coordinates, the connection coefficients
\begin{align*}
{\omega^k}_{ij} = \dee \theta^k( \nabla_i \partial_j ) 
\end{align*}
vanish identically. Such coordinates are known as the affine coordinates of the connection. Although the proof is not in itself very difficult, it requires some new concepts and properties, the introduction of which would burden the reader with more text than can be justified from their use in this dissertation. Interested readers are thus referred to the source material mentioned earlier for the details of this part of the proof.\\
It should be stressed that the results up to now hold for all flat affine connections. Riemann's theorem deals with the specific case of the Levi-Civita connection (\ref{eq: LCcon}). The interesting property of this connection is that it satisfies
\begin{align}
\label{eq: defLCcon}
\vec{X}( g(\vec{Y},\vec{Z}) ) = g(\nabla_{\vec{X}} \vec{Y},\vec{Z}) + g(\vec{Y}, \nabla_{\vec{X}} \vec{Z})
\end{align}
and so it preserves inner products between vector fields under parallel transport. But this means that when one chooses an orthonormal basis at one point of $\mathbb{M}$, parallel transport through the Levi-Civita connection will result in vector fields which are orthonormal everywhere. The drawing illustrates this for transport along a curve in the plane.
\begin{center}
\begin{tikzpicture}[scale=1.25]

\path[draw,-latex] (0.593,0.667) -- (1.065,0.831);
\path[draw,-latex] (0.593,0.667) -- (0.429,1.140);

\path[draw,-latex] (1.271,1.101) -- (1.697,1.363);
\path[draw,-latex] (1.271,1.101) -- (1.009,1.527);

\path[draw,-latex] (2.069,1.205) -- (2.481,1.488);
\path[draw,-latex] (2.069,1.205) -- (1.786,1.617);

\path[draw,-latex] (2.952,1.074) -- (3.382,1.330);
\path[draw,-latex] (2.952,1.074) -- (2.696,1.504);

\path[draw,-latex] (3.835,0.944) -- (4.281,1.172);
\path[draw,-latex] (3.835,0.944) -- (3.608,1.390);

\path[draw,-latex] (4.633,1.048) -- (5.066,1.298);
\path[draw,-latex] (4.633,1.048) -- (4.383,1.481);

\path[draw,-latex] (5.311,1.481) -- (5.680,1.819);
\path[draw,-latex] (5.311,1.481) -- (4.974,1.851);

\path[draw,domain=0:2*pi,smooth,variable=\t] plot( {\t*cos(20) - sin(20)*0.6*sin(\t r)},{0.6*sin(\t r)*cos(20) + \t*sin(20)} );

\end{tikzpicture}
\end{center}
By the above, the vector fields obtained in this way are partial derivatives with respect to some coordinates $\{\theta^i\}$. In those coordinates, the metric will thus take its Euclidean form (\ref{eq: gEuclid}). This concludes the proof of Riemann's theorem.\\
%\begin{center}
%\begin{tikzpicture}[scale=1.5]
%\path[draw,-latex] (0,0) -- (0,0.5);
%\path[draw,-latex] (0,0) -- (0.5,0);
%
%\path[draw,-latex] (0,1) -- (0.492,0.913);
%\path[draw,-latex] (0,1) -- (0.087,1.492);
%
%\path[draw,-latex] (1,0) -- (1.492,-0.087);
%\path[draw,-latex] (1,0) -- (1.087,0.492);
%
%\path[draw,-latex] (1,1) -- (1.470,0.829);
%\path[draw,-latex] (1,1) -- (1.171,1.470);
%
%\path[draw,-latex] (2,0) -- (2.470,-0.171);
%\path[draw,-latex] (2,0) -- (2.171,0.470);
%
%\path[draw,-latex] (2,1) -- (2.433,0.75);
%\path[draw,-latex] (2,1) -- (2.25,1.433);
%
%\path[draw,-latex] (3,0) -- (3.433,-0.25);
%\path[draw,-latex] (3,0) -- (3.25,0.433);
%
%\path[draw,-latex] (3,1) -- (3.383,0.679);
%\path[draw,-latex] (3,1) -- (3.321,1.383);
%\end{tikzpicture}\\
%Illustration of the above construction for an orthonormal frame on the plane. The two vectors at the bottom left form an orthonormal basis and similar bases are obtained at other points through parallel transport with the metric connection.
%\end{center}
Perhaps the best known application of connections consists in the construction of geodesics. Most people are familiar with geodesics as the shortest paths connecting two given points but this is not generally true as this is a particular property exclusive to the Levi-Civita connection. In general a geodesic is a curve to which the tangent vectors form a covariant constant vector field. Since such vector fields are said to be parallel to themselves, the geodesics are the natural generalisation of straight lines.\\
The curvature as measured by the $2$-forms ${\Omega^l}_k$ is the intrinsic curvature and, if applicable, is independent of the way the manifold is embedded in a larger manifold. As such it must be contrasted with extrinsic or embedding curvature, which depends on the way the manifold is embedded in a larger space and in particular on how the normal vector fields on the manifold behave as one moves over the manifold. An entire theory of embedding curvature can be set up and a short but clear summary can be found in Chapter 2 of Amari's book \cite{Amari1985}. Also Pressley's book \cite{Pressley} concerns itself with embedding curvature but it is limited to three dimensions and it uses the traditional notation, making it perhaps more difficult to see how to set up such a theory in an arbitrary number of dimensions. The difference between intrinsic and extrinsic curvature will become an important point later in this introduction, in particular in Chapter 3.\\
To draw an intuitive picture of the difference between the two kinds of curvature, some examples might be useful to keep in mind. A curve, which is a one-dimensional manifold, cannot have intrinsic curvature. This is because all $2$-forms, including the curvature $2$-forms ${\Omega^l}_k$, vanish on one-dimensional manifolds by the antisymmetry of the wedge product. A curve can have extrinsic curvature, however, and this is the case when a tangent vector field must change direction as one moves over the curve. This extrinsic curvature is measured by the inverse of the radius of curvature as it is known from more elementary texts. A sphere, on the other hand, is intrinsically curved. This means that when the sphere is embedded in a flat space like $\mathbb{R}^n$, the normal vectors to the sphere must necessarily change direction as one moves over the sphere and so the embedding curvature will not vanish globally, even though it may do so locally should the sphere be deformed appropriately.\\
In this dissertation several affine connections will appear and these will not all be metric connections. However, it is still possible to obtain interesting properties of affine connections beyond Riemann's famous result. In particular, it is possible to define the notion of dual connections through a relation generalising (\ref{eq: defLCcon}). When two connections $\nabla$ and $\nabla^*$ satisfy
\begin{align}
\label{eq: defdualcon}
{\vec{X}} g(\vec{Y},\vec{Z}) = g(\nabla_{\vec{X}} \vec{Y},\vec{Z}) + g(\vec{Y},\nabla_{\vec{X}}^*\vec{Z}),
\end{align}
again for all vector fields $\vec{X}$, $\vec{Y}$ and $\vec{Z}$, it is said these connections are dual with respect to the metric tensor $g$. Every connection $\nabla$ has a dual connection as expression (\ref{eq: defdualcon}) can be used as a definition of $\nabla^*$. Parallel transport of one vector field through a connection $\nabla$ and of the other vector field through the dual connection $\nabla^*$ will always preserve the inner product between the vector fields. The drawing illustrates this for a pair of orthogonal vector fields in the plane transported over a curve, each through a different connection. Note that neither of these connections preserves the length of the vectors under parallel transport. Since the original vectors are orthogonal, their transported counterparts are as well. In general the angle between the vectors need not be preserved either.
\begin{center}
\begin{tikzpicture}[scale=1.25]

\path[draw,-latex] (0.593,0.667) -- (1.038,0.822);
\path[draw,-latex] (0.593,0.667) -- (0.419,1.169);

\path[draw,-latex] (1.271,1.101) -- (1.646,1.331);
\path[draw,-latex] (1.271,1.101) -- (0.974,1.585);

\path[draw,-latex] (2.069,1.205) -- (2.404,1.435);
\path[draw,-latex] (2.069,1.205) -- (1.721,1.711);

\path[draw,-latex] (2.952,1.074) -- (3.272,1.265);
\path[draw,-latex] (2.952,1.074) -- (2.608,1.652);

\path[draw,-latex] (3.835,0.944) -- (4.139,1.099);
\path[draw,-latex] (3.835,0.944) -- (3.502,1.598);

\path[draw,-latex] (4.633,1.048) -- (4.906,1.205);
\path[draw,-latex] (4.633,1.048) -- (4.236,1.736);

\path[draw,-latex] (5.311,1.481) -- (5.528,1.680);
\path[draw,-latex] (5.311,1.481) -- (4.737,2.109);

\path[draw,domain=0:2*pi,smooth,variable=\t] plot( {\t*cos(20) - sin(20)*0.6*sin(\t r)},{0.6*sin(\t r)*cos(20) + \t*sin(20)} );

\end{tikzpicture}
\end{center}
Due to the symmetry of the metric, it holds that $(\nabla^*)^* = \nabla$ and it is possible to show that a pair of dual connections have curvatures which cannot vanish independently of each other \cite{Amari1985}. Such pairs of dual connections play an important role in information geometry but they are not so common in most applications of differential geometry. Therefore most introductory texts on differential geometry do not cover this topic.\\
Before finally concluding this introduction to differential geometry, it is important to devote attention to a last application of covariant derivatives. The previous seven pages considered the covariant derivative of a vector. At least as important is the notion of the covariant derivative of a $1$-form and the resulting definition of the Hessian.% The introduction of this idea is predicated entirely on ideas already discussed above.
Remember that the Leibniz rule for a covariant derivative, equation (\ref{eq: CovLeibniz}), implicitly made use of the property that the covariant derivative of a function is the same as the regular derivative, that is
\begin{align*}
\nabla_{\vec{v}} f = \vec{v}(f).
\end{align*}
One way to construct a differentiable scalar function is to apply a $1$-form to a vector field---both of which must have differentiable components. This leads to the scalar function $f = \alpha(\vec{X}) = \alpha_i X^i$. This form of the expression hints that a differentiation of $f$ can be performed using a variation on the Leibniz rule, 
\begin{align*}
\nabla_{\vec{v}}(f) = \nabla_{\vec{v}}( \alpha(\vec{X})) = (\nabla_{\vec{v}} \alpha)_i X^i + \alpha_i (\nabla_{\vec{v}} \vec{X})^i.
\end{align*}
The presence of a lower index on $\nabla_{\vec{v}}\alpha$ shows that this quantity itself is also a $1$-form, just as the covariant derivative of a vector field is again a vector field. Rearranging the above line and working out the resulting expression yields
\begin{align*}
(\nabla_{\vec{v}}\alpha)(\vec{X}) &= \nabla_{\vec{v}}(\alpha(\vec{X})) - \alpha( \nabla_{\vec{v}} \vec{X})\\
&= v^i \partial_i (\alpha_j X^j) - \alpha_k ( v^i \partial_i X^j + {\omega^k}_{ij} v^i X^j)\\
&= v^i [ \partial_i \alpha_j  - {\omega^k}_{ij} \alpha_k ] X^j.
\end{align*}
Since this must hold for any vector field $\vec{X}$, the covariant derivative of a $1$-form is given by
\begin{align*}
\nabla_{\vec{v}}\alpha = v^i [ \partial_i \alpha_j - {\omega^k}_{ij} \alpha_k] \dee \varphi^j.
\end{align*}
The covariant derivative has the same function for $1$-forms as for vectors: it performs a directional derivative in such a way that the result is independent of the chosen basis for the (co)tangent spaces. This is particularly useful when speaking about the Hessian in the context of manifolds. The Hessian is the generalisation of the second derivative of a scalar function. However, computing higher derivatives must take into account that not only the function is liable to change from point to point. Also the direction in which a partial derivative quantifies the change of a function need not remain the same. This change of direction is exactly what the connection coefficients express. Therefore, the Hessian of a function $f$ is defined as
\begin{align*}
\text{Hess}(f)(\vec{u},\vec{v}) &\define (\nabla_{\vec{u}} \dee f)(\vec{v}).
\end{align*}
Note that though $\text{Hess}(f)$ is a function taking two vectors as arguments and mapping them into a real number, it is not a $2$-form. If $f$ is continuously differentiable then $\text{Hess}(f)$ is a symmetric mapping---whereas a $2$-form would be antisymmetric. To stress the symmetry of the Hessian, a more commonly used notation for this object is
\begin{align}
\text{Hess}(f)(\partial_i,\partial_j) &= \nabla_i \nabla_j f\nonumber\\
&= \partial_i \partial_j f - {\omega^k}_{ij} \partial_k f.\label{eq: defHessian}
\end{align}
It is this notation that will be used throughout the other chapters.\\
As is stated in the first paragraphs of this chapter, a complete introduction to differential geometry would include a large amount of material not present in this introduction. In particular many theorems and properties are mentioned but not treated explicitly. Some of those theorems, such as the result on integrability that bears Frobenius' name and which is used as a step in the above proof of Riemann's theorem, are also important in other parts of this dissertation. Where such theorems are invoked they shall be mentioned when possible. For their proofs and the required background, the reader is again referred to introductory material.

\newpage

\chapter{Information geometry}
The largest source of inspiration for the work in this dissertation is the mathematical field of information geometry. In this part of the introduction, a brief overview of this discipline will be presented. The text of this chapter is based upon Amari's books \cite{Amari1985,AmariNagaoka}, supplemented with elements of landmark articles in this field of research.
\section{Parametrised statistical models}
Information geometry is the study of statistics through the methods of differential geometry. This can happen either through the use of known results from differential geometry to expand the existing knowledge of statistics or by employing geometric methods to facilitate computations in statistical problems. Some examples of such applications are found in \cite{Pistone,Ohara}. Of particular interest for this dissertation is the problem of parameter estimation. This problem occurs whenever one has quantitative data, which can be obtained from an experiment performed on a statistical sample, and one assumes that the true underlying distribution is an element of a parametrised set or a family of distributions. This assumption usually follows from a priori available information on the mechanism generating the data \cite{Fisher}.\\
The most familiar example of this problem in physics is probably encountered when one has measured the energies of particles in a gas and wishes to determine from this the temperature of the gas, assuming that the probability of a particle to have an energy $E$ when the gas has temperature $T$ is given by the Boltzmann-Gibbs distribution
\begin{align*}
p_T(E) = \frac{1}{\mathcal{Z}} \exp\left\{ - \frac{E}{k_B T}\right\}; \qquad \mathcal{Z} = \int_0^{\infty} \exp\left\{ -\frac{E}{k_B T}\right\}\rho(E) \dee E,
\end{align*}
where $\rho$ represents the density of states. Another familiar example is that of the normal distribution
\begin{align*}
p_{\mu,\sigma}(x) = \frac{1}{\sqrt{2\pi \sigma^2}} \exp\left\{ - \frac{(x-\mu)^2}{2\sigma^2}\right\}
\end{align*}
where $\mu$ represents the mean of the distribution and $\sigma$ is the standard deviation or second central moment which is used in many applications. In the rest of this introduction, the two-parameter Gaussian densities shall be used to illustrate the concepts discussed. The interested reader can find a more detailed discussion of multivariate Gaussian distributions and their information geometrical behaviour in \cite{Skovgaard}.\\
In the above examples, the parameters $T$, $\mu$ and $\sigma$ map homeomorphically onto the distributions they label. Hence, they can also be considered as coordinates for the family of distributions, which in turn can be viewed as a topological manifold. A manifold of statistical distributions (or equivalently: statistical measures) is called a statistical manifold, a term first introduced by Lauritzen \cite{Lauritzen}.
\section{Tangent spaces}
It is common in information geometry to look at a particular, less abstract representation for the tangent spaces than the one used in texts on differential geometry. However, this less abstract representation directly inherits all structure from the general setting. The intuitive choice is to take as the basis vectors of the tangent space $\tg{\theta}{M}$ not the partial derivative operators but rather the derivatives of a particular function of the probability density functions. The simplest choice is
\begin{align}
\label{eq: basevec}
\partial_i p_\theta(x) = \pdiff{}{\theta^i} p_\theta (x).
\end{align}
In the case of the Gaussian density functions, these derivatives become
\begin{align*}
\partial_\mu p_{\mu,\sigma}(x) &= p_{\mu,\sigma}(x) \left( \frac{x-\mu}{\sigma^2}\right)\\
\partial_\sigma p_{\mu,\sigma}(x) &= p_{\mu,\sigma}(x)\left( -\frac{1}{\sigma} + \frac{(x-\mu)^2}{\sigma^3}\right).
\end{align*}
Objects such as these are stochastic variables since they depend on $x$, which is an element of a measurable set. However, for reasons of convenience the practitioners of information geometry often choose a different representation of the tangent vectors. They choose as basis vectors the derivatives of a power-like function of the density function, called the $\alpha$-representation of the tangent space. These objects are given by the expressions, for some $\alpha \in \mathbb{R}$,
\begin{align*}
\partial_i \ell_\alpha(x) = \left\{ \begin{array}{l c l}
\dfrac{2}{1-\alpha}\partial_i \left(p_\theta(x)^{\frac{1-\alpha}{2}}-1\right),&\quad  &\alpha\neq 1,\\
\partial_i \ln p_\theta(x), & &\alpha = 1.
\end{array}\right.
\end{align*}
The previously mentioned basis vectors (\ref{eq: basevec}) are obtained as a special case by choosing $\alpha=-1$. The different representations of basis vectors may seem unconventional at first sight. However, they are the conventional basis vectors to tangent spaces of other manifolds, those consisting of the stochastic variables $x\mapsto \ell_\alpha(x)$. These manifolds are very similar to the statistical manifold and so they can be studied in lieu of it. Because of the similarity, it is also possible to study one manifold while using the tangent spaces of another. It is interesting to note that the functions $\ell_\alpha$ correspond to the $q$-logarithmic functions as introduced by Tsallis \cite{Tsallis} through the relation $\alpha = 1-2q$ \cite{AmariNagaoka,Naudts2004}.\\
Where the most natural representation of the tangent spaces corresponds to the case $\alpha=-1$, another very convenient representation is found when $\alpha=1$ is chosen. Not only does this representation have a clear relation to the log likelihood $\theta \mapsto\ln p_\theta(x)$, the tangent vectors in the $1$-representation all have vanishing expectation value, as
\begin{align*}
 \mathbb{E}_\theta [\partial_i \ln p_\theta ] = \int_X p_\theta(x) \pdiff{}{\theta^i} \ln p_\theta(x) \dee x = \pdiff{}{\theta^i} \int_X p_\theta(x) \dee x = 0.
\end{align*}
The tangent spaces in this representation are thus subsets of the set of stochastic variables whose expectation value vanishes in a sample described by $p_\theta$. Exactly which subset this is depends on $\theta$ as well as on the particular statistical manifold under consideration.\\
This latest property illustrates an important perspective of information geometry: a parametrised family of distributions over some measurable set $X$ is often explicitly thought of as a manifold embedded in the larger, convex set of all distributions over $X$. From that perspective it is can be seen that the tangent planes will depend on the choice of submanifold.\footnote{It is important to keep in mind this picture is only viable when the set $X$ has a finite number of elements. For measurable sets $X$ with an infinite number of elements, the simplex of all probability distributions over $X$ cannot be considered a manifold in the strict sense of the definition as it is given in the previous chapter.} I will not offer a detailed discussion of the other representations in this introduction as their treatment is more elegant for sets of measures which need not be normalised (see for example Chapter 2.6 of \cite{AmariNagaoka} for an introduction to this topic).\\
For ease of reference, the $\alpha{=}1$-representation is called the ``exponential representation" due to its relation with the logarithm and the $\alpha{=}{-}1$-re\-pre\-sen\-ta\-tion is referred to as the ``mixture representation".
\newpage
\section{The Riemannian metric}
In order to make the topological manifold $\mathbb{M}$ into a Riemannian manifold, a choice must be made for the inner product of the tangent spaces. Information geometry almost exclusively makes use of the Fisher information metric \cite{Fisher} for this purpose. This means the inner product of two tangent vectors $\vec{V}$ and $\vec{W}$ is defined to be equal to the covariance (in the statistical sense) of the stochastic variables $V$ and $W$ which correspond to the vectors, that is
\begin{align}
\label{eq: defFMbig}
g( \vec{V},\vec{W})|_\theta = \mathbb{E}_\theta[ VW] - \mathbb{E}_\theta[V]\mathbb{E}_\theta[W].
\end{align}
The most commonly used expression for the components of this tensor are found in the exponential representation, where they are given by
\begin{align}
g_{ij}(\theta) &= \mathbb{E}_\theta[(\partial_i\ln p_\theta)(\partial_j\ln p_\theta)]\nonumber\\
&= \int_X p_\theta(x) \left(\pdiff{}{\theta^i} \ln p_\theta(x)\right)\left( \pdiff{}{\theta^j} \ln p_\theta(x)\right)\dee x.\label{eq: defFM}
\end{align}
As an example, for a manifold of Gaussian distributions, the Fisher information matrix has components
\begin{align*}
g_{\mu\mu}(\mu,\sigma) = \frac{1}{\sigma^2}, \quad g_{\mu\sigma}(\mu,\sigma) = 0 \quad \text{and} \quad g_{\sigma\sigma}(\mu,\sigma) = \frac{2}{\sigma^2}.
\end{align*}
In his 1925 article \cite{Fisher}, Fisher first discussed desirable properties of estimators; functions of the data which are designed such that their values can be used to estimate the parameters of distributions describing the underlying data. He argues estimators must attain a fixed value when the sample size of the data is increased (``consistency'') and their variance multiplied by sample size must be minimal (``efficiency''). This efficiency is important as it lead Fisher to the introduction of the intrinsic precision of a probability density for the purpose of estimating a parameter $\theta$. In particular, for a distribution $p_\theta$ over a set $X$ he defines the quantity
\begin{align*}
g(\theta) = \int_X p_\theta(x) \left( \pdiff{}{\theta} \ln p_\theta(x)\right)^2\dee x,
\end{align*}
which is known as the Fisher information and of which (\ref{eq: defFM}) is the higher\-dimensional generalisation. Fisher himself describes this quantity as
\begin{quote}
\emph{the amount of information in a single observation belonging to such a distribution},
\end{quote}
by which he means the amount of information contained in this observation about the parameter(s) of that distribution \cite{Fisher}. The importance of the Fisher information to statistics, and to the problem of parameter estimation in particular, is contained in the Cram\'er-Rao theorem \cite{Rao}. Unbiased estimators for a parameter $\theta^k$ are stochastic variables $\hat{\theta}^k$ for which $\mathbb{E}_\theta[ \hat{\theta}^k ] = \theta^k$. The Cram\'er-Rao theorem states that after making $N$ observations of these estimators $\hat{\theta}^k$ in a population distributed according to $p_\theta$, it holds that
\begin{align*}
\left\{ \mathbb{E}_\theta[ \hat{\theta}^i \hat{\theta}^j ] - \mathbb{E}_\theta[ \hat{\theta}^i ] \mathbb{E}_\theta[  \hat{\theta}^j ]\right\} - \frac{1}{N}g^{ij}(\theta) \geqslant 0,
\end{align*}
where the inequality means that the expression on the left hand side represents the components of a positive definite matrix. The Fisher information matrix thus expresses the minimal variance these estimators may attain. Furthermore the theorem shows there exists an estimator which attains this lower bound on its covariance as the number of observations tends to infinity. Such an estimator is said to be maximally efficient.\\
In the same paper \cite{Rao} where he introduced the theorem which bears his name, Rao was the first to endow the statistical manifold with the Fisher information matrix as its Riemannian metric. This is possible as the covariance matrix (\ref{eq: defFM}) is positive definite and its components behave as those of a rank two tensor under a change of parameters---the general case of which follows from combining equations (\ref{eq: trafovec}) and (\ref{eq: defcomg})---
\begin{align*}
g_{ij}(\theta) &= \pdiff{\zeta^a}{\theta^i} \pdiff{\zeta^b}{\theta^j}\int_X p_\zeta(x) \left(\pdiff{}{\zeta^a}\ln p_\zeta(x) \right)\left(\pdiff{}{\zeta^b}\ln p_\zeta(x) \right)\dee x\\
&= \pdiff{\zeta^a}{\theta^i} \pdiff{\zeta^b}{\theta^j} g_{ab}(\zeta).
\end{align*}
An application of this transformation property is found in Bayesian probability. Since the Fisher information metric transforms as a rank two tensor, its volume form 
\begin{align*}
\text{vol}(\theta) = \sqrt{ \text{det}(g(\theta))}\text{ } \dee \theta^1 \wedge \ldots \wedge \dee \theta^n
\end{align*}
is unchanged under coordinate transformations. For this reason, Jeffreys suggested to use $\sqrt{\det{g}}$ as a non-informative (though possibly improper) prior on the parameter space \cite{Jeffreys}.\\
As a last property of the Fisher information matrix to be discussed here is that it also appears in the second (and thereby lowest) order term of the expansion of the Kullback-Leibler divergence
\begin{align}
\label{eq: defKLdiv}
D(p||p_\theta) &= \int_X p(x) \ln\left(\frac{p(x)}{p_\theta(x)}\right)\dee x,
\end{align}
which in the physical literature is better known as the relative entropy \cite{Naudts,KullbackLeibler}. More precisely, for $\delta \theta$ sufficiently small it is possible to expand this divergence function as a function of the parameters $\theta$ to obtain
\begin{align*}
&D_{KL}(p_\theta||p_{\theta+\delta \theta})\\
&= -\int_X p_\theta(x)\ln\left( \frac{p_{\theta+\delta\theta}(x)}{p_\theta(x)}\right) \dee x\\
&= - \int_X p_\theta(x) \ln\left( 1 + \frac{\delta p_\theta(x)}{p_\theta(x)}\right)\dee x\\
&= - \int_X p_\theta(x) \left( \frac{\delta p_\theta(x)}{p_\theta(x)} - \frac12 \left(\frac{\delta p_\theta(x)}{p_\theta(x)}\right)^2 + \mathcal{O}((\delta \theta)^3) \right) \dee x\\
&= \frac12 \int_X \frac{1}{p_\theta(x)} \left(\pdiff{}{\theta^j}p_\theta(x)\right)\left(\pdiff{}{\theta^k} p_\theta(x) \right)\dee x \text{ }\delta \theta^j \delta \theta^k + \mathcal{O}((\delta \theta)^3).
\end{align*}
The Fisher information thus expresses also the infinitesimal distance between nearby points on a manifold of probability distributions as expressed by the Kullback-Leibler divergence. This validates Rao's choice to endow statistical manifolds with the Fisher information matrix as their Riemannian metric.
\section{The affine connections}
Another important differential geometrical quantity is the affine connection with which a manifold can be endowed. The first attempt at investigating this structure for statistical manifolds was made by Rao. He studied the metric connection derived from the Fisher information metric and computed geodesic distances \cite{Rao}. However, a statistical interpretation of this metric connection was not immediately obvious \cite{Amari1985}. A breakthrough in the study of affine connections on statistical manifolds came in the 1970s with the work of Chentsov, Efron and Amari.\\
In his notably technical book \cite{Chentsov} and the preceding articles, Chentsov showed that the space of multinomial distributions admits only a single statistically invariant Riemannian metric---the Fisher information metric---and a unique family of statistically invariant affine connections. The statistical invariance means that the geometric quantities defined through the metric tensor and the affine connections remain unchanged when the underlying probability space is mapped into another one through a Markov process. Chentsov's work, though published in Russian in 1972, was only translated into English ten years later. As a consequence it remained unknown outside the Soviet Union for some time after its initial publication.\\
The study of these affine connections outside the Soviet Union started in 1975, when Bradley Efron published the paper ``Defining the Curvature of a Statistical Problem (with Applications to Second Order Efficiency)'' \cite{Efron}. In this article, Efron studies one-parameter families of distributions, seen as curves through the space of all distributions over some measurable set. He implicitly defines a connection this space such that one-parameter exponential families coincide with geodesics, which are curves for which the tangent vector field is covariant constant along the curve. A number of commentary texts have been published together with the article. The last two of these, written by Dawid and Reeds, are probably the most important.\\
Exponential families are parametrised sets of distributions over a (measurable) set $X$ for which there exist functions $\mathcal{H}_k : X \rightarrow \mathbb{R}$ and a function $\Phi: \mathbb{R}^n \rightarrow \mathbb{R}$ such that it is possible to write
\begin{align}
\label{eq: defexpfam}
p_\theta(x) = \exp\left\{ - \Phi(\theta) - \sum_{k=1}^n\theta^k \mathcal{H}_k(x)\right\}
\end{align}
with respect to some given measure \cite{Naudts,Efron}. The parameters $\theta$ are called the canonical parameters of the exponential family. The function $\Phi$ is called the Massieu function \cite{Massieu}. Since the distribution $p_\theta$ must be normalised, the function $\Phi$ satisfies
\begin{align*}
\Phi(\theta) = \ln \int_X \exp\left\{ - \sum_{k=1}^n \theta^k \mathcal{H}_k(x)\right\}\dee x.
\end{align*} In what follows it is always assumed that the values of the parameters $\theta$ are such that $\Phi(\theta) < \infty$.\\
The Gaussian distributions serving as examples in this chapter belong to the exponential family with parameters 
\begin{align*}
\theta^1 = \frac{1}{2\sigma^2}\quad \text{and}\quad \theta^2 = - \frac{\mu}{\sigma^2}
\end{align*}
and Hamiltonians
\begin{align*}
\mathcal{H}_1(x) = x^2 \quad \text{and}\quad \mathcal{H}_2(x) = x.
\end{align*}
This can be seen from rewriting the expression for the Gaussian density into an expression into the form (\ref{eq: defexpfam}).\\
The focus of Efron's paper is on one-dimensional subsets $S$ of exponential families, that is subsets for which it holds that
\begin{align*}
S = \{ p_\theta(x) | \theta = F(\eta), \eta \in N\subset \mathbb{R}\}.
\end{align*}
As Reeds points out in his commentary \cite{ReedsToEfron}, Efron implicitly makes a number of assumptions about the higher-dimensional exponential family. In particular, he takes this set to be a Euclidean space, endowed with a constant metric tensor equal to the value of the Fisher information metric at $\theta_0$, the true value of the parameter (or parameters) that one seeks to estimate. This assumption allows Efron to consider the one-dimensional submanifold $S$ as a curve through Euclidean space and to compute the extrinsic geometrical curvature thereof. This curvature is defined as the length of the vector
\begin{align}
\label{eq: curvEfron}
\diff{\vec{T}}{s}\bigg|_{\here},
\end{align}
where $\vec{T}$ is the tangent of unit length to the set $S$ and $s$ is the arc length of the curve, measured from a point which can be chosen arbitrarily. He then defines the ``statistical curvature'' of $S$ to be equal in value to this geometrical curvature. When the curvature (\ref{eq: curvEfron}) does not vanish, the one-dimensional family is said to be curved in the statistical sense as well as in the geometrical sense. Such a family is therefore called a curved exponential family.\\
The computation of geometrical curvature through formula (\ref{eq: curvEfron}) requires no less than three separate uses of the Riemannian metric (once for the definition of the arc length and twice to compute the length of a vector) as well as a covariant derivative (hidden as a regular derivative with respect to the arc length). This means Efron's assumptions of dealing with a Euclidian space and of the metric tensor having constant components in the coordinate system of canonical parameters are important.\\
The true interesting point of Efron's work lies not in his definition of the statistical curvature as a quantity itself but in the realisation that the square of this curvature plays a crucial role in the properties of the Fisher information metric \cite{Efron}. This establishes a connection between the properties of a statistical estimation problem and the geometry of the set of distributions that is used to model the data. The relation of Efron's work with affine connections is that (non-curved) one-dimensional exponential families coincide with geodesics through the larger space in which they are embedded. It was Dawid who identified the connection implicitly used by Efron in the definition of the geodesics \cite{DawidToEfron}. This connection would later be known as the ``exponential connection'' and is a member of the family identified by Chentsov. Dawid suggested also the ``mixture connection'' in his reply and remarks that both connections are torsionless as well as flat. It should be stressed that these are quantitatively different connections and not different representations of the same connection---despite what the naming similar to that of the exponential and mixture representations of the tangent spaces may suggest. Furthermore, Dawid briefly touches Chentsov's family of connections but he does not study their properties in his reply.\\
Another thing remarked by Reeds in his commentary is that the curvature used by Efron to define statistical curvature is the extrinsic or embedding curvature. This is a subtle but important point as it demonstrates the necessity of considering the exponential family to be embedded in a larger, Euclidean space. In information geometry, this space is the set of all distributions if it is finite dimensional. In the work presented in this dissertation, however, an analogous larger space cannot be specified in general, even though specific examples usually do allow for it. Reeds also initiates the work on a higher-dimensional generalisation of Efron's work. Many authors cite L.\ T.\ Madsen as having completed this task in her doctoral dissertation and the subsequent 1979 paper \cite{Madsen}. Unfortunately, this article seems to be very difficult to obtain, perhaps since it has not appeared in a peer reviewed journal but rather as a research report for the Danish Medical Society. For this reason, it is difficult to tell exactly which of the advances were made by dr.\ Madsen.\\
The work of Amari advanced that of Chentsov, Efron, Dawid and Madsen by introducing a differential geometric framework for the construction of higher-order asymptotic theory of statistical inference (see for instance \cite{AmariNagaoka,Amari1982,Amari2001}) in which the family of connections introduced by Chentsov plays an important role.\\
%The coefficients of these connections are given by (see for instance \cite{})
%\begin{align*}
%\Gamma^{(\alpha)}_{k,ij}(\theta) &= \frac{4}{1-\alpha^2} \int_X \left( \pdiff{}{\theta^k}p_\theta(x)^{\frac{1+\alpha}{2}} \right)\left( \pddiff{}{\theta^i}{\theta^j} p_\theta(x)^{\frac{1-\alpha}{2}} \right) \dee x\\
%&=\int_X p_\theta(x)^{\frac{-1+\alpha}{2}} \left( \pdiff{}{\theta^k} p_\theta(x)\right) \pdiff{}{\theta^i}\left( p_\theta(x)^{\frac{-1-\alpha}{2}} \pdiff{}{\theta^j}p_\theta(x)\right)\dee x\\
%&= \int_X \frac{1}{p_\theta(x)}\left( \pdiff{}{\theta^i} \right)
%\end{align*}
%where these expressions are related to the usual coefficients through the Fisher information metric
%\begin{align*}
%\Gamma^{(\alpha)}_{k,ij}(\theta) = I_{ks}(\theta) {(\Gamma^{(\alpha)})^s}_{ij}(\theta).
%\end{align*}
%The exponential connection of Efron is obtained by taking $\alpha =+1$ and the mixture connection appears for $\alpha=-1$. Even these two special cases may seem technical but they have a very simple interpretation. 
These affine connections are usually presented in a relatively technical way but they have very simple interpretations. This was pointed out by Dawid already in this reply to Efron's paper \cite{DawidToEfron} but this piece of knowledge is---unfortunately---often ignored in the rest of the introductory literature on the subject. Remember that affine connections serve to define covariant derivatives and thereby parallel transport. In the exponential representation, the elements of a tangent space $\tg{\theta}{M}$ are stochastic variables $V$ such that
\begin{align*}
\mathbb{E}_\theta[V] = \int_X p_\theta(x) V(x)\dee x = 0.
\end{align*}
However, there is no guarantee that the expectation value of $V \in \tg{\theta}{M}$ will also vanish when evaluated at another distribution, that is $\mathbb{E}_{\xi}[V] = 0$ cannot be guaranteed when $\xi \neq \theta$. Consequently, parallel transport need not preserve the vanishing of the above expectation value. The family of affine connections of information geometry solves this problem by defining parallel transport in such a way that the expectation value of a stochastic variable remains zero when it undergoes parallel transport. In particular, the exponential connection ($\alpha=1$) defines the operation $\Pi^{(1)}: \tg{\theta}{M} \rightarrow\tg{\xi}{M}$ by explicitly subtracting the expectation value in the end point:
\begin{align*}
\Pi^{(1)} V = V - \mathbb{E}_\xi[V].
\end{align*}
The mixture connection ($\alpha=-1$) on the other hand multiplies the statistic with the appropriate Radon-Nykodym derivative (see for instance \cite{Chentsov} or an introductory work on the matter), that is $\Pi^{(-1)} : \tg{\theta}{M} \rightarrow \tg{\xi}{M}$ works as
\begin{align*}
\Pi^{(-1)} V = V \frac{p_\theta}{p_\xi}.
\end{align*}
This means the expectation value equals
\begin{align*}
\mathbb{E}_{\xi}[\Pi^{(-1)} V] &= \int_X V(x) \frac{p_\theta(x)}{p_\xi(x)} p_\xi(x) \dee x\\
&= \int_X V(x) p_\theta(x) \dee x\\
&= \mathbb{E}_\theta[V]\\
&= 0.
\end{align*}
The other affine connections of the $\alpha$-family are simply linear combinations of the exponential and mixture connections, in particular
\begin{align*}
\nabla^{(\alpha)} = \frac{1+\alpha}{2} \nabla^{(1)} + \frac{1-\alpha}{2} \nabla^{(-1)}.
\end{align*}
A similar interpretation exists for the other representations of the tangent space but this requires the introduction of the notion of escort probability distributions \cite{AmariNagaoka}. This is closely related to the work of Tsallis and an extensive introduction of escort probabilities can be found in \cite{Naudts} but to treat this explicitly would fall outside the scope of this introduction.\\
Even though the exponential and mixture connections give rise to a path-independent definition of parallel transport and are thus flat, the rest of the $\alpha$-family has a constant but non-vanishing scalar curvature (also known as the Ricci scalar of the connection), which is given by \cite{Amari1985}
\begin{align*}
 R \define g^{ij} {\Omega^k}_{ikj} = \frac{1-\alpha^2}{4}.
\end{align*}
The metric connection associated with the Fisher information metric is obtained in the case $\alpha=0$ and so it is the most (positively) curved member of this family. As is elucidated in the previous chapter, the metric connection is the unique torsionless affine connection satisfying
\begin{align*}
\vec{X} (g(\vec{Y},\vec{Z})) = g(\nabla_{\vec{X}}\vec{Y},\vec{Z}) + g(\vec{Y},\nabla_{\vec{X}} \vec{Z}).\tag{\ref{eq: defLCcon} revisited}
\end{align*}
This property only holds when $\alpha=0$ and in general it can be shown that for any $\alpha \in \mathbb{R}$
\begin{align}
\vec{X} (g(\vec{Y},\vec{Z})) = g(\nabla^{(\alpha)} \vec{X},\vec{Y}) + g(\vec{X},\nabla^{(-\alpha)} \vec{Y}),
\end{align}
where $g$ represents the Fisher information metric. This shows that the $\alpha$- and $-\alpha$-connections are dual with respect to this metric. Dual connections play an important role in the more advanced topics of information geometry. One application is found in a generalisation of the Pythagorean property, which holds in a triangle where one leg is a geodesic for the exponential connection and the other leg is a geodesic for the mixture connection \cite{AmariNagaoka}.

\section{Divergence functions}
Due to the importance in information geometry of the relative entropy (\ref{eq: defKLdiv}), a central role in this dissertation shall be played by divergence functions, or as they are sometimes also called, contrast functions. They shall serve as the elementary structure from which all geometric objects shall be constructed, just as all the Riemannian geometry of statistical manifolds elucidated above can be derived from the divergence function of Kullback and Leibler.\\
The Kullback-Leibler distance is a function quantifying in a certain sense the difference between statistical distributions over a measurable set $X$. As was mentioned earlier in this chapter, it is given by the expression
\begin{align*}
D(q||p) = \int_X q(x) \ln\left( \frac{q(x)}{p(x)}\right)\dee x.
\end{align*}
It should be noted that this integral is always defined, even though its value may be infinite, as the measures from which $p$ and $q$ are derived are finite \cite{KullbackLeibler}. The relative entropy plays an important role not only in statistics but also in information theory. In fact, it was introduced by Kullback and Leibler as an abstraction of Shannon's entropy
\begin{align}
\label{eq: Shannon}
S(p) = -\int_X p(x) \ln p(x) \dee x,
\end{align}
which in turn was introduced for purposes of information theory in \cite{Shannon}, even though the expression also appears in the work of Boltzmann and Gibbs. The quantity 
\begin{align*}
\ln\left( \frac{q(x)}{p(x)}\right) 
\end{align*}
is a measure for the information obtained in the result of a measurement $x$ for discrimination between the hypotheses ``$x$ is distributed according to the distribution $q$'' and ``$x$ is distributed according to the distribution $p$''. The Kullback-Leibler divergence is therefore equal to the mean information for discrimination \cite{KullbackLeibler} and it is in this sense that it can be said to quantify the difference between its arguments.\\
A commonly used set of contrast functions are the $f$-divergences of Csisz\'ar \cite{Csiszar}. They are of the form
\begin{align*}
D_f(q||p) &= \int_X p(x) f\left( \frac{q(x)}{p(x)}\right) \dee x,
\end{align*}
where $f$ is a convex function for which $f(1) = 0$. The Csisz\'ar $f$-divergences are the largest class of statistically invariant divergence functions. Another often made choice is the set of Bregman divergences \cite{Bregman}. The most general definition of these divergences is not limited to statistical distributions but when restricted thereto, it is possible to write them in the form
\begin{align*}
D_F(q||p) &= \int_X \int_{p(x)}^{q(x)} [F^\prime(u) - F^\prime(p(x))] \dee u \dee x .
\end{align*}
where $F$ is a strictly convex function. These are also known as $U$-divergences, after their definition by Eguchi \cite{EguchiEtAl}. It can be shown that the Kullback-Leibler divergence is the only contrast function which belongs to both the classes of Csisz\'ar and Bregman divergences \cite{Amari1985}.\\
In order to gain more insight in general divergence functions, it is interesting to compare them to the more familiar metrics---not to be mistaken with metric tensors. A metric on a set $S$ is a function
\begin{align*}
d: S\times S \rightarrow \mathbb{R}^+
\end{align*}
which is zero everywhere on the diagonal of $S\times S$ and strictly positive elsewhere, is symmetric ($d(x,y) = d(y,x)$) and satisfies the triangular inequality
\begin{align*}
d(x,y) + d(y,z) \geqslant d(x,z).
\end{align*}
Divergence functions on $S$ on the other hand are only required to satisfy the first condition. An example of a symmetric divergence is the \emph{squared} Euclidean distance in $\mathbb{R}^3$, which satisfies the cosine rule instead of the triangular inequality:
\begin{align*}
d(x,y)^2 + d(y,z)^2 = d(x,z)^2 + 2d(x,y)d(y,z)\cos(\phi),
\end{align*}
with $\phi$ the angle between the legs of the triangle meeting at the point $y$.\\
Once an appropriate divergence function $D$ over a manifold $\mathbb{M}$ has been chosen, it can be used to define a differential geometric structure upon $\mathbb{M}$, see for example \cite{Eguchi1992,AmariCich}. In order for this to be possible, the divergence must be sufficiently many times continuously differentiable with respect to the coordinates of its arguments and this at least in a neighbourhood of the diagonal of $\mathbb{M}\times\mathbb{M}$. A lot of interesting work in this context has been due to Eguchi and collaborators \cite{EguchiEtAl,Eguchi1983,Eguchi1985} but also due to Amari and collaborators \cite{Amari1982,Amari2001,AmariCich}. In their research, they have investigated the geometric structure of statistical manifolds as well as more general manifolds endowed with divergence functions and applications thereof.\\
The first geometric structure to be introduced is, as usual, the metric tensor. Since divergence functions over a manifold $\mathbb{M}$ are zero everywhere on the diagonal of $\mathbb{M}\times \mathbb{M}$ and only there, it must automatically hold that the lowest order term in its Taylor expansion is the second order term, that is
\begin{align*}
D(m_{\theta}||m_{\theta+\delta \theta}) &= \frac12 \pddiff{}{\xi^i}{\xi^j} D(m_\theta || m_\xi ) \bigg|_{\xi = \theta} \delta \theta^i \delta \theta^j + \mathcal{O}((\delta \theta)^3).
\end{align*}
The coefficient of this lowest order term---without the factor $\tfrac12$---is the metric tensor induced by the divergence:
\begin{align}
\label{eq: definfogeomg}
g_{ij}(\theta) &= \pddiff{}{\xi^i}{\xi^j} D(m_\theta || m_\xi ) \bigg|_{\xi = \theta}.
\end{align}
This is a positive definite quantity as it is the matrix of second derivatives of a function in a local minimum. Despite this, it behaves properly under a coordinate transformation, as can be seen also from the alternative expression
\begin{align*}
g_{ij}(\theta) &= -\pddiff{}{\xi^i}{\theta^j} D(m_\theta || m_\xi ) \bigg|_{\xi = \theta},
\end{align*}
which can be shown to hold since the first derivatives of the divergence vanish identically on the diagonal.\\
Affine connections can be constructed from divergence functions just as well. In particular, it is possible to consider the pair of mutually dual connections $\nabla$ and $\nabla^*$ defined through the expressions \cite{AmariCich,Eguchi1983}
\begin{align*}
g_{ks}(\theta) {\omega^s}_{ij}(\theta) = -\frac{\partial^3}{\partial \theta^i \partial \theta^j \partial \xi^k} D(m_\theta||m_\xi)\bigg|_{\xi=\theta}
\end{align*}
and 
\begin{align}
\label{eq: definfogeomomega}
g_{ks}(\theta) {\varpi^s}_{ij}(\theta) = -\frac{\partial^3}{\partial \xi^i \partial \xi^j \partial \theta^k} D(m_\theta||m_\xi)\bigg|_{\xi=\theta}.
\end{align}
For more background regarding the geometry induced by divergence functions, the reader is referred to the work of Amari and Eguchi cited in the text above.

\section{Applications in thermodynamics}
The material introduced so far in this chapter is almost exclusively based upon the mathematical literature. However, it deserves mentioning that some of these geometrical aspects have also been adopted by the community of researchers in thermodynamics. Since the necessary background for the following chapters has already been introduced, this overview shall be kept fairly short. Nevertheless, the historical development of this particular field of research shows some interesting parallels with the goal of this dissertation. Indeed, both are instances where geometry is applied in an attempt to unearth the mathematical foundations of a physical theory of which the basics are less rigorous.\\
The first connection between differential geometry and thermodynamics was made by Constantin Carath\'eodory, who sought to establish an axiomatic basis for thermodynamics. He chose to express this in terms of differential geometry \cite{Caratheodory1909,Caratheodory1925}. In these papers he could phrase thermodynamics on sound mathematical principles, rather than on the more usual references to imaginary devices such as Carnot engines or to concepts such as the flow of heat.\footnote{It is interesting to note that Carath\'eodory introduces heat as a derived rather than as a fundamental quantity, which is the approach of conventional thermodynamics.} For this, he works on the topological manifold of thermodynamic states of the system. His rendering of the Second Law states \cite{Frankel,Pogliani}
\begin{quote}
\emph{In every neighbourhood of every equilibrium state $x$, there are states $y$ that are not accessible from $x$ via quasi-static adiabatic paths.}
\end{quote}
This formulation is weaker than Kelvin's better known one, which states that no cyclical process can exists which turns heat into its mechanical equivalent of work. Starting from this axiom, Carath\'eodory could derive thermodynamics as it was known in his time. His results therefore extend those of Helmholtz, who had already noticed that a definition of temperature or entropy does not require cyclical processes or ideal gasses \cite{Pogliani}.\\
The next step in building a geometric basis for thermodynamics and statistical mechanics comes from Tisza \cite{Tisza} and Griffiths and Wheeler \cite{GriffWheeler}, whose contributions may be deemed important more for steering geometric thermodynamics away from the difficult formalism of Carath\'eodory than for their actual results. A few years later, Weinhold published a series of papers regarding the metric geometry of equilibrium thermodynamics \cite{Weinhold}. His main contribution, at least in the context of this discussion, is to endow the equation of state surface appearing in the Gibbs formulation of equilibrium thermodynamics with a Riemannian metric. This is a completely different approach from that of Carath\'eodory, which is based primarily on differential $1$-forms over the set of states. Weinhold's metric has components given by
\begin{align*}
g_{ij} = \pddiff{U}{N^i}{N^j}
\end{align*}
where $U$ is the internal energy and the set of $N^i$ contains the independent conserved extensive quantities of the system (such as volume and particle numbers) and the entropy. This is a matrix containing quantities related to standard thermodynamic linear response functions such as compressibility and specific heat. Due to the convexity of the internal energy in single phase regions, the metric $g$ is positive definite. Application of the Cauchy-Schwarz and Bessel inequalities allowed Weinhold to derive many of the standard thermodynamical inequalities. He did not, however, compute distances between points of the equilibrium surface \cite{AndresenEtAl}.\\
Four years later, Ruppeiner published an interpretation of the metric structure and introduced an intrinsic rather than an extrinsic geometry \cite{Ruppeiner}. He does this not by endowing the equilibrium surface with a metric but rather by introducing the metric on an abstract manifold of equilibrium states. Apart from this difference, his metric tensor ($g^{(R)}$) is related to the one introduced by Weinhold ($g^{(W)}$) through a conformal transformation:
\begin{align*}
g^{(R)} = T g^{(W)}.
\end{align*}
The interpretation Ruppeiner gives to this metric tensor is very simple. It expresses the distance between neighbouring states in the sense that the more likely a fluctuation bringing one equilibrium state into another is to occur, the closer they are in the Ruppeiner metric. A related result is that the geodesic distance is related to the diffusion of the system through state space by fluctuations.\\
Furthermore, Ruppeiner argues that curvature exhibited by the metric connection is due to interactions in the fluid, as the interactionless ideal fluid gives rise to a geometry exhibiting no curvature. He goes on to expand upon this idea by arguing that the curvature is proportional to the cube of the correlation length of the system. He does this through a line of reasoning based on dimensional analysis and scaling relations. This leads to the identification of universal constants, a result supported by experimental evidence (see \cite{Ruppeiner} for the details).\\
The metric tensors introduced by Weinhold and Ruppeiner give rise to a measure of distance sometimes known under the name of thermodynamic length (see for instance \cite{Crooks} and references therein). Nevertheless, there is also criticism on the view that it is possible to see this metric tensor as suitable to define a meaningful distance between points on the equation-of-state surface. This was elucidated by Gilmore \cite{Gilmore1984}. He showed that also using the positive definite quantity introduced by Weinhold---which is the second fundamental form of this surface---as the metric tensor leads to constraints on the third derivatives of the internal energy of a system. Such constraints, however, are nowhere to be found in thermodynamics.\\
Another interesting result obtained by the application of geometrical considerations is found in other work by Gilmore \cite{Gilmore1985}. Basing himself on fluctuation theory just like Ruppeiner and applying the Cram\'er-Rao bound (which is nothing but a consequence of the Cauchy-Schwarz inequality from a geometrical point of view), he was able to obtain uncertainty relations for thermodynamical quantities in a system undergoing fluctuations. More precisely, if a system is in equilibrium with a reservoir with well-defined intensive variables, then variations in the measured values of the system's extensive variables will lead to variations in the estimation of the reservoir's corresponding intensive variables. Gilmore's result states that the product of the variances of the distributions of these quantities must be larger than a constant, for example
\begin{align*}
\Delta U \Delta \frac1T \geqslant k_B,
\end{align*}
where the right hand side here represents Boltzmann's constant. While this particular example was already known by Gibbs \cite{Gibbs}, Gilmore's line of reasoning can be applied to any conjugate pair of intensive or extensive variables. Furthermore, it is possible to show that these relations are equivalent to the stability criteria of equilibrium thermodynamics (see again \cite{Gilmore1985} for the details).

\newpage

\chapter{The data set model formalism}
This is the central chapter of this dissertation. In it, the development of the data set model formalism is outlined. First, the different elements which must be present in order for the formalism to be applicable are outlined and discussed in detail. Afterwards, the geometry of the data set models will be constructed and the general consequences investigated. Concrete examples and applications are examined in the next chapter.
\section{The elements of the formalism}
\subsection{The data sets}
Consider a set $\mathbb{X}$ of mathematical objects, called data sets, which one would like to model by representing them by the elements of a parametrised set. In principle $\mathbb{X}$ can be any set but the work set out herein will assume $\mathbb{X}$ to be endowed with a topology. This is truly an assumption and not a demand; the formalism can work without such structure. However, it is expected that when there is a topology on the set $\mathbb{X}$, as is often the case, one will desire that this topology is respected by the modelling process. This will be accommodated by the data set model formalism through for instance the demand that the mapping from data to model point is continuous. In order to make the discussion of this aspect meaningful, it is thus assumed that $\mathbb{X}$ has indeed been endowed with a topology.\\
In statistical physics, the data sets are elements of the simplex of distributions over a measurable set---they represent the empirical data obtained from an experiment \cite{Naudts}. Another often encountered example is a collection of measurements for which a functional relation is to be found, as is for example the case in linear regression. In a quantum mechanical setting, the data to be modelled could be data obtained through measurements performed on an ensemble of states. An example from machine learning could be a fingerprint acquired at the scene of a crime and of which the general class has already been determined but which still needs to be characterised to be stored efficiently in a database through which to search in future investigations.
\subsection{The model points}
The data sets in $\mathbb{X}$ are intended to be modelled by an element $m$ of a parametrised set, which will be called a model point. In order for the methods set out in this dissertation to be applicable, this parametrisation must be a ho\-meo\-mor\-phism. As such, the parameters can also play the role of coordinates and the model points will constitute a topological manifold. The symbol $\mathbb{M}$ will henceforth be used to denote the manifold of models---and no longer an arbitrary manifold. When the parameters are represented by $\theta$, the points of the manifold will often be denoted as $m_\theta$, in analogy with the parametrised distributions $p_\theta$. As it is the the case in information geometry, it is important to keep in mind that the number of parameters must be finite in order to practice differential geometry with the particular mathematical techniques used here. This finite number, which by definition is also the dimension of the manifold $\mathbb{M}$, is represented by the letter $n$ in this chapter. Examples of infinitely dimensional models would be the Hilbert space of some quantum systems such as the harmonic oscillator, models parametrised by response functions\ldots\ . \\
Perhaps the most common example of such a model is the family of first order polynomials
\begin{align*}
\{ f | f(x) = ax + b, \text{ }(a,b) \in \mathbb{R}^2\} 
\end{align*}
used to fit data points in a plane when practising linear regression. In a quantum setting, parametrised subsets of the Hilbert space provide an obvious choice for the models of the formalism. Earlier published work in this context considered the coherent states of the (one-dimensional) quantum harmonic oscillator, which is a two-dimensional family of the most classical states of that quantum system \cite{AnthonisNaudts1}. In the above example of the fingerprint, once the general class has been determined, the exact ridge pattern could be described by a limited number of continuously varying parameters indicating the positions of characteristic points of the pattern.
\subsection{The model map}
The actual process of modelling, determining which model point is chosen to represent a given data set, happens by means of the model map $\mu$. This map is required to be continuous (when a topology on $\mathbb{X}$ has been chosen) as a small change in the data set should never cause a large change in the model chosen to represent it.\\
Furthermore, the model map and the divergence function must be compatible. This condition will be elucidated in the upcoming discussion of the divergence. A consequence of this is, however, that the domain of $\mu$ may be limited to a subset of $\mathbb{X}$. Therefore it is more correct to identify $\mathbb{X}$ as the domain of $\mu$ and to take into account the possibility that this domain is in fact a subset of a larger set. For the sake of succinctness this remark will not be reiterated explicitly in those instances where the distinction is clear from context. A similar remark holds for the image of the model map; the geometric quantities that are introduced can only be defined on the (topological) closure of the image of $\mu$. In order for the data set model to be applicable, this closure must thus be a manifold in its own right.\\
The model map can be assumed not to be injective, as it would make the process of modelling redundant. The notation $\mu^{-1}(m)$ will therefore be used to indicate the collection of those data sets $x$ for which $\mu(x)=m$. These subsets of $\mathbb{X}$ upon which $\mu$ attains a constant value will be called fibres. For reasons of convenience the metaphor of fibre bundles is also used in saying a model point $m$ is the projection of a data set $x$, by which again it is meant that $\mu(x) = m$. It should be noted that this nomenclature is suggestive rather than rigorous. The set $\mathbb{X}$ can be likened to the total space of a fibre bundle, the manifold $\mathbb{M}$ to the base space and the model map $\mu$ to the projection but together these objects cannot be properly called a fibre bundle. The definition of a proper fibre bundle requires the fibres to be copies of the same space \cite{Frankel,Taubes} and this is not generally the case for the sets $\mu^{-1}(m)$.\\
In certain contexts, such as a physical experiment, one may not have access to the data sets itself but only to a limited list of observable quantities. When this occurs, the data sets may acquire extra structure. In the simplest case the data sets can only be described by a finite number of real numbers, in which case $\mathbb{X}$ will adopt a manifold structure. Also the divergence function may induce more properties upon the set $\mathbb{X}$. Though this may give rise to specific consequences, these were not investigated explicitly and hence this dissertation does not report upon results concerning such additional structure.
\subsection{The divergence}
A crucial element in the modelling process considered is a divergence function, which is consistently denoted as $D$. The value $D(x||m)$ is a measure of how well the model point $m \in \mathbb{M}$ describes the data set $x \in \mathbb{X}$ and where a smaller value indicates a better match. As such, it can be thought of as a ``badness of fit''. An alternative viewpoint is that $D(x||m)$ represents the cost of describing $x$ by means of $m$. Such perspective can be found treated in more detail for instance in the work of Tops\o e \cite{Topsoe1,Topsoe2,Topsoe3} but this idea is not actively entertained in this dissertation.\\
The divergence function must satisfy four conditions, most of which are in some way related to the properties of divergence functions as they were treated in the previous chapter. These conditions are:
\begin{enumerate}
\item The domain of $D$ contains $\text{Dom}(\mu)\times \text{Im}(\mu)$. In many practical problems the divergence will imply the model map and its properties, making this condition largely trivial.
\item The divergence is sufficiently many times continuously differentiable with respect to the parameters of its second argument. This demand must hold for each of the different coordinate systems used.
\item Given any $x \in \text{Dom}(\mu)$, the function
\begin{align*}
m \mapsto D(x||m)
\end{align*}
has a local minimum at $m = \mu(x)$. This is the condition of compatibility of the model map and the divergence that was mentioned earlier. That $\mu(x)$ provides a unique global minimum is only strictly required for the generalised Pythagorean theorem. Nevertheless, this will be assumed to hold as well.
\item For all $m_\theta \in \text{Im}(\mu)$, the function
\begin{align*}
x \mapsto \partial_i \partial_j D(x||m_\theta) 
\end{align*}
has a constant value on the fibre of $m_\theta$. The drawings illustrate this schematically for two divergence functions on a $1$-dimensional $\mathbb{M}$. For both drawings, two data points $x$ and $y$ within the same fibre are chosen. In the leftmost drawing, the curvature of both graphs in the minimum is the same, though this is not the case elsewhere, and the condition is satisfied. In the rightmost drawing, the curvature is not the same in the minimum even though the graphs may seem more similar---they are both parabolas---and the condition is not satisfied.
\begin{center}
\begin{tikzpicture}
\path[draw,white] (-5,0) circle (0.2);
\path[draw,->] (-4.25,0)--(-0.75,0);
\path[draw,->] (-4,-0.25)--(-4,2.5);
\path[draw] (-0.55,-0.3) node {$\theta$};
\path[draw] (-4.3,2.7) node {$D$};

\path[draw,domain=-4.25:-0.75,smooth,variable=\x] plot( \x,{1/2*(\x+2.5)^2*( 1 - exp( -3.5/((\x+2.5)^2+0.01) )) + 0.25 });
\path[draw,domain=-4.25:-0.75,smooth,variable=\x,dashed] plot( \x,{1/2*(\x+2.5)^2 + 0.5 });

\path[draw,->] (0.75,0)--(4.25,0);
\path[draw,->] (1,-0.25)--(1,2.5);
\path[draw] (4.45,-0.3) node {$\theta$};
\path[draw] (0.7,2.7) node {$D$};

\path[draw,domain=0.75:4.25,smooth,variable=\x] plot( \x,{1/4*(\x-2.5)^2+0.25});
\path[draw,domain=0.75:4.25,smooth,variable=\x,dashed] plot( \x,{1/2*(\x-2.5)^2+0.5});

\path[draw] (5,2.4)--(5.5,2.4) node[right] {$D(x||m_\theta)$};
\path[draw,dashed] (5,1.8)--(5.5,1.8) node[right] {$D(y||m_\theta)$};
\end{tikzpicture}
\end{center}
\end{enumerate}
From the assumption of a global minimum follows that the divergence is bounded from below, at least on $\text{Dom}(\mu)\times\text{Im}(\mu)$. Without loss of generality, this function can then be assumed to attain positive values only. In previous work \cite{AnthonisNaudts1,AnthonisNaudts2} a stronger condition was imposed but this turns out to be largely unnecessary. Furthermore, it is often convenient to impose extra conditions, depending on the presence of a topology for $\mathbb{X}$.
\begin{itemize}
\item[5.] When a topology for $\mathbb{X}$ has been specified, it will be assumed that the divergence is also continuous in its argument. (The continuity in the second argument is already implied by Condition 2.)
\item[6.] Given any $x \in \text{Dom}(\mu)$, there exists a neighbourhood $\mathcal{N}_x$ of $x$ wherein there exist data sets $x^{(l)}$ such that the numbers
\begin{align*}
\partial_k D(x^{(l)}||\mu(x)) 
\end{align*}
make up the components of an invertible matrix.\\
A stronger version of this condition is that in this neighbourhood, the function $y \mapsto \partial_k D(y||\mu(x)) \dee \theta^k$ is continuous and has an image homeomorphic to a subset of $\mathbb{R}^n$. Then there exist functions $X^l$ mapping a neighbourhood of 0 ($\in \mathbb{R}$) to $\mathcal{N}_x$ such that
\begin{align*}
\diff{}{\varepsilon} \partial_k D(X^l(\varepsilon)||\mu(x))\bigg|_{\varepsilon=0} = \delta_k^l.
\end{align*}
This is a construction similar to one used in the study of gradient flows on metric spaces---see for instance \cite{Ambrosio} for an introduction to that discipline.
\end{itemize}
At some points in this dissertation, it will be necessary to distinguish between the definition of a divergence function given here and the more restrictive definition that is presented in the introduction. To make the distinction where confusion may arise, the latter kind will sometimes be referred to as a ``proper divergence (function)''.\\
Together the sets and maps $\mathbb{X}$, $\mathbb{M}$, $\mu$ and $D$ make up what will called the data set model $(\mathbb{X},\mathbb{M},\mu,D)$. In what follows, the geometry of these models will be studied and it will be elucidated how this geometry gives rise to a method to easily determine whether or not a given parametrised family of distributions belongs to an exponential family.
\section{The geometry of data set models}
\subsection{Topology}
For the sake of completeness, it is worthwhile to consider for a moment the topologies of the set $\mathbb{X}$ and the manifold $\mathbb{M}$. Many, if not all, important functions in this dissertation have these two sets (or at least subsets thereof) as either their domain or as their co-domain and many of these functions can be demanded to be continuous. Examples thereof include the model map ($\mathbb{X} \rightarrow \mathbb{M}$), the divergence function ($\mathbb{X}\times \mathbb{M} \rightarrow \mathbb{R}$) and the coordinate functions ($\mathbb{M} \rightarrow \mathbb{R}^n$).\\
That $\mathbb{M}$ is a manifold is actually a property of its topology. In particular, the local homeomorphism of $\mathbb{M}$ with $\mathbb{R}^n$ is equivalent to stating that neighbourhoods of model points in $\mathbb{M}$ are, from a topological point of view, Tychonoff spaces \cite{Willard}.
\subsection{The Riemannian metric}
In order to induce a geometric structure on the manifold of models, a metric tensor must be constructed. This can be done in analogy with the Fisher information metric and the metric tensor derived from divergence functions. In those examples, the components of the metric tensor are second derivatives of the divergence. The geometric interpretation of this is a generalisation of Fisher's ideas as they were discussed in the previous chapter.\\
Fix a data set $x$ and consider the surface defined by the graph of the map
\begin{align*}
\theta \mapsto D(x||m_\theta)
\end{align*}
on a neighbourhood of the coordinates of $\mu(x)$. In order to obtain a good measure of information in Fisher's sense, it is required to look at how strongly peaked the graph is around its minimum as this encodes the sensitivity of the divergence for small variations in the chosen model point. This sharpness is nothing else than the extrinsic or embedding curvature of the surface in $\mathbb{R}^{n+1}$ and this can be quantified using the theory of surface geometry. In particular, it is encoded in the second fundamental form and it can be shown that in the minimum, the curvature is quantified by the values
\begin{align*}
\partial_i \partial_j D(x||\mu(x)).
\end{align*}
The logical choice for the metric tensor would therefore be an expression of the form
\begin{align}
\label{eq: defg}
 g_{ij}(\theta) = \partial_i \partial_j D(x||m_\theta) \quad \text{with} \quad \mu(x) = m_\theta.
\end{align} 
This expression is only well-defined, however, if the divergence satisfies Condition 4 set out earlier in this chapter. The tensor (\ref{eq: defg}) will be called the generalised Fisher information metric. Although the it is defined as a second derivative, which in general does not behave under coordinate transformations in the way a metric tensor ought to, no problems arise with this definition.
\begin{Invariance}
The definition (\ref{eq: defg}) for the metric tensor is invariant under coordinate transformations.
\end{Invariance}
\begin{proof}
Choose $x \in \mu^{-1}(m_\theta)$ and a transformation from the coordinates $\{\theta^i\}$ to the coordinates $\{\zeta^a\} = \{Z^a(\theta)\}$. Then the components of the metric tensor transform as
\begin{align*}
g_{ij}(\theta) &= \partial_i \partial_j D(x||m_\theta)\\
 &= \pdiff{Z^a}{\theta^i} \pdiff{Z^b}{\theta^j} \partial_a \partial_b D(x||m_{Z(\theta)}) + \pddiff{Z^a}{\theta^i}{\theta^j} \partial_a D(x||m_{Z(\theta)})\\
&= \pdiff{Z^a}{\theta^i} \pdiff{Z^b}{\theta^j} \partial_a \partial_b D(x||m_{Z(\theta)})\\\
&=  \pdiff{Z^a}{\theta^i} \pdiff{Z^b}{\theta^j} g_{ab}(Z(\theta)).
\end{align*}
This completes the proof.
\end{proof}
\noindent
The expression (\ref{eq: defg}) achieves this coordinate invariance without changing the second derivative into a Hessian, which would require a connection to be chosen. Since such a choice would necessarily be arbitrary at this point, it is a benefit of this approach that this scenario can be avoided. As such, a metric can be defined under relatively weak conditions; it will be shown shortly that the introduction of the affine connection places a stronger demand on the divergence.\\
The Kullback-Leibler divergence satisfies Condition 4 when $\mathbb{M}$ is an exponential family of probability distributions, since---with $\{\theta^i\}$ the canonical parameters---
\begin{align*}
\partial_i \partial_j D(p||p_\theta) &= -\pdiff{}{\theta^i} \int_X p(x) \pdiff{}{\theta^j} \ln p_\theta(x)\intdee x\\
&= \partial_i \partial_j \Phi(\theta).
\end{align*}
This is obviously a strong way to satisfy this condition, as the derivatives of the Massieu function $\Phi$ are completely independent of the distribution $p$. However, it is not necessary for the existence of a Riemannian geometry that such a strong condition is imposed. An example of a data set model outside the field of information geometry allowing for such a metric tensor is that of the grand canonical ensemble of bosonic particles, treated in detail in the next chapter. The Gumbel distributions discussed shortly thereafter are an example of a family of distributions which do not satisfy Condition 4 when the same Kullback-Leibler divergence is used.

\subsection{On the properties of exponential families}
After a metric tensor has been constructed for a data set model, the next geometric structure to be treated is the affine connection. In order to guide the development of the data set model formalism, it is useful to first discuss particular properties of exponential families.\\
As is mentioned in the introduction, Efron introduced the statistical cur\-va\-ture of one-dimensional submanifolds of exponential families. This was generalised to higher dimensions, first by Reeds and then by Madsen and Amari. A recurring property in their work is that the manifold of distributions belonging to an exponential family is flat when it is endowed with the exponential connection. (Amari has also shown similar results for generalised families, see \cite{Amari1985}.) This is an important element of the argument for the choice of affine connection for data set models.\\
Exponential families are parametrised sets of distributions that can be written in the canonical form
\begin{align}
\label{eq: expdist}
p_\theta(x) = \exp\{ -\Phi(\theta) - \theta^k \mathcal{H}_k(x) \}.
\end{align}
This means that for any distribution $p$, the second derivative of the Kullback-Leibler divergence evaluated in the pair $(p,p_\theta)$ satisfies
\begin{align*}
\partial_i \partial_j D(p||p_\theta) = \partial_i \partial_j \Phi(\theta),
\end{align*}
still in coordinates coinciding with the canonical parametrisation. Using arbitrary coordinates $\{\zeta^a\}$ and the notation of data set models, this becomes
\begin{align}
\label{eq: DallowsHess}
\nabla_a \nabla_b D(x||m_\zeta) = \nabla_a \nabla_b \Phi(\zeta)
\end{align}
for all $x \in \mathbb{X}$. That the Hessian of the divergence function does not depend its first argument is an important property since it enables the construction of an affine connection for the model manifold in the general formalism. As an additional advantage, it will be shown that the property (\ref{eq: DallowsHess}) cannot hold for an arbitrary connection. In fact that connection is unique---a claim the proof of which will be provided by the explicit construction of the coefficients in the next subsection.\\
Its uniqueness is not the only interesting property this connection exhibits however. It turns out that equation (\ref{eq: DallowsHess}) gives rise to a rich geometry. For this reason, data set models satisfying this equation are said to exhibit a Hessian structure. Another name could be ``Legendre structure''. This name is inspired by its use for very similar---though not necessarily identical---ideas in the literature \cite{Ohara,AmariChap,OharaMatsuzoeAmari}. The discussion of this structure and its properties provides the content of the next subsection. It is only afterwards that the connection will be constructed starting from equation (\ref{eq: DallowsHess}), thereby also showing its uniqueness. 

\subsection{The Hessian structure}
The property defining the Hessian structure for a data set model ($\mathbb{X},\mathbb{M},\mu,D$) is the relation (\ref{eq: DallowsHess}),
\begin{align*}
\nabla_a \nabla_b D(x||m_\zeta) = \nabla_a \nabla_b \Phi(\zeta).\tag{\ref{eq: DallowsHess} revisited}
\end{align*}
However, it is a sufficient condition that the left hand side of this expression is independent of the data set $x$. That there then always exists a function $\Phi$ satisfying equation (\ref{eq: DallowsHess}) is presented as a Theorem below. Note that this property implies a well-defined metric tensor on $\mathbb{M}$; it is a stronger version of Condition 4.\\
A most useful observation is that the Hessian of the divergence is equal to the metric tensor, which can be shown through a short computation. Indeed, choose $x\in\mu^{-1}(m_\zeta)$, then
\begin{align*}
\nabla_a \nabla_b D(x||m_\zeta) &= \partial_a \partial_b D(x||m_\zeta) - {\omega^k}_{ij}(\zeta) \partial_k D(x||m_\zeta)\\
&= \partial_a \partial_b D(x||m_\zeta)\\
&= g_{ab}(\zeta).
\end{align*}
Even though this argument only holds for data sets which have $m_\zeta$ as their projection on $\mathbb{M}$, the condition (\ref{eq: DallowsHess}) implies that 
\begin{align}
\nabla_a \nabla_b D(x||m_\zeta) &= g_{ab}(\zeta) \qquad \forall x  \label{eq: DallowsHesseq}
\end{align}
Another consequence of equation (\ref{eq: DallowsHess}) is that it imposes conditions on the properties of the connection $\nabla$. The first of these is that $\nabla$ is torsionless or, expressed through its coefficients
\begin{align}
\label{eq: notors}
{\omega^c}_{ab}(\zeta) = {\omega^c}_{ba}(\zeta),
\end{align}
which follows from the definition of the Hessian (\ref{eq: defHessian}) and the symmetry of both the matrix of second derivatives and of the components of the metric tensor. The other two properties can be determined by computing the partial derivative of equation (\ref{eq: DallowsHesseq}). Indeed, consider
\begin{align*}
\partial_a g_{bc} &= \partial_a( \partial_b \partial_c D(x||m_\zeta) - {\omega^d}_{bc} \partial_d D(x||m_\zeta))\\
&= \partial_a \partial_b \partial_c D(x||m_\zeta) - (\partial_a {\omega^d}_{bc}) \partial_d D(x||m_\zeta) - {\omega^d}_{cb} \partial_a \partial_d D(x||m_\zeta)\\
&= \partial_a \partial_b \partial_c D(x||m_\zeta) - (\partial_a {\omega^d}_{bc}) \partial_d D(x||m_\zeta)\\
&\qquad - {\omega^d}_{bc} [g_{ad} + {\omega^e}_{ad} \partial_e D(x||m_\zeta)].
\end{align*}
Subtracting this equation from itself with the indices $a$ and $b$ interchanged and rearranging the terms yields
\begin{align*}
0 &=  [\partial_a  {\omega^d}_{bc} - \partial_b {\omega^d}_{ac} + {\omega^d}_{ae} {\omega^e}_{bc} - {\omega^d}_{be} {\omega^e}_{ac}] \partial_d D(x||m_\zeta)\\
&\qquad +[\partial_a g_{bc} - \partial_b g_{ac} + g_{ad} {\omega^d}_{bc} - g_{bd} {\omega^d}_{ac}].
\end{align*}
Since this must hold for all data sets $x$, the two expressions between square brackets must vanish independently and as such the two other conditions are obtained. The best known of these,
\begin{align}
\label{eq: flat}
\partial_a {\omega^d}_{bc} - \partial_b {\omega^d}_{ac} + {\omega^d}_{ae} {\omega^e}_{bc} - {\omega^d}_{be} {\omega^e}_{ac} =0,
\end{align}
has already been identified in the introduction as expressing the connection to be flat and its consequences have been discussed there. The other condition,
\begin{align}
\label{eq: Codazzi}
\partial_a g_{bc} - \partial_b g_{ac} + g_{ad} {\omega^d}_{bc} - g_{bd} {\omega^d}_{ac}=0,
\end{align}
is of the same form as the Codazzi-Peterson equation, which originated as a condition on the second fundamental form of a two-dimensional surface embedded in $\mathbb{R}^3$ \cite{Frankel,Codazzi,Shima}. Since the generalised Fisher information metric is obtained as the second fundamental form of a surface---be it a different surface for every value of $\theta$---this property is not wholly unexpected. There is, however, also another interpretation. Using equation (\ref{eq: defdualcon}) to obtain the coefficients ${\varpi^c}_{ab}$ of the connection $\nabla^*$ dual to $\nabla$, (\ref{eq: Codazzi}) can be rewritten into
\begin{align*}
g_{cd} ({\varpi^d}_{ab} - {\varpi^d}_{ba})=0,
\end{align*}
which means the dual connection is also torsionless. The three conditions (\ref{eq: notors}), (\ref{eq: flat}) and (\ref{eq: Codazzi}) play an important role in the rest of this dissertation: they will directly or indirectly give rise to the rich properties of the Hessian structure and they will be used as a way to verify whether or not a parametrised family of distributions belongs to the exponential family. Their first use, however, is in showing the existence of a generalised Massieu function for data set models---a proof promised to the reader in the beginning of this subsection.
\begin{Hessian}
Given a data set model ($\mathbb{X},\mathbb{M},\mu,D)$, for which there exists a connection $\nabla$ such that 
\begin{align*}
\nabla_a \nabla_b D(x||m_\zeta)
\end{align*}
is independent of $x$, there exists a function $\Phi:\mathbb{R}^n \rightarrow \mathbb{R}$ such that its Hessian (with respect to $\nabla$) equals the Hessian of the divergence function.
\end{Hessian}
\begin{proof}
In the preceding discussion, the $x$-independence of $\nabla_a \nabla_b D(x||m_\zeta)$ was already shown to lead to the equality of this Hessian to the metric tensor. As such, it is only required to show that there exists a solution $\Phi$ to the system of differential equations
\begin{align*}
\partial_a \partial_b \Phi(\zeta) - {\omega^c}_{ab}(\zeta) \partial_c \Phi(\zeta) = g_{ab}(\zeta).
\end{align*}
As an intermediate step, consider the differential equations
\begin{align}
\partial_a \alpha_b(\zeta) - {\omega^c}_{ab}(\zeta) \alpha_c(\zeta) = g_{ab}(\zeta).\label{eq: ML}
\end{align}
This is a system of the Mayer-Lie type and it is integrable if and only if equations (\ref{eq: flat}) and (\ref{eq: Codazzi}) are satisfied.\footnote{The detailed proof of this statement is rather long and technical. The interested reader is therefore referred to Frankel's excellent book \cite{Frankel}, which contains the argument in full.} Since these conditions were shown to be a consequence of this theorem's premise, a solution $\alpha: \mathbb{R}^n \rightarrow \mathbb{R}^n$ to the system (\ref{eq: ML}) is thus found to exist.\\
The only step required to complete the proof is to show that $\alpha_b = \partial_b \Phi$ for some function $\Phi$. Since the metric tensor is symmetric and the connection $\nabla$ is torsionless by assumption, it follows from (\ref{eq: ML}) that 
\begin{align*}
\partial_a \alpha_b(\zeta) - \partial_b \alpha_a(\zeta) = 0.
\end{align*}
Invoking Poincar\'e's lemma shows the existence of a potential for $\alpha\define \alpha_c \dee \zeta^c$, which is exactly what is needed to complete the proof.
\end{proof}
\noindent
As a consequence of this theorem, it immediately follows that
\begin{align*}
g_{ab}(\zeta) &= \nabla_a \nabla_b D(x||m_\zeta) = \nabla_a \nabla_b \Phi.
\end{align*}
A metric that can be written as the Hessian of a generalised Massieu function is called a Hessian metric (see also \cite{Shima}). Since the metric is a positive definite tensor, this also shows that $\Phi$ is a strictly convex function when expressed in the affine coordinates of $\nabla$. (In the remainder of this subsection, the notation $\theta^i$ will consistently be used for these affine coordinates.) Without any additional constraints, the Massieu function thus has a well-defined Legendre-Fenchel transform $S$, defined by
\begin{align*}
 S(U) = \inf_\theta \{ \Phi(\theta) + \theta^k U_k\}%\quad \text{with} \quad U_k = -\pdiff{\Phi}{\theta^k}
\end{align*}
and where the convention common in the physics literature has been used.\footnote{Another possible convention would use $\sup_\theta \{ \theta^k U_k - \Phi(\theta)\}$ to obtain a second convex function, whereas $S$ as introduced in the main body of the text is concave.} The function $S$ can be called the generalised thermodynamic entropy in analogy with statistical mechanics.\\
From the equation (\ref{eq: DallowsHesseq}) and Theorem 2 follows that the divergence of a data set model exhibiting a Hessian structure takes a very simple form when expressed in the affine coordinates of the connection. Indeed, in such coordinates it is possible to write 
\begin{align*}
%\label{eq: twoeq}
\partial_i \partial_j D(x||m_\theta) = g_{ij}(\theta) = \partial_i \partial_j \Phi(\theta).
\end{align*}
This leads to the simple expression
\begin{align}
\label{eq: divbreg}
D(x||m_\theta) = \Phi(\theta) + \theta^k q_k(x) - \sigma(x),
\end{align}
where the values of the functions $q_k$ and $\sigma$ depend on the particular choice for $\Phi$, which is determined only up to an affine term in the $\theta$-coordinates. However, the functions $q_k$ must have constant values on the different fibres. This is the case since data sets $x$ contained in the fibre of $m_\theta$ must satisfy
\begin{align*}
%\label{eq: bigdeal}
\partial_k D(x||m_\theta) = 0 \quad \text{and so}\quad \partial_k \Phi(\theta) = - q_k(x).
\end{align*}
Introduce the function $u : \mathbb{R}^n \rightarrow \mathbb{R}^n$ through
\begin{align*}
u_k(\theta) = q_k(x) \quad \text{where} \quad \mu(x) = m_\theta.
\end{align*}
Since the functions $q_k$ are constant on the fibres, the function $u$ is well-defined. It is also a homeomorphism as it is defined through the derivative of a strictly convex function. This function $u$ is the key to interpreting the precise meaning of the generalised Fisher information metric. After all,
\begin{align}
\label{eq: whygmatters}
g_{ij}(\theta) = \partial_i \partial_j \Phi(\theta) = - \partial_i u_j(\theta).
\end{align}
Because of the definition of the function $u$, the Fisher information thus quantifies how sensitive the values of the functions $q_k$ are to a change in data set, or more precisely how sensitive these values are to a change in the fibre containing the data set. Perhaps an easier way to think about this is that the inverse of this tensor expresses how sensitive the parameters of a data set's projection is to a small change in the observed values of the $q_k$. This generalises the meaning of the Fisher information metric in statistics, where it has this same interpretation for exponential families, with the expectation values of the Hamiltonians taking the role of the $q_k$ \cite{Naudts}. It is important to remember that this is a purely geometric statement, concerned only with the sensitivity of the parameters to a change in the values of the functions $q_k$. It does not support statements about the accuracy---how close to or how far from the true value of the parameters a given choice of model point is.\\
An obvious question to ask in any theory pertaining to statistics is whether or not a generalisation of the Cram\'er-Rao bound \cite{Rao} can be formulated. At this point an elegant answer to that question can be provided. It is indeed possible to produce an inequality reminiscent of this important result, although its interpretation as a bound on the variances of estimators does not generally hold. Instead, it is possible to place a bound on the derivatives of the function $u$, which reduces to the Cram\'er-Rao bound in the case of exponential families of probability distributions.\\
Since the metric tensor is positive definite, it must satisfy the inequality of Cauchy-Schwarz,
\begin{align*}
(v^i g_{ij} w^j)^2 \leqslant (v^i g_{ij} v^j) (w^i g_{ij} w^j),
\end{align*}
for all vectors $v$ and $w$. Using the equalities (\ref{eq: whygmatters}) to replace the components of the metric, it is found after some rearranging that
\begin{align}
\label{eq: CramerRao}
-v^i v^j \partial_i u_j \geqslant \frac{(v^i w^j \partial_i \partial_j \Phi)^2}{ w^i w^j g_{ij} }.
\end{align}
The mathematical form of this inequality is very similar to a general expression of the Cram\'er-Rao bound for multivariate exponential families and even for deformed exponential families \cite{Naudts}. Indeed, in the former case it holds that
\begin{align*}
-\partial_i u_j(\theta) &= -\partial_i \mathbb{E}_{\theta}[\mathcal{H}_j]\\
%&= -\partial_i \int_X \mathcal{H}_j(x) p_\theta(x)\dee x\\
%&= \int_X \mathcal{H}_i(x) [ \partial_j \Phi(\theta) + \mathcal{H}_j(x)] p_{\theta}(x) \dee x\\
%&= \int_X \mathcal{H}_i(x) [ -\mathbb{E}_{p_{\theta}}[\mathcal{H}_j] + \mathcal{H}_j(x)] p_{\theta}(x) \dee x\\
&= \mathbb{E}_{\theta}[\mathcal{H}_i \mathcal{H}_j] - \mathbb{E}_{\theta}[\mathcal{H}_i]\mathbb{E}_{\theta}[\mathcal{H}_j],
\end{align*}
where $\mathbb{E}_\theta[\mathcal{H}_i]$ is the expectation value of the Hamiltonian $\mathcal{H}_i$ with respect to $p_\theta$. Substituting this expression back into (\ref{eq: CramerRao}) yields the promised result.
%\newpage
\subsection{Computing the connection coefficients}
The construction of the connection coefficients for a data set model exhibiting a Hessian structure will proceed from equation (\ref{eq: DallowsHesseq}). This formulation is easier to use in practice than (\ref{eq: DallowsHess}) since one has access to the metric tensor in principle as soon as the data set model has been specified, whereas the generalised Massieu function $\Phi$ is harder to find.\\
The sixth condition on the divergence implies the existence of data sets $x^{(c)}$ such that
\begin{align*}
\partial_d D(x^{(c)}||m_\zeta) 
\end{align*}
form the components of an invertible matrix $A$. By employing these data sets it is possible to isolate the connection coefficients from equation (\ref{eq: DallowsHesseq}). Indeed, choose $x\in \mu^{-1}(m_\theta)$, then the Hessian structure implies
\begin{align*}
0 &= \nabla_a \nabla_b D(x^{(c)}||m_\zeta) - \nabla_a \nabla_b D(x||m_\zeta)\\
&= \partial_a \partial_b D(x^{(c)}||m_\zeta) - {\omega^d}_{ab}(\zeta) \partial_d D(x^{(c)}||m_\zeta) - g_{ab}(\zeta)\\
&=  \partial_a \partial_b D(x^{(c)}||m_\zeta) - {\omega^d}_{ab}(\zeta) {A_d}^c(\zeta) - g_{ab}(\zeta).
\end{align*}
From this it follows easily that 
\begin{align*}
{\omega^c}_{ab}(\zeta) =  \sum_{d=1}^n {( A^{-1}(\zeta))^c}_d [ \partial_a \partial_b D( x^{(d)}||m_\zeta) - g_{ab}(\zeta)].
\end{align*}
In many practical applications a more careful choice of $x^{(c)}$ can be made such that $A$ is a diagonal matrix. The expression for the connection coefficients can then be written in the simpler form---without using Einstein summation---
\begin{align}
\label{eq: disdefomega}
{\omega^c}_{ab}(\zeta) =\frac{\partial_a \partial_b D(x^{(c)}||m_\zeta) - g_{ab}(\zeta)}{\partial_c D(x^{(c)}||m_\zeta)}.
\end{align}
%where no Einstein summation is used.\\
In order to be able to make a link with the definition of connections in the existing literature, however, requires use of the curves $X^{(a)}$ appearing as a consequence of the stronger form of Condition 6. Note that these curves will in general have an implicit parameter dependency---as their definition requires the choice of a model point $\mu(X^{(a)}(0))$---even though the notation does not indicate this. Starting again from equation (\ref{eq: DallowsHesseq}) and deriving with respect to the parameter of the curve $X^{(a)}$ shows
\begin{align}
0 &= \diff{}{\varepsilon} \nabla_a \nabla_b D(X^c(\varepsilon)||m_\zeta) \bigg|_{\varepsilon=0}\nonumber\\
&= \diff{}{\varepsilon} \partial_a \partial_b D(X^c(\varepsilon)||m_\zeta) \bigg|_{\varepsilon=0} - {\omega^d}_{ab}(\zeta) \diff{}{\varepsilon} \partial_d D(X^c(\varepsilon)||m_\zeta) \bigg|_{\varepsilon=0}\nonumber\\
&= \diff{}{\varepsilon} \partial_a \partial_b D(X^c(\varepsilon)||m_\zeta) \bigg|_{\varepsilon=0} - {\omega^c}_{ab}(\zeta).\label{eq: defomegaexp}
\end{align}
This definition is more technical than it needs to be for practical purposes. In particular, it is again sufficient that the expression
\begin{align*}
\diff{}{\varepsilon}\partial_b D(X^c(\varepsilon)||m_\theta)\bigg|_{\varepsilon=0}
\end{align*}
is an invertible matrix, instead of specifically the identity, so the connection coefficients can be determined uniquely.\\
It is instructive to remark that the connection introduced by applying the the definitions (\ref{eq: disdefomega}) or (\ref{eq: defomegaexp}) to a statistical model endowed with the relative entropy as its divergence can be identified as the exponential connection of Efron and Reeds. To be precise, it is the construction of the connection that generalises the construction of the exponential connection in information geometry. Using the same definition for different divergence functions than yields different connections. For example, using the Kullback-Leibler divergence with its arguments interchanged would yield the mixture connection \cite{Amari1985,AmariNagaoka}.\\
The absence of a derivative in expression (\ref{eq: disdefomega}) compared to expression (\ref{eq: defomegaexp}) may give the impression that the discrete method is easier and less likely to cause mathematical problems. This need not be the case, however. For a general data set model, there is no guarantee that the connection coefficients defined through the expression (\ref{eq: disdefomega}) are independent of the choice of the data sets $x^{(c)}$. This will only be the case when the data set model exhibits a Hessian structure.

%As $g_{ij} = -\partial_i u_j$, this means $-\partial^i \theta^j = g^{ij}$. Since the right hand side of this last equation is a positive definite matrix, it satisfies the inequality of Cauchy-Schwarz,
%\begin{align*}
%( v_i g^{ij} w_j)^2 \leqslant (v_i g^{ij} v_j) (w_i g^{ij} w_j)
%\end{align*}
%for all numbers $\{v^i,w^j\}$. By choosing $v_i = \delta_i^k$, $w_j = \delta_j^l$, one obtains
%\begin{align*}
%(\partial^k \theta^l)^2  \leqslant g^{kk} g^{ll} .
%\end{align*}
%Conversely, by applying the inequality of Cauchy Schwarz with the metric tensor itself, 
%\begin{align*}
%(v^i w^j \partial_i \partial_j \Phi)^2 \leqslant (-v^i v^j \partial_i u_j ) (w^i w^j g_{ij}) 
%\end{align*}

\section{A generalised Pythagorean theorem}
For some data set models, the analysis of their geometric properties can be simplified considerably if a specific property is satisfied. In particular, this property allows one to reproduce much of the geometry of proper divergence functions as they are discussed in Chapter 3.\\
A sufficient demand for a model to satisfy Condition 4 as well as for a connection to exist is that for all $(x,m) \in \text{Dom}(D)$ it holds that the difference
\begin{align}
\label{eq: defmanifoldD}
D(x||m) - D(x||\mu(x))
\end{align}
does not depend on the chosen data set $x$, only on its projection $\mu(x)$. If for a data set model ($\mathbb{X},\mathbb{M},\mu,D$) the function 
\begin{align}
\label{eq: Dfixx}
m \mapsto D(x||m)
\end{align}
has its unique global minimum in $\mu(x)$ for all $x \in \text{Dom}(\mu)$, this property determines a proper divergence function on the manifold $\mathbb{M}$. Abusing notation, it is then possible to define
\begin{align}
\label{eq: defmanifoldD4real}
D(\mu(x)||m) \define D(x||m) - D(x||\mu(x)).
\end{align}
It is quickly verified that this construction does indeed give rise to a proper divergence function. Since $\mu(x)$ is assumed to be the unique point where the function (\ref{eq: Dfixx}) reaches its global minimum, it holds that $D(x||m) \geqslant D(x||\mu(x))$ and as a consequence $D(\mu(x)||m) \geqslant 0$ with equality only if $\mu(x) = m$. When $\mu(x) = m$, the difference between $D(x||m)$ and $D(x||\mu(x))$ vanishes trivially and so the newly defined divergence equals zero everywhere on the diagonal of $\mathbb{M}\times \mathbb{M}$.\\
If the left hand side of equation (\ref{eq: defmanifoldD4real}) is well defined on the domain of $D$, this equality will be referred to as the generalised Pythagorean theorem. An illustration is given in the drawing, where the divergence between points is represented by the squared length of the dashed line between them.
%\newpage
\begin{center}
%\begin{float}
\begin{tikzpicture}
\path[draw] (0,0) arc(115:90:9) arc(60:90:8 and 6) arc(90:115:9) arc(90:60:8 and 6) ;
\path[fill] (-1,1.25) circle (0.05);
\path[fill] (-1,3) circle (0.05);
\path[draw,dashed] (-1,1.25) -- (-1,3);
\path[draw,dashed] (-1,1.25) arc (85:65:5 and 4);
\path[fill] (0.7,0.9) circle (0.05);
\path[draw,dashed] (-1,3) arc (40:17:9 and 6);

\path[draw] (-1.3,3.2) node {$x$};
\path[draw] (-1.5,1.05) node {$\mu(x)$};
\path[draw] (1,0.65) node {$m$};

\path[draw] (3.5,1.4) node {$\mathbb{M}$};
\end{tikzpicture}
%\end{float}
\end{center}
\vspace{0.25cm}
\noindent
In some examples, such as in information geometry, it is the case that $\mathbb{M} \subset \mathbb{X}$, making it possible that the divergence function is actually defined on $\mathbb{X}\times\mathbb{X}$ as a whole. Such divergence functions may satisfy a stronger version of the generalised Pythagorean theorem. An example of this is the Kullback-Leibler divergence, for which it holds that
\begin{align*}
D(p||r) = D(p||q) + D(q||r)
\end{align*}
when $q$ is the projection of $p$ on a particular choice of submanifold containing the distribution $r$ \cite{AmariNagaoka}. However, some of these divergences will fail to satisfy the generalised Pythagorean theorem on $\mathbb{X} \times \mathbb{M}$ yet condition (\ref{eq: defmanifoldD}) may still hold for all data sets and model points. The latter condition is thus weaker than the former. In that case it is important to formulate the Pythagorean theorem through expression (\ref{eq: defmanifoldD4real}) rather than using the original expression for the divergence in all three terms.\\
For data set models satisfying the property (\ref{eq: defmanifoldD4real}), it automatically also holds that the metric tensor can have its components written as
\begin{align*}
g_{ij}(\theta) &= \partial_i \partial_j D(x||m_\theta)\\
&= \partial_i \partial_j\left[ D(x||\mu(x)) +  D(\mu(x)||m_\theta)\right]\\
&= \partial_i \partial_j D(\mu(x)||m_\theta).
\end{align*}
Indeed, when $x$ is an element of the fibre of $m_\theta$, both arguments of the expression $\partial_i \partial_j D(\mu(x)||m_\theta)$ coincide. This in turn is the definition (\ref{eq: definfogeomg}) for the metric tensor derived from proper divergence functions presented earlier.\\
Also the definition of the connection coefficients (\ref{eq: defomegaexp}) can be expressed in terms of the proper divergence. Following an analogous line of reasoning as for the metric tensor, one obtains
\begin{align*}
{\omega^k}_{ij}(\theta) &= \diff{}{\varepsilon} \partial_i \partial_j D(X^k(\varepsilon)||m_\theta)\bigg|_{\varepsilon=0}\\
%&= \diff{}{\varepsilon} \partial_i \partial_j D(X^k(\varepsilon)||\mu(X^k(\varepsilon)))\bigg|_{\varepsilon=0} + \diff{}{\varepsilon} \partial_i \partial_j D(\mu(X^k(\varepsilon))||m_\theta)\bigg|_{\varepsilon=0}\\
&= \diff{}{\varepsilon} \partial_i \partial_j D(\mu(X^k(\varepsilon))||m_\theta)\bigg|_{\varepsilon=0}.
\end{align*}
To show the correspondence with the definition (\ref{eq: definfogeomomega}) for the connection coefficients derived from proper divergence functions, fix $\theta$ and define $\Theta^{(k)}$ by
\begin{align*}
\mu(X^k(\varepsilon)) = m_{\Theta^{(k)}(\varepsilon)}, \quad X^k(0) = m_\theta.
\end{align*}
Then it is possible to write
\begin{align*}
\delta_k^l %&= \diff{}{\varepsilon}\partial_k D(\mu(X^l(\varepsilon))||m_\theta)\bigg|_{\varepsilon=0}
&= \diff{}{\varepsilon}\partial_k D(m_{\Theta^{(l)}(\varepsilon)}||m_\theta)\bigg|_{\varepsilon=0}\\
&= \pdiff{}{\xi^s} \partial_k D(m_{\xi}||m_\theta)\big|_{\xi=\theta} \diff{\Theta}{\varepsilon}^{(l)s}\bigg|_{\varepsilon=0}\\
%&=  - \pdiff{}{\theta^s}\pdiff{}{\theta^k} D(m_{\xi}||m_\theta)\big|_{\xi=\theta} \diff{\Theta}{\varepsilon}^{(l)s}\bigg|_{\varepsilon=0}\\
&= - g_{sk}(\theta) \diff{\Theta}{\varepsilon}^{(l)s}\bigg|_{\varepsilon=0}.
\end{align*}
This knowledge can be used to compute the connection coefficients in an analogous way as above:
\begin{align*}
{\omega^k}_{ij}(\theta) &= \diff{}{\varepsilon} \partial_i \partial_j D(X^k(\varepsilon)||m_\theta)\bigg|_{\varepsilon=0}\\
&= \left(\pdiff{}{\xi^s} \partial_i \partial_j D(m_\xi||m_\theta)\big|_{\xi=\theta}\right) \diff{\Theta}{\varepsilon}^{(k)s}\bigg|_{\varepsilon=0}\\
&= g^{ks}(\theta)\left[ -  \pdiff{}{\xi^s} \partial_i \partial_j D(m_\xi||m_\theta)\bigg|_{\xi=\theta}\right].
\end{align*}
This is indeed the definition found in the literature and a reported upon in the previous chapter, in particular in equation (\ref{eq: definfogeomomega}).

\section{Identifying exponential families}
%\subsection{Verification of exponential family}
This section contains two easily obtained results concerning exponential families of probability distributions. In particular, it is shown that a statistical model belongs to the exponential family if and only if it exhibits a Hessian structure. This is useful since this allows for the establishment of both positive and negative results through a straightforward step-by-step process. For exponential families, it is shown that the canonical parameters can be found by solving a system of linear differential equations. I do not claim originality of these results, however. Especially the second one I expect to be contained in at least some books on differential geometry as a method to find the affine coordinates of a connection. The first result on the other hand, genuinely appears to be missing form the important reference works---both on exponential families and on Hessian structures---as well as from the available historically important papers cited in the introduction. Since both properties will be used extensively in the treatment of the examples in the next chapter, it is thus instructive to (re-)derive them here.\\
It was shown earlier in this chapter that exponential families exhibit a Hessian structure. The converse is also true: the presence of a Hessian structure in a statistical model only occurs for an exponential family---assuming that the divergence is chosen to be that of Kullback and Leibler. To show this, assume a statistical model with a Hessian structure derived from the relative entropy. Since the connection is necessarily flat and torsionless, it allows for affine coordinates $\theta$. In these coordinates the Hessian of the divergence reads
\begin{align*}
\partial_i \partial_j D(p||p_{\theta}) = - \partial_i \partial_j \int_X p(x) \ln p_\theta(x) \dee x
\end{align*}
and by assumption both sides are of this equality are independent of the distribution $p$. But this is only possible when
\begin{align*}
x \mapsto \partial_i \partial_j \ln p_\theta(x)
\end{align*}
is a function of $\theta$ only. With a bit of foresight, this function can be identified as $-\Phi$. Then it follows that
\begin{align*}
\ln p_{\theta}(x) = -\Phi(\theta) - \theta^k \mathcal{H}_k(x)
\end{align*}
for properly chosen functions $\mathcal{H}_k$. (A term independent of $\theta$ could be absorbed in the definition of the measure $\dee x$ and can thus be discarded.) This shows the result that was promised to the reader. Due to its practical importance, it is summarised in its own theorem:
\begin{Identify}
A statistical model belongs to the exponential family if and only if it exhibits a Hessian structure when it is endowed with the Kullback-Leibler divergence (\ref{eq: defKLdiv}).
\end{Identify}
\noindent
A recurring property in the work of Efron, Reeds and Amari is that exponential families of probability distributions are flat when endowed with the exponential connection \cite{Amari1985,Efron,ReedsToEfron}. Since the connection of the data set model geometry generalises this exponental connection, Theorem 3 can be seen as the data set model generalisation of that information geometric property and its converse.\\
A useful corollary pertains to the canonical parameters of the exponential families. When employed as coordinates for the model manifold, they serve as the affine coordinates for the flat and torsionless (exponential) connection. This means that these parameters $\theta^i$ can be found as a function of arbitrary coordinates $\zeta^a$ by studying the connection. More specifically, the connection coefficients are in general defined through (\ref{eq: defomega}) and this relation can be transformed from general coordinates $\zeta^a$ to the affine coordinates $\theta^i = \Theta^i(\zeta)$. This looks like
%\begin{align*}
%0 &= \nabla_i\big( (\partial_j Z^b) \partial_b \big)\\
%&= (\partial_i \partial_j Z^b) \partial_b + (\partial_j Z^b) (\partial_i Z^a) \nabla_a \partial_b\\
%&= (\partial_i \partial_j Z^c) \partial_c + (\partial_j Z^b) (\partial_i Z^a) {\omega^c}_{ab} \partial_c.
%\end{align*}
%Conversely,
\begin{align*}
{\omega^c}_{ab} \partial_c &= \nabla_a \partial_b\\
&= \nabla_a \big( (\partial_b \Theta^j) \partial_j)\\
&= (\partial_a \partial_b \Theta^j) \partial_j + (\partial_b \Theta^j) (\partial_a \Theta^j) (\nabla_i\partial_j)\\
&= (\partial_a \partial_b \Theta^j) \partial_j + (\partial_b \Theta^j) (\partial_a \Theta^j) {\omega^k}_{ij} \partial_k,
\end{align*}
where letters from the beginning of the alphabet refer to the arbitrary $\zeta$-coordinates and letters from the middle of the alphabet belong to the affine coordinates $\theta$. The coefficients ${\omega^k}_{ij}$ thus vanish identically by assumption. Computing the $\theta^j$-component of the left hand side and using $\dee \theta^j(\partial_c) = \partial_c \Theta^j$ reveals 
\begin{align}
\label{eq: flatcon}
\partial_a \partial_j \Theta^j(\zeta) = {\omega^c}_{ab}(\zeta) \partial_c \Theta^j(\zeta).
\end{align}
This is a system of linear differential equations which may be solved to find the function $\zeta \mapsto \Theta(\zeta)$ expressing the affine coordinates as a function of arbitrary coordinates.
\newpage
\section{Discussion}
It is important as well as instructive to compare the data set model formalism with the theory of information geometry upon which it is based.\\
First amongst these differences is of course the fact that probability theory has been deliberately excluded from the building blocks of the formalism. As is explained in the introduction, this was done in order to identify the crucial link between information theory and the geometry of the formalism, rather than relying on the much-walked path of basing information theory on probability theory. The possibility of using properties ultimately based on probability---on purpose or by accident---is always present when studying information geometry. After all, despite its geometric nature, it still relies in part on properties of the objects it deals with to show theorems and additional properties, as a read-through of the literature referred to in the previous chapter will indicate. Naturally, this cannot be taken as a criticism of information geometry in itself, which has many useful applications in a statistical context which rely on exactly those properties.
%However, contamination of the more general data set model formalism by statistics must be avoided if the goals set forth in this dissertation are to be pursued in earnest. It is also needed to respect the goal of generalisation: to find simpler, more natural and more elegant perspectives on a subject matter and its properties.\\
\\
A striking contrast with existing methods of information geometry, and perhaps the most promising aspect of the data set model formalism in terms of applications, is the explicit option to model data through qualitatively different mathematical objects. In information geometry, as well as in the study of proper divergence functions, the divergence functions take both arguments from the same set or from a set and a subset thereof, the latter of which then serves as the model for the former. By taking into account the possibility that both arguments of the divergence are qualitatively different a much wider array of possible models can be described and can have their divergence-implied geometry constructed and studied. This general formalism strongly resembles that of pattern recognition in machine learning, see for instance \cite{Eguchi2006} for a good comparison between pattern recognition and information geometry. This generality does, however, come at the price of losing some structure which does exist in other formalisms and contexts. The family of affine connections introduced by Chentsov and Amari is an example of such a structure that could not be replicated.\\
People familiar with information geometry may perceive the emphasis of the data set model formalism on intrinsic geometry rather than the extrinsic geometry as an important contrast. Nevertheless, this offers a great advantage in terms of generality as not every model manifold has an obvious embedding in some larger set. This embedding is required to enable the study of an extrinsic geometry. Removing the need for an embedding also opens up the possibility of considering the entire simplex of everywhere strictly positive probability distributions over a measurable set as the model manifold. Doing this is rather unnatural in the usual formalism of information geometry since there this simplex serves as the space in which the model is embedded and from which it derives its geometry.\\
There are also some strong similarities between the data set model formalism and information geometry. The first one is obviously that it is possible to reconstruct the metric tensor and the exponential connection from the latter theory in the former, at least under certain conditions. The reconstruction of a family of connections has not been achieved, however. The metric tensor is used in quantifying information in a way which is very analogous to its use in information geometry: both are a measure for the sensitivity of the modelling process to changes in the data---or more precisely: changes in the fibre containing the data.\\
It also proved possible to reconstruct a number of useful properties. The most obvious of these is probably the vanishing curvature of statistical manifolds corresponding to exponential families. This result is well-known from information geometry, as it was one of the starting points for the study of connections in this field. The fact that exponential families can be identified by means of their flatness and the correspondence of canonical parameters and affine coordinates was already implied in the work of Amari, who demonstrated this fact through direct computation \cite{Amari1985,AmariNagaoka}. However, the above work shows that a similar result can be obtained in a more general context by identifying all data set models which give rise to a Hessian metric as exponential families.\\
The generalised Pythagorean theorem is also interesting when comparing the data set model formalism the study of proper divergence functions. It makes clear that the data set model geometry is a true extension of the differential geometry induced by proper divergence functions on a manifold.

\newpage

\chapter{Examples and applications}
This last chapter is devoted to the illustration of the data set formalism. A few familiar examples from statistics and statistical physics will be presented to show how the formalism developed in the previous chapter behaves when applied to these models. There are a total of six examples, each of which demonstrates one or more important properties of the data set model formalism.
%The first example is the simplest and is concerned with a statistical model of normal distributions. As such this example serves to illustrate how the basics of information geometry fit the data set model formalism. Furthermore, it is also the first example showcasing an exponential family with its accompanying flat connection and its affine coordinates serving as the canonical parameters.\\
%The second example treats linear regression, the well-known method of fitting a first order polynomial to data points in the plane. Using this technique one minimises the sum of the squares of the vertical distances between the data points and the intended fit. As such this sum of squares is adopted as a divergence function between data and model, which gives rise to a model map yielding the familiar fitting formulas. It is shown, however, that this approach does not satisfy the conditions for the data set model formalism. The analysis is then continued through a different divergence function yielding the same model map and a Euclidean geometry. This example shows that the data set model formalism is capable of performing parameter estimation on data and models which are not probability distributions.\\
%The third example is more elaborate than the previous ones. It is concerned with a semi-classical model for non-interacting bosonic particles. The data consists out of the occupation numbers for the states in which the particles may be placed. The model is the statistical model of grand canonical distribution  
\section{The normal distributions}
\subsection{Using the relative entropy}
In this first example, the Gaussian probability densities (or normal distributions) will be covered as a model for probability densities over $\mathbb{R}$. This family is also used as an example in Chapter 3. The results found there are mostly rederived, now along the lines of the data set model formalism as set out in the previous chapter. It is also illustrated how the canonical coordinates of the exponential can be found in a straightforward manner by constructing the affine connection.\\
The data sets represent an empirical distribution over the real numbers with finite first and second moments. In information geometry this set would be treated as an infinitely dimensional manifold of which the normal distributions are a submanifold. The data set model formalism, however, does not employ this additional structure. The task at hand in this example is then to find the best fitting member of the two-parameter family of normal distributions, given by the expression
\begin{align*}
p_{\mu,\sigma}(x) = \exp\left\{ - \ln \sqrt{ 2\pi \sigma^2} - \frac{(x-\mu)^2}{2\sigma^2}\right\}.
\end{align*}
These distributions form a two-dimensional manifold for which $\mu$ and $\sigma > 0$ are used as parameters. (The model map will not be named explicitly in this example and so the symbol $\mu$ is free to be used for the mean of the normal distribution, as is conventional.) Some sources use $\sigma^2$ as the second parameter, leading to some minor differences between the results obtained there and the ones derived here.\\
The best fit will be sought by minimising the Kullback-Leibler divergence 
\begin{align}
D(p||p_{\mu,\sigma}) &= \int_{-\infty}^{+\infty} p(x) \ln \frac{p(x)}{p_{\theta}(x)} \dee x\label{eq: KLex1}\\
&= - S(p) + \ln\sqrt{2\pi\sigma^2} + \frac{1}{2\sigma^2}\mathbb{E}_p[(x-\mu)^2],\nonumber
\end{align}
where $S$ represents the Shannon entropy (\ref{eq: Shannon}) of the probability distribution $p$. The model map here is entirely implied by the divergence: there is only a single normal distribution minimising this divergence for any of the empirical distributions serving as the first argument and this global minimum is also a local minimum. The derivatives themselves satisfy
\begin{align*}
\partial_\mu D(p||p_{\mu,\sigma}) &= -\frac{1}{\sigma^2}\mathbb{E}_p[x-\mu],\\
\partial_\sigma D(p||p_{\mu,\sigma}) &= \frac{1}{\sigma} - \frac{1}{\sigma^3} \mathbb{E}_p[(x-\mu)^2].
\end{align*}
The condition that these expressions vanish simultaneously yields the well known-expressions
\begin{align}
\mathbb{E}_p[x] &= \mu,\label{eq: FMimu}\\
\mathbb{E}_p[(x-\mu)^2] &= \sigma^2 \label{eq: SMiss}.
\end{align}
In order to determine the Fisher information metric, it is required to know the matrix of second derivatives of the relative entropy. These satisfy the expressions
\begin{align*}
\partial^2_\mu D(p||p_{\mu,\sigma}) &= \frac{1}{\sigma^2},\\
\partial_\sigma \partial_\mu D(p||p_{\mu,\sigma}) &= \frac{2}{\sigma^3} \mathbb{E}_p[x-\mu],\\
\partial^2_\sigma D(p||p_{\mu,\sigma}) &= - \frac{1}{\sigma^2} +  \frac{3}{\sigma^4}\mathbb{E}_p[(x-\mu)^2].
\end{align*}
Both expectation values appearing in this matrix can be rewritten as functions of the parameters only. This is achieved by using the conditions (\ref{eq: FMimu}) and (\ref{eq: SMiss}). As such it is implied that the metric is well-defined and has components
\begin{align*}
g_{\mu\mu}(\mu,\sigma) = \frac{1}{\sigma^2}, \quad g_{\mu\sigma}(\mu,\sigma) = 0, \quad g_{\sigma\sigma}(\mu,\sigma) = \frac{2}{\sigma^2}.
\end{align*}
The next step in  constructing the data set model geometry is to determine the connection coefficients. This requires the identification of distributions for which all but one of the derivatives of the divergence vanish. Let us first turn our attention to finding $p^{(\mu)}$, which is used to find the connection coefficients $ {\omega^\mu}_{ij}$. Such a distribution must satisfy 
\begin{align*}
\partial_\mu D(p^{(\mu)}||p_{\mu,\sigma}) \neq 0\quad\text{and}\quad \partial_\sigma D(p^{(\mu)}||p_{\mu,\sigma}) &= 0.
\end{align*}
It is thus necessary to find a distribution with a mean different from $\mu$ but with its expectation value of $(x-\mu)^2$ equal to $\sigma^2$. Take the mean of $p^{(\mu)}$ to be equal to $\mu + \delta \mu$. This can be used to compute the connection coefficients ${\omega^\mu}_{ij}$ through equation (\ref{eq: disdefomega}). The result is 
\begin{align*}
{\omega^\mu}_{\mu\mu}(\mu,\sigma) &= 0,\\
{\omega^\mu}_{\mu\sigma}(\mu,\sigma) %&= \frac{\partial_\mu\partial_\sigma D(p^{(\mu)}||p_{\mu,\sigma}) - g_{\mu\sigma}}{\partial_\mu D(p^{(\mu)}||p_{\mu,\sigma})}\\
&= \frac{2\sigma^{-3} (\delta \mu) - 0}{-\sigma^{-2}(\delta \mu)} = -\frac{2}{\sigma},\\
{\omega^\mu}_{\sigma\sigma}(\mu,\sigma) &= 0.
\end{align*}
The other three independent connection coefficients ${\omega^\sigma}_{ij}$ can be computed in an analogous way. This requires the choice of a data set $p^{(\sigma)}$, which must have $\mu$ for its mean but $(\sigma + \delta \sigma)^2$ as its second central moment. Substituting this distribution again into equation (\ref{eq: disdefomega}) reveals the expressions for the last three independent connection coefficients to be
\begin{align*}
{\omega^\sigma}_{\mu\mu}(\mu,\sigma) &= 0,\\
{\omega^\sigma}_{\mu\sigma}(\mu,\sigma) &= 0,\\
{\omega^\sigma}_{\sigma\sigma}(\mu,\sigma)  %&=\frac{\partial^2_\sigma D(p^{(\sigma)}||p_{\mu,\sigma}) - g_{\sigma\sigma}}{\partial_\sigma D(p^{(\sigma)}||p_{\mu,\sigma})}\\
&= \frac{-\sigma^{-2} + 3\sigma^{-4}(\sigma + \delta \sigma)^2 - 2\sigma^{-2}}{\sigma^{-1} - \sigma^{-3}(\sigma + \delta \sigma)^2} = -\frac{3}{\sigma}.
\end{align*}
It can be verified quickly that these coefficients are indeed the correct ones in order to write the metric tensor as the Hessian of the divergence function, regardless of the chosen empirical distribution $p$. Take, as an example, the component $g_{\sigma\sigma}$. A straightforward computation yields
\begin{align*}
&\nabla_\sigma \nabla_\sigma D(p||p_{\mu,\sigma})\\
&= \partial^2_\sigma D(p||p_{\mu,\sigma}) - {\omega^\sigma}_{\sigma\sigma}(\mu,\sigma)\partial_\sigma D(p||p_{\mu,\sigma})\\
&=  - \sigma^{-2} +  3\sigma^{-4} \mathbb{E}_p[(x-\mu)^2] - \left( -3\sigma^{-1}\right)\left(\sigma^{-1} - \sigma^{-3}\mathbb{E}_p[(x-\mu)^2]\right)\\
&=  \frac{2}{\sigma^2}\\
&= g_{\sigma\sigma}(\mu,\sigma)
\end{align*}
as desired. This data set model can thus be concluded to feature a Hessian structure.\\
It is shown in the text of the previous chapter that this Hessian structure implies the connection to be flat and torsionless, as well to satisfy the Codazzi-Peterson-like equation (\ref{eq: Codazzi}). This can be verified explicitly, but it is more instructive to find the affine coordinates of the connection. The function $(\mu,\sigma) \mapsto \Theta^i(\mu,\sigma)$ expressing these affine coordinates must satisfy three partial differential equations, given by equation (\ref{eq: flatcon}). In this example, these equations look like
\begin{align*}
\partial^2_\mu \Theta^i(\mu,\sigma) &= 0,\\
\partial_\mu \partial_\sigma \Theta^i(\mu,\sigma) &= - 2 \sigma^{-1} \partial_\mu \Theta^i(\mu,\sigma),\\
\partial^2_\sigma \Theta^i(\mu,\sigma) &= - 3 \sigma^{-1} \partial_\sigma \Theta^i(\mu,\sigma).
\end{align*}
From the first of these equations, it follows that
\begin{align*}
\Theta^i(\mu,\sigma) &= A^i(\sigma) \mu + B^i(\sigma),
\end{align*}
where the functions $A^i$ and $B^i$ still need determining. Substitution into the third partial differential equations yields
\begin{align*}
\mu \partial^2_\sigma A^i + \partial_\sigma^2 B^i = - 3\mu \sigma^{-1} \partial_\sigma A^i - 3\sigma^{-1} \partial_\sigma B^i.
\end{align*}
Since this must hold for all values of $\mu$, one obtains two independent equations for $A^i$ and $B^i$. The solutions of these demand that both $A^i$ and $B^i$ are proportional to $\sigma^{-2}$. The second of the above partial differential equations is then also satisfied. It is now possible to choose the constants $A^i$ and $B^i$ in such a way as to obtain the well-known canonical parameters
\begin{align*}
\theta^1 = \frac{1}{2\sigma^2} \quad \text{and}\quad \theta^2 = - \frac{\mu}{\sigma^2}.
\end{align*}
\newpage
\subsection{Using a different divergence function}
The simplicity of this example offers a good opportunity to study the effects of using a different divergence function---in this case a divergence function different from the relative entropy (\ref{eq: KLex1}). The set $\mathbb{X}$ of data, the model manifold $\mathbb{M}$ and the model map $\mu$ will thus remain unchanged. An obvious choice for the new divergence $D^\prime$ is the expression
\begin{align}
D^\prime(p||p_{\mu,\sigma}) &= \frac{1}{2\mu_0^2} \left( \mu - \mathbb{E}_p[x]\right)^2 + \frac{1}{4\sigma_0^4} \left( \mu^2 + \sigma^2 - \mathbb{E}_p[x^2]\right)^2\label{eq: Dprime},
\end{align}
where $\mu_0$ and $\sigma_0$ are strictly positive constants which may be arbitrary otherwise. These constants are introduced for dimensional reasons but they can also be used to change the relative sensitivity of the model map with respect to the values of $\mathbb{E}_p[x]$ and $\mathbb{E}_p[x^2]$. The derivatives of this divergence satisfy 
\begin{align*}
\partial_\mu D^\prime(p||p_{\mu,\sigma}) &= \frac{1}{\mu_0^2}( \mu - \mathbb{E}_p[x]) + \frac{\mu}{\sigma_0^4}( \mu^2 + \sigma^2 - \mathbb{E}_p[x^2]),\\
\partial_\sigma D^\prime(p||p_{\mu,\sigma}) &= \frac{\sigma}{\sigma_0^4}(\mu^2 + \sigma^2 - \mathbb{E}_p[x^2]).
\end{align*}
For these two expressions to vanish simultaneously, it is required that 
\begin{align*}
\mathbb{E}_p[x] &= \mu,\\
\mathbb{E}_p[x^2] &= \sigma^2 + \mu^2.
\end{align*}
By a straightforward computation it can be seen that these two conditions are equivalent to (\ref{eq: FMimu}) and (\ref{eq: SMiss}). This means the model maps implied by the divergence functions (\ref{eq: KLex1}) and (\ref{eq: Dprime}) coincide as intended.\\
To compute the metric tensor $g^\prime$, the matrix of second derivatives of the divergence is again needed. Another straightforward computation yields
\begin{align*}
\partial^2_\mu D^\prime(p||p_{\mu,\sigma}) &= \frac{1}{\mu_0^2} + \frac{3\mu^2  + \sigma^2 - \mathbb{E}_p[x^2]}{\sigma_0^4},\\
\partial_\mu \partial_\sigma D^\prime(p||p_{\mu,\sigma}) &= \frac{2\mu\sigma}{\sigma_0^4},\\
\partial^2_\sigma D^\prime(p||p_{\mu,\sigma}) &= \frac{\mu^2 + 3\sigma^2 - \mathbb{E}_p[x^2]}{\sigma_0^4}.
\end{align*}
In order to determine the metric tensor, the distribution $p$ needs to be chosen in the fibre of $p_{\beta,\mu}$. This means $\mathbb{E}_p[x^2] = \mu^2 + \sigma^2$ and so the metric tensor has components
\begin{align*}
g_{\mu\mu}^\prime(\mu,\sigma) &= \frac{2\mu^2}{\sigma_0^4} + \frac{1}{\mu_0^2}, \quad g_{\mu\sigma}^\prime(\mu,\sigma) = \frac{2\mu\sigma}{\sigma_0^4},\quad g_{\sigma\sigma}^\prime(\mu,\sigma) = \frac{2\sigma^2}{\sigma_0^4}.
\end{align*}
Remark that the sum of squares divergence (\ref{eq: Dprime}) yields a more complicated metric tensor than the apparently more complex Kullback-Leibler divergence. It is also easy to verify that the connection constructed from the divergence $D^\prime$ is not the metric connection, since the latter does not vanish---a fact that can be checked equation (\ref{eq: defLCcon}). These complications are all in some way a consequence of the presence of the parameter $\mu$ in the second term---a presence which is in itself necessitated by the expression for the second (non-central) moment of a normal distribution, which refers to both the mean $\mu$ and the variance $\sigma^2$.\\
Since the matrix of second derivatives does not coincide with the metric tensor, it is necessary to compute the expressions (\ref{eq: disdefomega}) to find the connection coefficients ${\varpi^c}_{ab}$ of $\nabla^\prime$. Since only two of the second derivatives of the divergence $D^\prime$ depend on the arbitrary distribution and then only on the second moment of this distribution, only ${\varpi^\sigma}_{\mu\mu}$ and ${\varpi^\sigma}_{\sigma\sigma}$ may be different from zero. The probability distribution $p^{(\sigma)}$ required to compute these coefficients must have a second moment different different from $\sigma^2 + \mu^2$---say $(\sigma + \delta \sigma)^2 + \mu^2$. Using again the definition (\ref{eq: disdefomega}) yields for the first coefficient
\begin{align*}
{\varpi^\sigma}_{\mu\mu}(\mu,\sigma) &= \frac{\partial_\mu^2 D^\prime(p^{(\sigma)}||p_{\mu,\sigma}) - g_{\mu\mu}(\mu,\sigma)}{\partial_\sigma D^\prime(p^{(\sigma)}||p_{\mu,\sigma})}\\
&= \frac{3\mu^2 + \sigma^2 - [(\sigma+\delta \sigma)^2+\mu^2] - 2\mu^2}{\sigma(\mu^2 + \sigma^2 - [(\sigma+\delta \sigma)^2 + \mu^2])}\\
%&= \frac{\sigma^2 - (\sigma+\delta \sigma)^2}{\sigma(\sigma^2 - (\sigma+\delta\sigma)^2)}\\
&= \frac{1}{\sigma}.
\end{align*}
The second one is computed in a very analogous way to obtain
\begin{align*}
{\varpi^\sigma}_{\sigma\sigma}(\mu,\sigma) &= \frac{\partial_\sigma^2 D^\prime(p^{(\sigma)}||p_{\mu,\sigma}) - g_{\sigma\sigma}(\mu,\sigma)}{\partial_\sigma D^\prime(p^{(\sigma)}||p_{\mu,\sigma})}\\
&= \frac{\mu^2 + 3\sigma^2 - [(\sigma + \delta \sigma)^2 + \mu^2] -2\sigma^2}{\sigma(\mu^2 + \sigma^2 - [(\sigma+\delta \sigma)^2 + \mu^2])}\\
&= \frac{1}{\sigma}.
\end{align*}
As such, the connection is well-defined but it is a different connection than the one obtained when the relative entropy is used. Computing the curvature tensor explicitly allows one to verify that this connection is flat. However, finding the affine coordinates of this connection will have this as a corollary. To find these special coordinates, the differential equations (\ref{eq: flatcon}) can be used and for this particular example they take the form
\begin{align*}
\partial_\mu^2 \Theta^i(\mu,\sigma) &= \sigma^{-1} \partial_\sigma \Theta^i(\mu,\sigma),\\
\partial_\mu\partial_\sigma \Theta^i(\mu,\sigma) &= 0,\\
\partial_\sigma^2 \Theta^i(\mu,\sigma) &= \sigma^{-1} \partial_\sigma \Theta^i(\mu,\sigma).
\end{align*}
The second of these equations demands that
\begin{align*}
\Theta^i(\mu,\sigma) = A^i(\mu) + B^i(\sigma).
\end{align*}
Inserting this in the third equation yields a differential equation for $B^i$,
\begin{align*}
\sigma \partial^2_\sigma B_i(\sigma) = \partial_\sigma B^i(\sigma) 
\end{align*}
and thus $B^i(\sigma) = B^i_0 \sigma^2$. The first equation is then satisfied only if
\begin{align*}
\partial_\mu^2 A^i(\mu) &= 2 B_0^i.
\end{align*}
A particular solution $A^i = A^i_0 \mu$ is obtained when $B^i=0$, whereas the general solution is $A^i = B^i_0 \mu^2$. As such two canonical parameters are revealed to be
\begin{align*}
\eta^1 = \mu\quad \text{and} \quad \eta^2 = \mu^2 + \sigma^2.
\end{align*}
These are the so-called ``expectation parameters'' of the family of normal distributions \cite{Amari1985}.\\
Note that this treatment never used the explicit expression for the model distributions. A similar ``sum of squares'' divergence function may thus be defined in a very broad range of cases. The approach does have the disadvantage of lacking interpretation, however. To take a concrete example, the family of Gumbel distributions---discussed in details in the last example---could be adopted as a model when endowed with this same divergence (\ref{eq: Dprime}) as they too can be distinguished uniquely by their first and second moments. In fact this last example shows, \emph{mutatis mutandis}, that this form of the divergence will always yield a flat and torsionless connection---at least when the squares in (\ref{eq: Dprime}) contain sufficiently well-behaved expressions. 

\newpage
\section{Linear regression}
Perhaps the simplest and best known-way of fitting a functional relation to data points is linear regression. The input data in this example takes the form of a set of couples $(x_j,y_j)$ which are believed to exhibit a functional relationship in principle but which have been contaminated by some form of noise. It should be noted that nothing in this example is new---this is merely an illustration that a wide array of problems fits into the data set model formalism.\\
The set $\mathbb{X}$ contains all collections $S$ consisting of $N_S$ couples $(x_j,y_j)\in \mathbb{R}^2$. Remark that these data sets $S$ must are not required to contain the same number of couples, however they must satisfy
\begin{align*}
N_S \sum_j x_j^2 - \bigg( \sum_j x_j\bigg)^2 \neq 0.
\end{align*}
The model points are the first order polynomials of the form
\begin{align*}
f_{a,b}(x) = ax + b, \quad a,b\in \mathbb{R}.
\end{align*}
The numbers $a$ and $b$, indicating the slope and intercept of the polynomial function, are also used as the coordinates for the model manifold $\mathbb{M}$. The obvious choice of divergence, which must indicate the best fitting model, is to adopt the quantity which must be minimised in the least squares method,
\begin{align}
\label{eq: leastsquares}
D(S||f_{a,b}) = \frac12 \sum_j \left(y_j - ax_j - b\right)^2. 
\end{align}
Minimising this function may happen by setting equal to zero the derivatives of this divergence function, which yields
\begin{align*}
\partial_a D(S||f_{a,b}) &= -\sum_j (y_j - a x_j - b) x_j,\\
\partial_b D(S||f_{a,b}) &= -\sum_j (y_j - a x_j - b).
\end{align*}
Through a straightforward computation, it is found that these derivatives vanish simultaneously when
\begin{align*}
a &= \frac{N_S \sum_j y_j x_j - \sum_j x_j \sum_i y_i}{N_S \sum_j x_j^2 - \left( \sum_j x_j\right)^2}\quad\text{and}\quad b = \frac{1}{N_S} \sum_j (y_j - ax_j).
\end{align*}
These are of course nothing but the regular expressions for the slope and intersect of the best fit to the data $S$. Thus, the data set model formalism contains this aspect of the least squares linear regression method.\\
The next interesting quantity is the metric tensor. To compute this quantity, knowledge of the matrix of second derivatives of the divergence is needed. The independent components are given by
\begin{align*}
\partial_a^2 D(S||f_{a,b}) &= \sum_j x_j^2,\\
\partial_a \partial_b D(S||f_{a,b}) &= \sum_j x_j,\\
\partial^2_b D(S||f_{a,b}) &= N_S.
\end{align*}
This matrix depends only on the data and not on the parameters $a$ or $b$. This is not in itself a problem, as it might in principle be possible to express these quantities only as a function of those parameters by restring these expressions to data sets for which the best fit is a given line. However, it is clear that this will not be possible: the best fit is determined also by the $y$-values of the couples in the data set and there is no mention of those numbers in the matrix above. As a consequence, the subsets of $\mathbb{X}$ on which the above expressions are constant are not related to the fibres of the model. This shows that the divergence (\ref{eq: leastsquares}) does not satisfy the necessary conditions for a divergence in the data set model formalism.\\
Another divergence which yields the same model map was already mentioned in an earlier publication \cite{AnthonisNaudts1}. It is given by
\begin{align*}
D_\lambda(S||f_{a,b}) &= \frac{1}{2\sum_{j,k} (x_i - x_k)^2} \lambda^2 \sum_{j,k} [y_j - y_k - a (x_j - x_k)]^2\\
&\qquad + \frac{1}{2\sum_{j,k} (x_i - x_k)^2} \sum_{j,k} [x_j y_k - x_k y_j - b (x_j - x_k)]^2,
\end{align*}
where $\lambda > 0$ is an arbitrary constant introduced for dimensional reasons. The matrix of second derivatives of this divergence is particularly simple, with constant components wholly independent of the data. This means this matrix necessarily coincides with the metric tensor, which is characterised by
\begin{align*}
g_{aa}(a,b) = \lambda^2,\quad g_{ab}(a,b) = 0, \quad g_{bb}(a,b) = 1.
\end{align*}
Since both objects coincide, all connection coefficients will vanish and the coordinates already in use are affine coordinates for this connection. The model manifold is therefore concluded to be endowed with a flat and torsionless connection.

\newpage
\section{The grand canonical ensemble for identical particles}
%\subsection{Construction and study of the geometry}
In this first non-trivial example, the goal is to model a system of non-interacting bosonic particles which may occupy states $\{j\}$ with corresponding energy levels $\{\varepsilon_j\}$. This is a well-known and extensively studied model. Hence, no surprising results are to be expected. Nevertheless, such a familiar example may be interesting for the reader as it applies the developed formalism to a system he or she may be already be familiar with.\\
The set $\mathbb{X}$ of all data sets contains possible outcomes of an experiment to measure the number $n_j$ of bosons occupying the state $j$. These will be denoted by $(n_i)_i$ or just $n$ where the context makes confusion unlikely to occur. (The letter $n$ is not used for the dimension of $\mathbb{M}$ in this example. A similar remark holds for $\mu$ introduced soon.) It will follow implicitly from the discussion that not all possible configurations $n$ are in the domain of the model map. Such configurations are excluded from the onset. The model points making up the manifold $\mathbb{M}$ are the distributions of the grand canonical ensemble. That is, they are the probabilities of the states $j$ having occupations $n_j$ and they are given by
\begin{align} 
p_{\beta,\mu}(\{n_i\}) &= \frac{1}{\mathcal{Z}(\beta,\mu)} \prod_j \exp\{ - \beta n_j (\varepsilon_j - \mu)\}\nonumber\\
&= \exp\left\{ - \ln \mathcal{Z}(\beta,\mu) - \beta \sum_j n_j\varepsilon_j + \beta \mu \sum_j n_j\right\},
\label{eq: pGCE}
\end{align}
where $\mathcal{Z}$ is the partition function of the system---also serving as a normalisation factor---, $\beta$ represents the inverse temperature and $\mu$ the chemical potential. It is a well-known result of statistical physics, see for example \cite{Naudts}, that 
\begin{align*}
\mathcal{Z}(\beta,\mu) &= \left( \prod_j \sum_{n_j=0}^\infty \right) \exp\left\{ - \beta \sum_i n_i( \varepsilon_i- \mu)\right\}\\
&= \prod_j \frac{1}{1 - \exp\{ - \beta (\varepsilon_j - \mu)\}},
\end{align*}
where the expression between large parenthesis is an abuse of notation to indicate a multitude of sums. It is assumed that the remaining product converges. This is not a trivial assumption but it is not native to the data set model formalism and so no excessive attention will be given to this problem.  From expression (\ref{eq: pGCE}) it is clear that the model distributions belong to the exponential family with parameters $\theta^1 = \beta$ and $\theta^2 = - \beta \mu$. However, it is more instructive to use $\beta$ and $\mu$ as parameters or coordinates of the model manifold as this will illustrate that the geometric properties of the model are indeed independent of the choice of parameters.\\
The geometry of a data set model is completely determined by the chosen divergence. The choice made here is given by the expression
\begin{align}
D(n||p_{\beta,\mu}) &=  \ln \mathcal{Z}(\beta,\mu) - \sum_i n_i \left(  - \beta \varepsilon_i + \beta \mu \right).
\label{eq: Dgce}
\end{align}
For a realistic experimental outcome there is a highest occupied energy level and a finite number of particles, thereby avoiding additional mathematical difficulties with this definition.\\
It first needs to be verified whether or not equation (\ref{eq: Dgce}) does indeed define a divergence. The expression is continuous and continuously differentiable as a function of the parameters when $\beta > 0$ and $\mu \notin \{\varepsilon_j\}$. Since from a physical point of view this model only makes sense when $\beta > 0$ and $\mu < \min_j \varepsilon_j$, no difficulties are expected to be encountered in a practical setting.\\
The derivative of the divergence (\ref{eq: Dgce}) with respect to the inverse temperature $\beta$ is given by the expression
\begin{align}
\partial_\beta D (n||p_{\beta,\mu}) &= \partial_\beta \ln \mathcal{Z}(\beta,\mu) + \sum_i n_i \varepsilon_i - \mu \sum_i n_i\nonumber\\
&=  -\sum_j \frac{\varepsilon_j - \mu}{\exp\{\beta(\varepsilon_j - \mu)\}-1}  + \sum_i n_i \varepsilon_i - \mu \sum_i n_i. \label{eq: dDbeta}
\end{align}
Through an analogous computation, it is found that
\begin{align}
\partial_\mu D(n||p_{\beta,\mu}) &= \partial_\mu \ln \mathcal{Z}(\beta,\mu) - \beta\sum_i n_i \nonumber\\
&= \beta \sum_j\frac{1}{ \exp\{\beta(\varepsilon_j - \mu)\}-1} - \beta\sum_i n_i. \label{eq: dDmu}
\end{align}
These two derivatives must vanish simultaneously if $n$ is to be contained in the fibre of the grand canonical distribution $p_{\beta,\mu}$. Rewriting the above expressions shows this is equivalent to demanding that the total energy and total number of bosonic particles can be expressed through the well-known relations
\begin{align*}
\sum_i n_i \varepsilon_i &= \sum_j \frac{\varepsilon_j }{\exp\{\beta(\varepsilon_j - \mu)\}-1},\\
\sum_i n_i &= \sum_j \frac{1}{\exp\{ \beta(\varepsilon_j - \mu)\} - 1}.
\end{align*}
Whether or not a given data set $n$ is acceptable depends on whether or not these equations have solutions for $\beta$ and $\mu$ for this data set. The second derivatives are also important to the data set model formalism. A straightforward computation yields
\begin{align*}
\partial_\beta^2 D(n||p_{\beta,\mu}) &= \sum_j \frac{(\varepsilon_j - \mu)^2}{(\exp\{\beta(\varepsilon_j  -\mu)\} -1)^2} \exp\{\beta(\varepsilon_j - \mu)\},\\
\partial_\beta\partial_\mu D(n||p_{\beta,\mu}) &= \sum_j \frac{1}{\exp\{\beta(\varepsilon_j - \mu)\}- 1} - \sum_i n_i\\
&\qquad - \beta \sum_j \frac{\varepsilon_j - \mu}{(\exp\{\beta(\varepsilon_j - \mu)\}-1)^2} \exp\{\beta(\varepsilon_j- \mu)\},\\
\partial^2_\mu D(n||p_{\beta,\mu}) &= \beta^2 \sum_j \frac{1}{(\exp\{\beta(\varepsilon_j - \mu)\}-1)^2} \exp\{\beta(\varepsilon_j - \mu)\}.
\end{align*}
The mixed derivative still depends on the data set $n$. However, this dependency only involves the sum of all occupation numbers and can thus be rewritten using the equalities right above when $n$ is in the fibre of $p_{\beta,\mu}$. Furthermore, this will cancel out the first sum in the expression, thereby simplifying the derivative. This means the metric tensor can easily be written down as
\begin{align}
g(\beta,\mu) = \sum_j \frac{\exp\{\beta(\varepsilon_j - \mu)\}}{(\exp\{\beta(\varepsilon_j - \mu)\} - 1)^2} \left( \begin{array}{c c}
(\varepsilon_j - \mu)^2 & - \beta(\varepsilon_j - \mu)\\
-\beta(\varepsilon_j - \mu) & \beta^2
\end{array}\right).\label{eq: ggce}
\end{align}
This metric can be shown to be positive definite through direct computation. For any vector $\vec{v} \in \tg{\beta,\mu}{M}$ it holds that
\begin{align*}
g(\vec{v},\vec{v}) = \sum_j \frac{\exp\{\beta(\varepsilon_j - \mu)\}}{(\exp\{\beta(\varepsilon_j - \mu)\}-1)^2} \left( v^\beta (\varepsilon_j - \mu) - \beta v^\mu  \right)^2 \geqslant 0.
\end{align*}
The next step in the search for the canonical coordinates of the model distributions is to determine the connection. Since only the mixed derivative of the divergence depends on the data set $n$, it can quickly be seen that 
\begin{align*}
{\omega^\beta}_{\beta\beta} = {\omega^\mu}_{\beta\beta} = {\omega^\beta}_{\mu\mu} = {\omega^\mu}_{\mu\mu} = 0.
\end{align*}
The two remaining independent coefficients, ${\omega^\beta}_{\beta\mu}$ and ${\omega^\mu}_{\beta\mu}$, are only slightly harder to find. $\mathbb{X}$ is a discrete set, but since it can be thought of as embedded in $\mathbb{R}^N$ with $N$ (at least) the number of available energy levels if this is finite, it is possible to use either the expression (\ref{eq: defomegaexp}) to compute the coefficients or the discretised version (\ref{eq: disdefomega}). It is the latter expression that will be used here. This means fixing values for the parameters $\beta$ and $\mu$ and then finding appropriate data sets with which to compute the coefficients of the connection. The computations of connection coefficients each include two data sets: one data set $n$ which is contained in the fibre of $p_{\beta,\mu}$ and another one denoted as $n^{(k)}$ for which 
\begin{align*}
\partial_l D(n^{(k)}||p_{\beta,\mu}) = \delta_l^k,
\end{align*}
or another invertible matrix if this is more practical. Since the derivatives of the divergence only depend on the data sets  through the quantities
\begin{align*}
\sum_i n_i \quad \text{and} \quad \sum_i n_i \varepsilon_i,
\end{align*}
which are constant in the fibres, it is not too difficult to find these ``off-fibre'' data sets. The condition for $n^{(\beta)}$ is that
\begin{align*}
\partial_\beta D(n^{(\beta)}||p_{\beta,\mu}) \neq 0 \quad\text{and} \quad \partial_\mu D(n^{(\beta)}||p_{\beta,\mu}) = 0.
\end{align*}
From the second of the conditions, it can be seen that
\begin{align*}
 \sum_i n_i^{(\beta)} = \sum_i n_i = \sum_j \frac{1}{\exp\{\beta(\varepsilon_j -\mu)\} - 1}.
\end{align*}
This is enough to compute the the connection coefficient ${\omega^\beta}_{\beta\mu}$, given by
\begin{align*}
{\omega^\beta}_{\beta\mu} &= \frac{\partial_\beta\partial_\mu D(n^{(\beta)}||p_{\beta,\mu}) - \partial_\beta\partial_\mu D(n||p_{\beta,\mu})}{\partial_\beta D(n^{(\beta)}||p_{\beta,\mu}) }\\
&= \frac{\sum_i n_i^{(\beta)} - \sum_i n_i}{\partial_\beta D(n^{(\beta)}||p_{\beta,\mu}) }\\
&= 0.
\end{align*}
This means only the coefficient ${\omega^\mu}_{\beta\mu}$ remains to be determined. The conditions on the data set $n^{(\mu)}$ can be expressed as
\begin{align}
\partial_\mu D(n^{(\mu)}||p_{\beta,\mu}) \neq 0 \quad\text{and} \quad \partial_\beta D(n^{(\mu)}||p_{\beta,\mu}) = 0.\label{eq: defnmu}
\end{align}
The first condition implies
\begin{align*}
\partial_\mu D(n^{(\mu)}||p_{\beta,\mu}) &= \beta \sum_j \frac{1}{\exp\{\beta(\varepsilon_j - \mu)\} - 1} - \beta\sum_i n_i^{(\mu)}\\
&= \beta\left( \sum_i n_i - \sum_i n_i^{(\mu)}\right)
\end{align*}
must differ from zero. Since $n^{(\mu)}$ is arbitrary within the constraints (\ref{eq: defnmu}), the expression between parentheses can be chosen to be equal to $1$. This suffices to determine the final (and only) independent connection coefficient, being
\begin{align*}
{\omega^\mu}_{\beta\mu} &= \frac{\partial_\beta\partial_\mu D(n^{(\mu)}||p_{\beta,\mu}) - \partial_\beta\partial_\mu D(n||p_{\beta,\mu})}{\partial_\mu D(n^{(\mu)}||p_{\beta,\mu}) }\\
&= \frac{-\sum_i n_i^{(\mu)} + \sum_i n_i}{\beta}\\
&= \frac{1}{\beta}.
\end{align*}
By simple substitution it can be verified that this connection will indeed make the Hessian of the divergence independent of its first argument and equal to the metric found before, as it is given in expression (\ref{eq: ggce}).\\
To find the canonical parameters making up the affine coordinates for this model manifold, attention is turned towards equation (\ref{eq: flatcon}). The partial differential equations to be solved in this example are
\begin{align*}
\partial^2_\beta \Theta^i &= 0,\\
\partial_\beta \partial_\mu \Theta^i &= \beta^{-1} \partial_\mu \Theta^i,\\
\partial^2_\mu \Theta^i &=0.
\end{align*}
From the first and the last of these equations, it follows that
\begin{align*}
\Theta^i(\beta,\mu) = A^i \mu \beta + B^i \mu + C^i \beta + D^i,
\end{align*}
where the Roman capital letters are constants. The second differential equation demands that $A^i = \beta^{-1}( A^i \beta + B^i)$ and thus $B^i=0$. The other three constants may be chosen freely and with a proper choice one obtains the canonical parameters mentioned earlier,
\begin{align*}
\theta^1 = \beta \quad\text{and}\quad \theta^2 = -\beta \mu.
\end{align*}
Since the connection is flat and its coefficients can be expressed in a fairly simple way in the coordinates $\beta$ and $\mu$, it is worthwhile to see what the geodesics and covariant constant vector fields would look like in the coordinate system determined by $\beta$ and $\mu$.\\
Given a vector $\vec{v}$ at one particular point of the manifold $\mathbb{M}$, it is possible to construct a covariant constant vector field everywhere through parallel transport. Along an arbitrary curve parametrised by $t$, the covariant constant vector field satisfies the equations
\begin{align*}
\diff{v^\beta}{t} = 0\quad\text{and}\quad \diff{v^\mu}{t} = -\frac{1}{\beta} \diff{\mu}{t} v^\beta - \frac{1}{\beta} \diff{\beta}{t} v^\mu
\end{align*}
The first equation implies that $v^\beta$ has the same value everywhere on $\mathbb{M}$. Because this component is constant, it is possible to rewrite the second equation as
\begin{align*}
\diff{}{t}\left( \frac{\beta v^\mu}{v^\beta}\right) = - \diff{\mu}{t}.
\end{align*}
This differential equation has for its solution 
\begin{align*}
v^\mu(t) = \frac{\mu_0 - \mu(t)}{\beta(t)} v^\beta,
\end{align*}
where $\mu_0$ is an integration constant which can be determined by choosing the value of the vector field at a given point. An example of a covariant constant vector field for this connection is depicted in the drawing. This may not look to the reader as a vector field which is parallel with itself everywhere but this is in fact the case. The origin of the possible confusion is that a coordinate system (and thus a coordinate frame) is chosen for the illustration in which the connection coefficients do not vanish.
\begin{center}
\begin{tikzpicture}
\path[draw,->] (-0.25,0) -- (4.5,0);
\path[draw,->] (0,-1.25) -- (0,3.25);
\path[draw] (4.8,-0.2) node {$\beta$};
\path[draw] (-0.2,3.55) node {$\mu$};

\path[draw] (-0.05,1.25) -- (0.05,1.25);
\path[draw] (0,1.25) node[anchor=east] {$\mu_0$};

\foreach \x in {0.5,1,...,4}
	\foreach \y in {-1,-0.5,...,3}
		{
		\path[draw,-latex] (\x,\y) -- (\x + 0.2, {\y + (1.25-\y)/\x*0.2});
		}
\end{tikzpicture}
\end{center}
In order to get a better understanding of the behaviour of this connection, also the geodesics can be considered. The general geodesic equations are given by the two non-linear differential equations
\begin{align*}
\ddiff{\beta}{t} = 0 \quad \text{and} \quad \ddiff{\mu}{t} = - \frac{2}{\beta}\diff{\beta}{t}\diff{\mu}{t}.
\end{align*}
%A particular solution of this system has $\beta$ a constant and then $\mu(t) = Ct+D$ for suitable constants $C$ and $D$. These are the $\beta$-coordinate curves, which naturally coincide with the $\theta^1$-curves as both coordinates are identical. All other geodesics can be described by taking $\beta(t) = At + B$.
From the first equation it follows that $\beta(t) = At + B$. This allows the second differential equation to be rewritten into the form
\begin{align*}
\ddiff{\mu}{t}(t) = - 2 \frac{A}{At+B}\diff{\mu}{t}(t).
\end{align*}
A particular solution is achieved when $A=0$, making $\beta$ constant and $\mu(t) = Ct+D$. Otherwise, this equation can be solved by elementary methods such as separation of variables and order reduction to obtain the result
\begin{align*}
\mu(t) = \mu(t_0) + \dot{\mu}(t_0) \left( \frac{1}{At_0 + B} - \frac{1}{At + B}\right).
\end{align*}
This family of curves excludes the $\beta$- or $\theta^1$-curves, which are a particular solution to the geodesic equations corresponding to $A=0$. The drawing below shows a selection of canonical coordinate curves---all of which are geodesics---as they appear in the original coordinate system of the parameters $\beta$ and $\mu$.
%To obtain the $\theta^2$ coordinate curves, it is required to compute the proper values of $\mu(t_0)$ and $\dot{\mu}(t_0)$ for each of the desired curves. This can happen through the (inverse) transformation formulas for the canonical coordinates
%\begin{align*}
%\beta(t) = \theta^1(t) \quad \text{and} \quad \mu(t) = -\frac{\theta^2(t)}{\theta^1(t)}
%\end{align*}
%and for the vector components
%\begin{align*}
%\left( \begin{array}{c} \dot{\beta}(t) \\ \dot{\mu}(t)\end{array}\right) &= \left( \begin{array}{c c} 1 &  0 \\ \frac{\theta^2(t)}{(\theta^1(t))^2} & -\frac{1}{\theta^1(t)}\end{array}\right) \left( \begin{array}{c} \dot{\theta}^1(t)\\ \dot{\theta}^2(t)\end{array}\right).
%\end{align*}
%Using $\beta$ as the parameter $t$ simplifies these equations greatly, as does taking $t_0 = \beta_0 = 1$. For a $\theta^2$-curve, this means $\mu(1) = -\theta^2$ and $\dot{\mu}(1) = \theta^2$.
\begin{center}
\begin{tikzpicture}
\path[draw,->] (0,0) -- (5.8,0);
\path[draw,->] (0,-2.2) -- (0,2.3);

\path[draw] (6,-0.2) node {$\beta$};
\path[draw] (-0.2,2.5) node {$\mu$};

\path[clip] (-0.5,2.2) -- (5.5,2.2) -- (5.5,-2.2) -- (-0.5,-2.2) -- cycle;

\foreach \b in {1,2,...,5}
	{
	\path[draw,domain=-2.2:2.2,gray,smooth,variable=\m] plot( \b,\m);
	}
	
\foreach \n in {-12,-11,...,12}
	{
	\path[draw,domain=0.1:5.5,gray,smooth,variable=\b] plot(\b,{\n/\b});
	}
	
\foreach \n in {-2,-1,0,1,2}
	{
	\path[draw] (-0.05,\n) -- (0.05,\n);
	\path[draw] (-0.05,\n) node[anchor=east] {\small \n};
	}
	
\foreach \n in {1,2,...,5}
	{
	\path[fill,white] ({\n-0.125},-0.1) rectangle ({\n+0.125},-0.5);
	\path[draw] (\n,-0.05) node [anchor=north] {\small \n};
	}
\end{tikzpicture}
\end{center}

\newpage
\section{The von Mises-Fisher distributions}
This example is divided into two separate modelling problems. In both cases, the manifold is a subset of an exponential family. This is very instructive as the geometry of these submanifolds as induced by the divergence is exactly what is expected from a submanifold of a Euclidean space---even though the containing space is not endowed with a Euclidean metric.
\subsection{Fixing the width of the distribution}
The von Mises-Fisher distributions are families of distributions on the sphere \cite{FisherDist}. That is, they belong to the set of distributions $p$ for which
\begin{align}
\label{eq: risone}
\sum_i \mathbb{E}_p[x_i]^2 = 1.
\end{align}
In particular, the von Mises-Fisher distributions take the exponential form
\begin{align}
\label{eq: defVMF}
p_{\kappa,\mu}(x) = \exp\{ - \Phi(\kappa) + \kappa \mu^i x_i \} \quad &\text{where} \quad \sum_i (\mu^i)^2 = 1.
\end{align}
This notation employs the traditional parametrisation but it is easy to see that this is an exponential family by taking as the canonical parameters $\theta^i = -\kappa \mu^i$, analogous to the situation for the grand canonical distribution of bosonic particles treated previously. It is not the intention to repeat the above example with a few extra degrees of freedom. Instead the computations here will be restricted to three-dimensional data and---more importantly---only a subset of the von Mises-Fisher distributions will be considered as the model manifold $\mathbb{M}$. In particular the parameter $\kappa$, which is a measure for the width of the distribution, will be held constant. The manifold $\mathbb{M}$ thus obtained is topologically equivalent to the $2$-sphere, which can be seen by employing a parametrisation by spherical polar coordinates\footnote{The usual remarks for the use of this parametrisation apply, that is when $\theta=0$ or $\theta=\pi$, the coordinate $\varphi$ is not well-defined and so $\theta$ and $\varphi$ do not form a proper coordinate system. As this will not hinder the treatment of this example too much, this parametrisation will be used nonetheless.} $\theta$ and $\varphi$, 
\begin{align*}
p_{\theta,\varphi}(x) = \exp\{ -\Phi(\kappa) + \kappa (\sin(\theta)\cos(\varphi) x_1 + \sin(\theta)\sin(\varphi) x_2 + \cos(\theta) x_3)\}.
\end{align*}
The choice for the symbol $\theta$ is made in accordance with the usual names of the spherical polar coordinates and it should not be mistaken for a canonical coordinate of an exponential family.\\
The set $\mathbb{X}$ of data sets is the space of all possible probability distributions on the sphere as defined by equation (\ref{eq: risone}). The model manifold $\mathbb{M}$ is the two dimensional set of von Mises-Fisher distributions with a fixed value of $\kappa$. As the divergence $D$ it proves convenient to opt for the Kullback-Leibler divergence function,
\begin{align*}
D(p||p_{\kappa,\mu}) &= - S(p) - \int_X p(x) \left[ - \Phi(\kappa) + \kappa \mu^i x_i \right]\dee x\\
&= - S(p) + \Phi(\kappa)\\
&\qquad - \kappa ( \sin(\theta)\cos(\varphi) \mathbb{E}_p[x_1] + \sin(\theta) \sin(\varphi) \mathbb{E}_p[x_2] + \cos(\theta)\mathbb{E}_p[x_3]),
\end{align*}
where as usual $S(p)$ is the Shannon entropy (\ref{eq: Shannon}) of the probability distribution $p$. The derivatives of this divergence are given by
\begin{align*}
\partial_\theta D(p||p_{\theta,\varphi}) &= -\kappa( \cos(\theta)\cos(\varphi) \mathbb{E}_p[x_1] + \cos(\theta) \sin(\varphi) \mathbb{E}_p[x_2] - \sin(\theta)\mathbb{E}_p[x_3]),\\
\partial_\varphi D(p||p_{\theta,\varphi}) &= -\kappa( -\sin(\theta)\sin(\varphi) \mathbb{E}_p[x_1] + \sin(\theta)\cos(\varphi) \mathbb{E}_p[x_2]).
\end{align*}
Though it is formally possible to solve these equations together with the normalisation condition (\ref{eq: risone}) to find which distributions $p$ make up the fibres of the von Mises-Fisher distribution $p_{\theta,\varphi}$, there is an easier way. The divergence $D(p||p_{\theta,\varphi})$ as defined above is minimal when the expression $\mu^i \mathbb{E}_p[x_i]$ attains its maximal value. Since this can be viewed as the inner product of two vectors of unit norm, the maximum is obtained when the two vectors are equal, that is
\begin{align*}
\mathbb{E}_p[x_1] &= \sin(\theta)\cos(\varphi),\\
\mathbb{E}_p[x_2] &= \sin(\theta)\sin(\varphi),\\
\mathbb{E}_p[x_3] &= \cos(\theta).
\end{align*}
Note that there is also another set of distributions $p$ for which the derivatives of $D(p||p_{\theta,\varphi})$ vanish---those distributions for which the values $\mathbb{E}_p[x_i]$ take the negative of the values above. This is a local maximum of the divergence function, however, and hence it is no candidate for the required solution. The second derivatives of the divergence are found to equal
\begin{align*}
\partial^2_\theta D(p||p_{\theta,\varphi}) &= \kappa( \sin(\theta)\cos(\varphi) \mathbb{E}_p[x_1] + \sin(\theta) \sin(\varphi) \mathbb{E}_p[x_2] + \cos(\theta)\mathbb{E}_p[x_3]),\\
\partial_\theta\partial_\varphi D(p||p_{\theta,\varphi}) &= \kappa( \cos(\theta)\sin(\varphi) \mathbb{E}_p[x_1] - \cos(\theta) \cos(\varphi) \mathbb{E}_p[x_2]),\\
\partial^2_\varphi D(p||p_{\theta,\varphi}) &= \kappa( \sin(\theta)\cos(\varphi) \mathbb{E}_p[x_1] + \sin(\theta)\sin(\varphi) \mathbb{E}_p[x_2]).
\end{align*}
Combined with the expressions for $\mathbb{E}_p[x_i]$ that make up the conditions for $p$ to be an element of the $p_{\theta,\varphi}$-fibre, knowledge of these second derivatives allows for the metric tensor to be expressed. Its components read
\begin{align*}
g_{\theta\theta}(\theta,\varphi) &= \kappa,\quad
g_{\theta\varphi}(\theta,\varphi) = 0,\quad
g_{\varphi\varphi}(\theta,\varphi) = \kappa \sin^2(\theta).
\end{align*}
This is the metric tensor of a $2$-sphere with radius $\kappa^{1/2}$. This is a clear example of a model where the geometry introduced by the divergence coincides with the geometry that is expected from the manifold $\mathbb{M}$ itself, even though the manifold itself is not isomorphic to $\mathbb{R}^n$. However, this does not ensure the connection $\nabla$ will be the metric connection on such a sphere. In fact, at this point there is no guarantee that a connection as introduced in the previous chapter would even exist in this example. A formal investigation is thus required.\\
The determination of the connection coefficients will proceed by first finding curves $\varepsilon \rightarrow X^k(\varepsilon)$ through the space $\mathbb{X}$. Both the first and second derivatives of the divergence function in this example depend only on the expectation values of the random variables $x_i$, which means it suffices to focus on these expectation values. However, these values are not independent: they too lie on a sphere just as the parameters $\mu^i$ do. The use of this knowledge reveals that the result that one family of desirable curves can be obtained (as functions of a parameter $\psi$) through the expressions
\begin{align*}
\mathbb{E}_p[x_1] &= \sin(\theta-\psi)\cos(\varphi),\\
\mathbb{E}_p[x_2] &= \sin(\theta-\psi)\sin(\varphi),\\
\mathbb{E}_p[x_3] &= \cos(\theta-\psi).
\end{align*}
Analogously, the second family of curves is obtained (as functions of $\xi$):
\begin{align*}
\mathbb{E}_p[x_1] &= \sin(\theta)\cos(\varphi-\xi),\\
\mathbb{E}_p[x_2] &= \sin(\theta)\sin(\varphi-\xi),\\
\mathbb{E}_p[x_3] &= \cos(\theta).
\end{align*}
When either of these parameters are equal to zero, the curves find themselves in a point representing a fibre of the model $p_{\theta,\varphi}$ as intended. Using these curves over the sphere of the expectation values, it is found that
\begin{align*}
\pddiff{D}{\psi}{\theta}\bigg|_{\psi=\xi=0} = \kappa, &\quad \pddiff{D}{\psi}{\varphi}\bigg|_{\psi=\xi=0} = 0,\\
\pddiff{D}{\xi}{\theta}\bigg|_{\psi=\xi=0} = 0, &\quad \pddiff{D}{\xi}{\varphi} \bigg|_{\psi=\xi=0} = \kappa \sin^2(\theta).
\end{align*}
This yields exactly the metric tensor, even though this is a coincidental consequence of the choice of the $X^k$ rather than a general property---another choice of curves would have been equally valid but would have yielded different expressions. While the result is not the Kronecker-delta one would desire under ideal circumstances, the invertibility of the metric tensor means these curves can indeed be used to define the connection coefficients. The following step is to use the above expressions for the curves through the expectation value sphere in the expressions for the second derivatives of the divergence. The resulting quantities can be derived with respect to $\psi$ and $\xi$ to obtain the coefficients $\omega_{k,ij} = g_{ks}{\omega^s}_{ij}$. The usual coefficients can then be found through a multiplication with the matrix inverse of the metric tensor. There are six derivatives to compute. These are---again suppressing some straightforward function arguments in order to reduce the burden of notation---
\begin{align*}
\pdiff{}{\psi}\pdiff{{}^2 D}{\theta^2}\bigg|_{\psi=\xi=0} &= 0, &\pdiff{}{\xi}\pdiff{{}^2 D}{\theta^2}\bigg|_{\psi=\xi=0} &= 0,\\
\pdiff{}{\psi}\pddiff{ D}{\theta}{\varphi}\bigg|_{\psi=\xi=0} &= 0,& \pdiff{}{\xi}\pddiff{ D}{\theta}{\varphi}\bigg|_{\psi=\xi=0} &= \kappa \cos(\theta)\sin(\theta),\\
\pdiff{}{\psi}\pdiff{{}^2 D}{\varphi^2}\bigg|_{\psi=\xi=0} &= - \kappa \sin(\theta)\cos(\theta),& \pdiff{}{\xi}\pdiff{{}^2 D}{\varphi^2}\bigg|_{\psi=\xi=0} &= 0.
\end{align*}
Since the matrix multiplication that needs to be performed involves a square matrix, this operation is straightforward and the two independent non-van\-ish\-ing connection coefficients appear as
\begin{align*}
{\omega^\theta}_{\varphi\varphi}(\theta,\varphi) = - \sin(\theta)\cos(\theta)\quad \text{and}\quad {\omega^\varphi}_{\theta\varphi}(\theta,\varphi) = \frac{\cos(\theta)}{\sin(\theta)}.
\end{align*}
These coincide with the coefficients of the metric connection on a spherical surface (see for example \cite{Arfken}). It is not necessary to compute the curvature tensor to see if it vanishes: the reader will no doubt agree that the familiar metric connection of a spherical surface will not be found to exhibit flatness.\\
The example is thus a very instructive one. It shows that for a properly chosen divergence function, the geometry introduced by the formalism of data set models will coincide with the geometry expected from the choice of the model manifold $\mathbb{M}$, something which may come unexpected from readers familiar with information geometry. After all, a prominent role there is played by connections which are not metric. Also, since this connection is not flat, it is an example of a data set model where a connection exists but where the metric is not of the Hessian type.
\newpage
\subsection{A cylindrical submanifold}
With the knowledge obtained in the previous example regarding von Mises-Fisher distributions, it is easy to consider a different submanifold of the three-dimensional family (\ref{eq: defVMF}). In particular, this example will treat a submanifold homeomorphic to the half-cylinder, that is the two-parameter family of distributions
\begin{align*}
p_{\varphi,\lambda}(x) = \exp\{ - \Xi(\kappa,\lambda) + \kappa( \cos(\varphi)x_1 + \sin(\varphi)x_2) - \lambda x_3\},
\end{align*}
where $\kappa$ is again a positive constant and $\lambda>0$ is now an independent parameter. This distribution is essentially a product of the one-dimensional von Mises(-Fisher) distribution and the exponential distribution. The motivation for computing this very similar example is to see if here a flat connection follows. This might be expected from the analogy in the previous example, where the geometry of the model manifold coincided with the sphere with which it is homeomorphic. Cylinder mantles are known to have flat metric connections---a fact which follows from their ability to be unrolled onto $\mathbb{R}^2$ without distorting the intrinsic geometry of the surface.\\
The data sets which will be modelled by these new cylindrical von Mises-Fisher distributions are those distributions $p$ for which
\begin{align*}
\mathbb{E}_p[x_1]^2 + \mathbb{E}_p[x_2]^2 = 1\quad\text{and}\quad \mathbb{E}_p[x_3] > 0.
\end{align*}
Unlike in the previous example, it is actually necessary to know the normalisation function $\Xi$ in order to complete the computations. The distribution $p_{\varphi,\lambda}$ must be normalised and thus
\begin{align*}
1 &= \int_X \exp\{-\Xi(\kappa,\lambda) + \kappa(\cos(\varphi)x_1 + \sin(\varphi)x_2) - \lambda x_3\} \dee x\\
&= \exp\{-\Xi(\kappa,\lambda)\} \int_{S^1} \exp\{ \kappa( \cos(\varphi)x_1 + \sin(\varphi) x_2)\} \dee \ell \\
&\qquad \times \int_0^{\infty} \exp\{ - \lambda x_3\} \dee x_3\\
&= \exp\{-\Xi(\kappa,\lambda)\} \frac{2\pi I_0(\kappa)}{\lambda},
\end{align*}
where $I_0$ is the modified Bessel function of order 0 \cite{Arfken}. This means
\begin{align*}
\Xi(\kappa,\lambda) = \ln\left(\frac{2\pi I_0(\kappa)}{\lambda}\right).
\end{align*}
The remaining part of the computation is analogous to the treatment above and so less details will be supplied so as not to burden the reader. The choice of divergence is again the Kullback-Leibler divergence, which has derivatives in this example given by
\begin{align*}
\partial_{\varphi} D(p||p_{\varphi,\lambda}) &= - \kappa( -\sin(\varphi) \mathbb{E}_p[x_1] + \cos(\varphi)\mathbb{E}_p[x_2]),\\
\partial_\lambda D(p||p_{\varphi,\lambda}) &= -\frac{1}{\lambda} + \mathbb{E}_p[x_3].
\end{align*}
The distributions $p$ in the $p_{\varphi,\lambda}$-fibres are thus those distributions satisfying
\begin{align*}
\mathbb{E}_p[x_1] &= \cos(\varphi),\\
\mathbb{E}_p[x_2] &= \sin(\varphi),\\
\mathbb{E}_p[x_3] &= \lambda^{-1}.
\end{align*}
The second derivatives of the divergence are equal to
\begin{align*}
\partial^2_\varphi D(p||p_{\varphi,\lambda}) &= \kappa( \cos(\varphi) \mathbb{E}_p[x_1] + \sin(\varphi)\mathbb{E}_p[x_2]),\\
\partial_\varphi\partial_\lambda D(p||p_{\varphi,\lambda}) &= 0,\\
\partial^2_\lambda D(p||p_{\varphi,\lambda}) &= \lambda^{-2}.
\end{align*}
On the fibres these expressions are constants and so the metric tensor takes the form
\begin{align*}
g(\varphi,\lambda) = \left(\begin{array}{c c} \kappa & 0 \\ 0 & \lambda^{-2} \end{array}\right).
\end{align*}
The $g_{\varphi\varphi}$-component, as well as the off-diagonal components of this metric are what are to be expected for a cylinder with radius $\kappa^{1/2}$. The $\lambda^{-2}$ serving as the $g_{\lambda\lambda}$-component may seem unexpected for a cylinder, but this should come as no surprise given that the model distributions attach an exponential probability density to the random variable $x_3$.\\
In order to compute the connection coefficients, proper curves must be found along which to compute the derivatives serving as the definition of the connection coefficients---just like in the previous example, preference is given to this method for practical purposes. Since only the second derivative with respect to $\varphi$ of the divergence depends on the distribution $p$ all coefficients except ${\omega^\lambda}_{\varphi\varphi}$ and ${\omega^\varphi}_{\varphi\varphi}$ can already be seen to vanish. The first of these two will also vanish. Indeed, the appropriate derivative must be taken along a curve which is parametrised by $\mathbb{E}_p[x_3]$---as that is the only $p$-dependency of $\partial_\lambda D(p||p_{\varphi,\lambda})$---but $\partial^2_\varphi D(p||p_{\varphi,\lambda})$ does not depend on this expectation value, making the derivative in (\ref{eq: defomegaexp}) vanish. In order to determine the last coefficient, ${\omega^\varphi}_{\varphi\varphi}$, take curves parametrised by $\xi$ such that along this curve
\begin{align*}
\mathbb{E}_p[x_1] = \cos(\varphi-\xi), \quad \mathbb{E}_p[x_2] = \sin(\varphi-\xi), \quad \mathbb{E}_p[x_3] = \text{cte},
\end{align*}
analogous to what happened in the spherical example. Then, along this curve, the second $\varphi$-derivative of the divergence takes the form
\begin{align*}
\partial^2_\varphi D(p||p_{\varphi,\lambda}) &= \kappa( \cos(\varphi) \cos(\varphi-\xi) + \sin(\varphi)\sin(\varphi-\xi)).
\end{align*}
Deriving this expression with respect to $\xi$ and evaluating in $\xi=0$ shows that also this coefficient vanishes. This means that the chosen parameters $\varphi$ and $\lambda$ are indeed the affine coordinates of the connection and thus the canonical parameters of the distribution.\\
Since all the connection coefficients vanish identically, it is not necessary to compute the curvature tensor. However, it is possible to check whether or not the Codazzi-Peterson-like equation (\ref{eq: Codazzi}) is satisfied. As the computation is already taking place in an affine coordinate system, it is sufficient to verify whether or not
\begin{align*}
\partial_\varphi g_{\lambda\varphi} = \partial_\lambda g_{\varphi\varphi} \quad \text{and}\quad \partial_{\lambda}g_{\varphi\lambda} = \partial_\varphi g_{\lambda\lambda}.
\end{align*}
A quick peek at the coefficients of the metric tensor teaches us that all four of these quantities vanish and so both equalities are satisfied. This would mean that there does indeed exist a Massieu function $\Phi$, the Hessian of which is the metric tensor. This may come as a surprise since the cylinder has a periodic nature in one direction and so it seems impossible that there exists a properly behaved convex function everywhere on the cylinder mantle. The answer to this is of course that there is no such function. After all, the coordinate $\varphi$ is only a proper coordinate if a curve parametrised by $\lambda$ is removed from the half-cylinder mantle and the argument leading to the existence of the Massieu function made use of Poincar\'e's lemma, which holds locally rather than globally. When the points with coordinates $(\pi,\lambda)$ are removed, a proper coordinate is obtained and then
\begin{align*}
\Phi(\varphi,\lambda) = \frac12 \kappa \varphi^2 - \ln \lambda
\end{align*}
is indeed a convex function of which the matrix of second derivatives equals the metric tensor. This is a consequence of the local nature of differential geometry and one should be careful for this caveat in applications.

\newpage
\section{The Gumbel distributions}
The last example in this chapter considers the Gumbel distributions \cite{Gumbel} as models for empirical probability distributions. As such this example is very similar to the first one. The biggest difference here is that the chosen statistical model is not an exponential family. This means that the construction of the geometrical quantities of the data set model formalism will fail. In this way, it will be confirmed that the model is indeed not an exponential family.\\
The Gumbel distributions form a two-parameter family of probability densities which are often employed to model the distribution of the minimal or maximal values of a number of statistical samples. The Gumbel distributions, whose domain is the set of real numbers, can be written as
\begin{align*}
p_{\alpha,\mu}(x) &= \exp\{ \ln\alpha - \alpha(x-\mu) - e^{-\alpha(x-\mu)}\}.
\end{align*}
The parameters are $\alpha$ and $\mu$. The first one of these, is a strictly positive parameter determining the shape of the distribution. For reasons of notational convenience, the choice for $\alpha$ differs from the one in the literature, where $\beta = \alpha^{-1}$ is more commonly used. The effect of varying the parameter $\alpha$, for $\mu=0$, is sketched in the illustration directly below. It can be seen that larger values of $\alpha$ indicate a sharper distribution.
\begin{center}
\begin{tikzpicture}[scale=1.5]
\path[draw,->] (-2.2,0)--(3.1,0);
\path[draw,->] (0,-0.075)--(0,3);

\path[draw] (3.3,-0.25) node {$x$};
\path[draw] (-0.5,3.2) node {$p(x)$};

\path[draw] (1,0.075)--(1,-0.075);
\path[draw] (2,0.075)--(2,-0.075);
\path[draw] (-1,0.075)--(-1,-0.075);
\path[draw] (-2,0.075)--(-2,-0.075);

\path[draw] (-2.1,-0.25) node {$-2$};
\path[draw] (-1.1,-0.25) node {$-1$};
\path[draw] (0,-0.25) node {$0$};
\path[draw] (1,-0.25) node {$1$};
\path[draw] (2,-0.25) node {$2$};

\path[draw,domain=-2:3,samples=200,dashed]  plot (\x,{2.5*3*exp(-3*\x-exp(-3*\x)) });
\path[draw,domain=-2:3,samples=200]  plot (\x,{2.5*2.2*exp(-2.2*\x-exp(-2.2*\x)) });
\path[draw,domain=-2:3,samples=200,dotted]  plot (\x,{2.5*1.4*exp(-1.4*\x-exp(-1.4*\x)) });

\draw[color=white] (-2,2)--(-3,2);%fengshui

\path[draw] (3,3) node {\small $\alpha=1{,}4$};
\path[draw] (3,2.7) node {\small $\alpha=2{,}2$};
\path[draw] (3,2.4) node {\small $\alpha=3{,}0$};
\path[draw,dotted] (3.6,3)--(4,3);
\path[draw] (3.6,2.7)--(4,2.7);
\path[draw,dashed] (3.6,2.4)--(4,2.4);
\end{tikzpicture}
\end{center}
The parameter $\mu$ represents the mode of the distribution---and not the mean, as is common for Guassian distributions. The mean of the Gumbel distribution equals $\mu + \alpha^{-1}\gamma$, where $\gamma$ is the Euler-Mascheroni constant. The effect of changing $\mu$ would be to perform a horizontal shift of the distribution over a distance $\mu$ and so no distributions with non-zero $\mu$-values are included in the illustration.\\
It is tempting to choose as the set $\mathbb{X}$ again all possible distributions over the real numbers. Unfortunately this is untenable as this would mean certain expectation values appearing in the computation below will not exist. This is not a problem specific to the data set model formalism and so it will be ignored by assuming that all distributions $p\in\mathbb{X}$ are such that the expectation values mentioned below do indeed exist.\\
Due to the appearance of the exponential functions in this expression, an obvious choice for the divergence is again that of Kullback and Leibler. This divergence, with a suitable probability distribution $p$ as its first argument and a Gumbel distribution as its second argument, takes the form
\begin{align}
D(p||p_{\alpha,\mu}) &= - S(p) - \int_X p(x) \ln p_{\alpha,\mu}(x) \intdee x\nonumber\\
&= - S(p) - \int_X p(x) \left[ \ln\alpha - \alpha(x-\mu) - e^{-\alpha(x-\mu)}\right]\dee x,\label{eq: DGumb}
\end{align}
where $S$ represents the Shannon entropy (\ref{eq: Shannon}). In order to construct the differential geometry induced by this divergence on the manifold of Gumbel distributions, the derivatives with respect to the parameters are necessary. They are given by the expressions
\begin{align*}
\partial_\alpha D(p||p_{\alpha,\mu}) &= - \frac{1}{\alpha} + \mathbb{E}_p[(x-\mu)] - \mathbb{E}_p[ e^{-\alpha(x-\mu)}(x-\mu)],\\
\partial_\mu D(p||p_{\alpha,\mu}) &= -\alpha + \alpha \mathbb{E}_p[e^{-\alpha(x-\mu)}].
\end{align*}
This means a distribution $p$ is in the fibre of the Gumbel distribution $p_{\alpha,\mu}$ if and only if it simultaneously satisfies the equations
\begin{align}
\mathbb{E}_p[e^{-\alpha(x-\mu)}] = 1 \quad \text{and} \quad \mathbb{E}_p[ \alpha (x-\mu)\{ 1 - e^{-\alpha(x-\mu)}\}] = 1. \label{eq: cond}
\end{align}
Even though this is not a trivial computation, it can be verified that when $p=p_{\alpha,\mu}$ these equations are indeed satisfied. The second derivatives of the divergence (\ref{eq: DGumb}) with respect to the parameters are given by the expressions
\begin{align*}
\partial^2_\alpha D(p||p_{\alpha,\mu}) &= \frac{1}{\alpha^2} + \mathbb{E}_p[ (x-\mu)^2 e^{-\alpha(x-\mu)}],\\
\partial_\alpha\partial_\mu D(p||p_{\alpha,\mu}) &= -1 + \mathbb{E}_p[e^{-\alpha(x-\mu)}] - \alpha \mathbb{E}_p[(x-\mu)e^{-\alpha(x-\mu)}],\\
\partial^2_\mu D(p||p_{\alpha,\mu}) &= \alpha^2 \mathbb{E}_p[ e^{-\alpha(x-\mu)}].
\end{align*}
It is clear that these expressions in general depend on the chosen probability density $p$. To define a metric, it is sufficient that these expressions become $p$-independent when $p$ is in the fibre of $p_{\alpha,\mu}$. However, this is not the case, as is shown by a counterexample. It is only required to show that the first of the derivatives,
\begin{align*}
\partial^2_\alpha D(p||p_{\alpha,\mu}) &= \frac{1}{\alpha^2} + \mathbb{E}_p[ (x-\mu)^2 e^{-\alpha(x-\mu)}],
\end{align*}
cannot be expressed without reference to $p$, even when using the conditions (\ref{eq: cond}) imposed by the vanishing of the first derivatives of the divergence. This shows there exists no metric tensor as defined in the general theoretical outset of the previous chapter. An obvious choice of distribution to provide a counterexample, be it in the largest part for computational convenience, is the family of exponential distributions with density functions
\begin{align*}
p_\lambda(x) = \left\{ \begin{array}{r c l} \lambda e^{-\lambda x} & &x \geqslant 0,\\
 0 & \phantom{if}& x < 0.\end{array}\right.
\end{align*}
Fixing a value of $\lambda$, it is possible to determine which fibre of the manifold of Gumbel distributions contains $p_\lambda$. This computation comprises most of the work needed in demonstrating the counterexample.\\
The equations (\ref{eq: cond}) impose relations between the parameters $\alpha$ and $\mu$ on one hand and $\lambda$ on the other, in particular these are
\begin{align}
1 &= \int_0^{\infty} \lambda e^{-\lambda x - \alpha (x-\mu) } \dee x \nonumber\\
&= \frac{\lambda e^{\alpha\mu}}{\lambda + \alpha}\label{eq: condbis1}
\end{align}
and 
\begin{align}
1 &= \alpha\lambda \int_0^\infty (x-\mu) \{1 - e^{-\alpha (x-\mu)}\} e^{-\lambda x}\dee x\nonumber\\
&= \alpha\lambda \int_0^\infty (x-\mu) e^{-\lambda x} \dee x - \alpha \lambda e^{\alpha \mu} \int_0^\infty (x-\mu) e^{-(\alpha+\lambda)x} \dee x\nonumber\\
&= \frac{\alpha}{\lambda} - \alpha \mu - \frac{\alpha \lambda e^{\alpha\mu}}{(\alpha+\lambda)^2}  + \frac{\alpha \lambda \mu e^{\alpha \mu}}{\alpha + \lambda}.\label{eq: condbis1bis}
\end{align}
This relation can be simplified considerably by using (\ref{eq: condbis1}) to yield
\begin{align*}
1 &= \frac{\alpha}{\lambda} - \alpha \mu - \frac{\alpha}{\alpha + \lambda} + \alpha \mu\\
%&= \frac{\alpha(\alpha+\lambda) - \lambda \alpha}{\lambda (\alpha+\lambda)}\\
&= \frac{\alpha^2}{\lambda(\alpha + \lambda)}.
\end{align*}
Essentially a quadratic equation in $\alpha$, this condition can be solved for its positive root
\begin{align*}
\alpha &= \frac{\lambda +\sqrt{\lambda^2 + 4\lambda^2}}{2} =  \frac{1 + \sqrt{5}}{2} \lambda.
\end{align*}
Negative $\alpha$-values such as the one obtained by choosing the minus sign are excluded by definition of the Gumbel distributions and so this solution does not need to be given any attention. By substituting this result in relation (\ref{eq: condbis1}), an expression for $\mu$ in terms of $\lambda$ can be obtained as
\begin{align*}
\mu &= \frac{1}{\alpha} \ln \left( \frac{\lambda + \alpha}{\lambda}\right) = \frac{1}{\lambda} \cdot \frac{2}{1+\sqrt{5}} \ln \left( \frac{3 + \sqrt{5}}{2}\right).
\end{align*}
Hence the values of the parameters $\alpha(\lambda)$ and $\mu(\lambda)$ of a Gumbel distribution upon which $p_\lambda$ is mapped are known. It is now possible to show that the second derivative of the divergence (\ref{eq: DGumb}) with respect to $\alpha$ is not constant on the fibres. This derivative contains the term
\begin{align*}
\mathbb{E}_{p_\lambda}[(x-\mu)^2 e^{-\alpha(x-\mu)} ] &= \int_{-\infty}^{+\infty} p_\lambda(x) (x-\mu)^2 e^{-\alpha(x-\mu)} \dee x\\
%&= \lambda \int_0^\infty (x-\mu)^2 e^{-\lambda x - \alpha(x-\mu)} \dee x\\
&= \lambda e^{\alpha \mu} \int_0^\infty (x-\mu)^2 e^{-(\lambda + \alpha)x} \dee x\\
&= \lambda e^{\alpha \mu} \left[ \frac{2}{(\lambda + \alpha)^3} -  \frac{2\mu}{(\lambda + \alpha)^2} +  \frac{\mu^2}{\lambda+\alpha}\right]\\
&\approx \frac{0,5}{\alpha^2} \qquad (\text{using } \alpha(\lambda) \text{ and }\mu(\lambda)).
\end{align*}
Since the Gumbel distribution $p_{\alpha,\mu}$ is an element of its own fibre, it must yield this same result when computing this expectation value if the metric tensor is to exist. A direct computation shows that
\begin{align*}
\mathbb{E}_{p_{\alpha,\mu}}[(x-\mu)^2 e^{-\alpha(x-\mu)} ] &= \alpha \int_{-\infty}^{\infty} (x-\mu)^2 \exp\{- 2\alpha(x-\mu) - e^{-\alpha(x-\mu)} \}\dee x\\
&= \frac{1}{\alpha^2} \int_{-\infty}^{\infty} u^2 e^{-2u - e^{-u}} \dee u\\
&= \frac{1}{\alpha^2} \int_0^\infty (\ln t)^2 t e^{-t} \dee t\\
&= \frac{1}{\alpha^2}\left(\gamma^2 - 2\gamma + \frac{\pi^2}{6} \right)\\
&\approx \frac{0,82}{\alpha^2}.
\end{align*}
It is concluded that---for this data set model---the suggested expression for the metric is not constant along fibres. This does not mean the manifold cannot be endowed with a metric tensor at all. For instance it remains possible to use as a metric tensor the matrix of second derivatives of the divergence restricted to $\mathbb{M}\times \mathbb{M}$, evaluated on the diagonal. Using the equations (\ref{eq: cond}) to simplify the expressions, this would yield
\begin{align*}
g(\alpha,\mu) \approx \left(\begin{array}{c c}
\frac{1,82}{\alpha^2} & \gamma \\
\gamma & \alpha^2
\end{array}\right).
\end{align*}
This metric tensor only contains information about the manifold of Gumbel distributions and, as shown above, is not compatible with the structure of the fibres of the full data set model.\\
It is instructive to see if a connection could exist for this model. As the metric is not well-defined, there is no Hessian structure. The spherical submanifold of the von Mises-Fisher distribution shows that such structure does not need to be present in order to define an affine connection, however. \\
Once again, the first step in looking for a connection is to single out curves through the set $\mathbb{X}$ of distributions. The derivatives of the divergence are proportional to the expectation values
\begin{align*}
\mathbb{E}_p[e^{-\alpha(x-\mu)}] \quad\text{and}\quad \mathbb{E}_p[(x-\mu)\{ 1 - e^{-\alpha(x-\mu)}\}]
\end{align*}
and thus it is convenient to take these values as the parameters for the curves appearing in the strong version of Condition 6. Indeed, differentiating the derivatives of the divergence with respect to these parameters would yield the desired Kronecker-delta as a result. However, when trying to express the second derivatives of the divergence as a function of these parameters, in order to perform the necessary differentiation, a problem arises. Again it is the second derivative with respect to $\alpha$ which causes difficulties. As is stated above, this quantity is given by the expression
\begin{align*}
\partial^2_\alpha D(p||p_{\alpha,\mu}) &= \frac{1}{\alpha^2}+ \mathbb{E}_p[(x-\mu)^2 e^{-\alpha(x-\mu)}].
\end{align*}
Since this quantity is not constant within fibres, different choices of curves $X^k$ satisfying all required conditions are still possible but which yield different results for the connection coefficients. Just as is the case with the metric, a certain choice of connection could be made but there exists no choice compatible with the fibre structure.

\newpage

\chapter{Conclusion and outlook}
\section{Conclusion}
This doctoral research seeks to contribute to the formulation of an abstract theory of information that is not based on probability. Instead, the mathematical foundation is provided by differential geometry. The resulting framework is called the data set model formalism.\\
Pursuing information theory without probability may seem counter-intuitive to many readers, not in small part due to the way this discipline is traditionally treated in textbooks. It is nevertheless an idea which has been advocated by a number of authors in the past. Recent interest in this question is motivated by numerous attempts in the literature to base quantum theory on informational principles. This endeavour may be facilitated by the availability of a sufficiently general and abstract perspective on information theory. Also advances in experimental quantum physics could in time vindicate the development of a formalism such as the one presented in this dissertation. These new results enabled by the recently mastered ability to perform weak measurements, may in time argue in favour of---or even demand---the formulation of a novel description of quantum information theory. In that case, having at hand a mathematical foundation such as the one provided by the data set model formalism could be beneficial to the involved research community.\\
The inspiration for this approach is found in information geometry, a field concerned with the description of probability theory and statistical models through differential geometry. The framework developed here is a proper generalisation of information geometry in the sense that the latter can be re-derived as a special case of the former. The most obvious way in which the abstraction is exercised is by dropping the demand that the data and the models are probability distributions. A related and in all likelihood more important feature is that the models need not be a subset of the data and may even be qualitatively different mathematical objects. Despite these differences, the construction of a geometric structure strongly reminiscent of information geometry is performed in this dissertation.\\
The geometric structure of the data set model formalism is derived from a generalised divergence function which quantifies how well a data set is described by a given model point. More in particular, a Riemannian metric and an affine connection can be constructed under suitable conditions. The metric is a generalisation of the Fisher information metric and can be employed to express the sensitivity of the inferred parameters to measurable functions of the data. Consequently, when the data set model under consideration is an exponential family of probability distributions, the Cauchy-Schwarz inequality for the metric tensor reduces to the well-known Cram\'er-Rao bound. The affine connection is flat and torsionless and it is a generalisation of the exponential connection taking up a prominent role in information geometry. Of central importance to the data set model geometry is a Hessian structure, where the metric can be written as the Hessian of a generalised Massieu function. The point of view that the formalism is a proper generalisation of the existing literature is further reinforced by discussing a Pythagorean theorem, which provides the link required to derive the geometry of divergence functions as it has been developed by other researchers from the data set model geometry.\\
The theoretical discussion is concluded by establishing a straightforward technique to determine whether or not a statistical model belongs to the exponential family and to determine the canonical parameters. While these notions are not expected to be original, they can be quite useful for someone interested in exponential families and the new formalism allows them to be applied in an elegant fashion.\\
The last chapter of this dissertation is devoted to working out a number of examples. Each of these illustrates one or more prominent aspects of the data set model formalism touched upon in the preceding paragraphs. The examples vary from familiar probability theory and information geometry to a linear regression method. A particular example interesting to physicists is found third in that chapter. It concerns systems of non-interacting bosonic particles and it allows, given experimentally observed occupation numbers of the energy levels of that system, to find the grand canonical distribution function which best describes the state of the system (at the time of measurement). Those readers more attracted to differential or information geometry may have their interest sparked in particular by the two examples in the fourth section. Two statistical models, both submanifolds of the von Mises-Fisher distributions, are studied there and have their data set model geometry constructed. The metric tensor and the affine connection obtained in this way are exactly the ones that would be expected if the submanifolds had been embedded in Euclidean space. This is a remarkable result since the containing space is not endowed with a Euclidean metric.
\newpage
\section{Outlook}
A number of interesting questions regarding data set models remain open. One question is whether a natural construction for a one-parameter family of affine connections exists for data set models, as is the case in information geometry. A better understanding of the remarkable results of the von Mises-Fisher examples may yield insights in properties of the geometry of data set models. Should these results hold in general, they could serve to detect statistical models which are submanifolds of exponential families. It may also enable a further expansion of the data set model formalism in order to treat also models with a connection which exhibits curvature.\\
From the perspective of applications, the data set model approach may become a fruitful technique in quantum information theory. Recent---still unpublished---work shows that the work of Petz on positive-operator valued measures (see for instance \cite{Petz}) is also encompassed by the formalism. It even offers a ground for the belief that a further extension and a mathematical simplification of that research may be possible. Whereas I personally hold the opinion that the preceding suggestion is the application which looks the most promising---at least at the time of writing---the great flexibility offered by the data set model opens up the possibility for plenty of applications. As the examples of linear regression and of the non-interacting bosons show, the data set model formalism is very suitable to function as a mathematical framework for very general fitting procedures. For this reason it could also be useful to researchers in machine learning. Their discipline is concerned with a highly varied collection of different types of what are essentially modelling problems. In particular, a significant part of their field makes use of probability theory whereas an equally important part does not. This community is thus familiar with many powerful techniques which are employed in the study of the former type, which could be extended to the latter type using the data set model formalism as a unifying framework.

\newpage
%\null\thispagestyle{empty}
\begin{center}{\color{white}Page intentionally left blank}\end{center}
\newpage

\chapter*{Curriculum Vitae and Academic Overview}
\addcontentsline{toc}{chapter}{Curriculum Vitae and Academic Overview}
%vspace{0.5cm}
\textbf{\Large Ben Anthonis}\\
\newline
Born on April $2^{\text{nd}}$ 1986, Lier (Belgium)\\
\\
\textbf{\large Contact Information}\\
Departement Fysica\\
Universiteit Antwerpen (Campus Drie Eiken)\\
Universiteitsplein 1\\
B-2610 Wilrijk\\
Tel: +32 (0)3 265 24 76\\
e-mail: ben.anthonis@uantwerpen.be\\
\\
\textbf{\large Education}
\begin{itemize}
\item 2004--2007: Bachelor in Physics at Universiteit Antwerpen (Belgium)
\item 2007--2009: Master in Physics at Universiteit Antwerpen (Belgium)
\item 2009--2014: PhD in science and teaching assistant at Universiteit Antwerpen (Belgium)
\end{itemize}
\textbf{\large List of publications}
\begin{itemize}
\item Verhulst T., \underline{Anthonis B.} and Naudts J.: \emph{Analysis of the N=4 Hubbard ring using a counting operator}, Phys.\ Lett.\ A \textbf{373} p.2109--2113 (2009)
\item Naudts J., Verhulst T.\ and \underline{Anthonis B.}: \emph{Counting operator analysis of the discrete spectrum of some model Hamiltonians}, Phys.\ Lett.\ A \textbf{373}, p.3419--3422 (2009)
\item Naudts J.\ and \underline{Anthonis B.}: \emph{Data set models and exponential families in statistical physics and beyond}, Mod.\ Phys.\ Lett.\ B \textbf{26}, 1250062 (2012)
\item Naudts J.\ and \underline{Anthonis B.}: \emph{The Exponential Family in Abstract Information Theory}, published in Geometric Science of Information, Lecture Notes in Computer Science \textbf{8085} (eds.: Nielsen F.\ and Barbaresco F.), Springer Berlin Heidelberg, p.265--272 (2013)
\end{itemize}
\newpage
\textbf{\large Educational assignments}
\begin{itemize}
\item 2009--2014: Exercise classes for `Fysica I' to the first-year students in Bio-engineering and Chemistry,
\item 2009--2014: Exercise classes for `Fysica II' to the second-year students in Bio-engineering and Chemistry,
\item 2012--2013: Exercise classes for `Fysica voor biomedisch onderzoek' to the first-year students in Biomedical Sciences,
\item Co-supervision of the bachelor theses of Quinten Collier and Amy De Schutter (2009--2010), Inge van Hooydonk (2010--2011) and Romeo Van Snick (2011--2012),
\item Co-supervision of the master theses of Geert Van Wauwe (2011--2012) and Inge van Hooydonk (2012--2013),
\item 2011--2014: Ombudsperson for the Bachelor in Physics program at Universiteit Antwerpen.
\end{itemize}

\end{document}